\documentclass[usenatbib]{mn2e}
\usepackage[centering,totalwidth=480pt,totalheight=680pt]{geometry}
\usepackage{psfig}
\usepackage[dvips]{graphicx}

\usepackage{amsmath}

\newcommand{\kms}{\rm\ km\ s^{-1}}
\newcommand{\LCDM}{$\Lambda$CDM}
\newcommand{\msun}{\mbox{${\rm M}_{\odot}$}}
\newcommand{\mvir}{\mbox{$M_{\rm vir}$}}
\newcommand{\rvir}{\mbox{$r_{\rm vir}$}}
\newcommand{\rcool}{\mbox{$r_{\rm cool}$}}
\newcommand{\tcool}{\mbox{$t_{\rm cool}$}}
\newcommand{\vvir}{\mbox{$V_{\rm vir}$}}
\newcommand{\cnfw}{\mbox{$c_{\rm NFW}$}}
\newcommand{\mbh}{\mbox{$M_{\rm BH}$}}
\newcommand{\etabh}{\mbox{$\eta_{\rm rad}$}}
\newcommand{\HI}{\mbox{${\rm H}_{\rm I}$}}
\newcommand{\Htwo}{\mbox{${\rm H}_{2}$}}

\title[Co-evolution of Galaxies, Black Holes, and AGN]{A Semi-Analytic Model for the Co-evolution of Galaxies, Black Holes, and Active Galactic Nuclei}

\author[Somerville et al.] {
\parbox[t]{\textwidth}{ 
Rachel S. Somerville$^{1}$,\thanks{E-mail: somerville@mpia.de}
Philip F. Hopkins$^{2}$, 
Thomas J. Cox$^{2}$, 
Brant E. Robertson$^{3,4,5}$,
Lars Hernquist$^{2}$
}
\vspace*{6pt} \\
$^1$ Max-Planck-Institut f\"{u}r Astronomie, K\"{o}nigstuhl 17, Heidelberg 
D-69117 Germany\\
$^2$ Harvard-Smithsonian Center for Astrophysics, 60 Garden Street, Cambridge, MA 02138\\
$^3$ Kavli Institute for Cosmological Physics, and the Department of  
Astronomy and Astrophysics, University of Chicago, \\
933 East 56th Street, Chicago, IL, 60637, USA\\
$^4$  Enrico Fermi Institute, 5640 South Ellis Avenue, Chicago, IL,  
60637, USA\\
$^5$ Spitzer Fellow
\\
}

\begin{document}

\newcommand{\refs}{{\bf (ref)}}
\newcommand{\notice}[1]{{\it [{#1}]}}
\newcommand{\plotone}[1]
           {\centering \leavevmode \psfig{file=#1,width=\columnwidth,clip=}}
\newcommand{\plotside}[1]
           {\centering \leavevmode \psfig{file=#1,width=\textwidth,clip=}}
\newcommand{\plottwo}[2]
           {\centering \leavevmode \psfig{file=#1,width=\columnwidth,clip=}
                            \hfill \psfig{file=#2,width=\columnwidth,clip=}}
\newcommand{\plotfull}[1]
           {\centering \leavevmode \psfig{file=#1,width=\textwidth,clip=}}
\newcommand{\plotwhole}[1]
           {\centering \leavevmode
             \psfig{file=#1,width=\textwidth,height=8.2in,clip=}}

\def\lesssim{\lower.5ex\hbox{$\; \buildrel < \over \sim \;$}}
\def\gtrsim{\lower.5ex\hbox{$\; \buildrel > \over \sim \;$}}

\maketitle

\begin{abstract}
We present a new semi-analytic model that self-consistently traces the
growth of supermassive black holes (BH) and their host galaxies within
the context of the \LCDM\ cosmological framework. In our model, the
energy emitted by accreting black holes regulates the growth of the
black holes themselves, drives galactic scale winds that can remove
cold gas from galaxies, and produces powerful jets that heat the hot
gas atmospheres surrounding groups and clusters. We present a
comprehensive comparison of our model predictions with observational
measurements of key physical properties of low-redshift galaxies, such
as cold gas fractions, stellar metallicities and ages, and specific
star formation rates.  We find that our new models successfully
reproduce the exponential cutoff in the stellar mass function and the
stellar and cold gas mass densities at $z\sim0$, and predict that star
formation should be largely, but not entirely, quenched in massive
galaxies at the present day. We also find that our model of
self-regulated BH growth naturally reproduces the observed relation
between BH mass and bulge mass.  We explore the global formation
history of galaxies in our models, presenting predictions for the
cosmic histories of star formation, stellar mass assembly, cold gas,
and metals. We find that models assuming the ``concordance''
\LCDM\ cosmology overproduce star formation and stellar mass at high
redshift ($z\gtrsim 2$). A model with less small-scale power predicts
less star formation at high redshift, and excellent agreement with the
observed stellar mass assembly history, but may have difficulty
accounting for the cold gas in quasar absorption systems at high
redshift ($z\sim 3$--4).
\end{abstract}


\section{Introduction}
\label{sec:intro}

It is now well-established that the Cold Dark Matter paradigm for
structure formation \citep{bfpr:84}, in its modern, dark
energy-dominated (\LCDM) incarnation, provides a remarkably successful
paradigm for interpreting a wide variety of observations, from the
cosmic microwave background fluctuations at $z\sim 1000$
\citep{spergel:03,spergel:07}, to the large-scale clustering of
galaxies at $z\sim 0$
\citep{percival:02,tegmark:04,eisenstein:05}. Other successful
predictions of the \LCDM\ paradigm include the cosmic shear field as
measured by weak gravitational lensing
\citep{heymans:05,bacon:05,hoekstra:06}, the small-scale power
spectrum as probed by the Lyman-alpha forest
\citep{desjacques:05,jena:05}, and the number densities of
\citep{borgani:01}, and baryon fractions within, galaxy clusters
\citep{allen:04,white:93}.

\LCDM\ as a paradigm for understanding and simulating galaxy formation
has had more mixed success. In this picture, as originally proposed by
\citet{wr:78} and \citet{bfpr:84}, galaxies form when gas cools and
condenses at the centers of dark-matter dominated potential wells, or
``halos''. More detailed calculations, using semi-analytic and
numerical simulations of galaxy formation, have shown that this
framework does provide a promising qualitative understanding of many
features of galaxies and their evolution
\citep[e.g.,][]{kwg:93,cafnz:94,sp:99,kauffmann:99,cole:00,spf:01,baugh:06}. However,
it has been clear for at least a decade now that there is a
fundamental tension between certain basic predictions of the
\LCDM-based galaxy formation paradigm and some of the most fundamental
observable properties of galaxies. In this paper we focus on two
interconnected, but possibly distinct, problems: 1) the
``overcooling'' or ``massive galaxy'' problem and 2) the star
formation ``quenching'' problem.

The first problem is manifested by the fact that both semi-analytic
and numerical simulations predict that a large fraction of the
available baryons in the Universe rapidly cools and condenses, in
conflict with observations which indicate that only about $\sim 5$--10
\% of the baryons are in the form of cold gas and stars
\citep{bell:03a,fukugita_peebles:04}.  This is due to the fact that
gas at the densities and temperatures characteristic of dark matter
halos is expected to cool rapidly, and is related to the classical
``cooling flow'' problem
\citep{fabian_nulsen:77,cowie_binney:77,mathews:78}. Direct
observations of the X-ray properties of hot gas in clusters similarly
imply that this gas should have short cooling times, particularly near
the center of the cluster, but the condensations of stars and cold gas
at the centers of these clusters are much smaller than would be
expected if the hot gas had been cooling so efficiently over the
lifetime of the cluster \citep[for reviews
see][]{fabian:94,peterson:06}. Moreover, X-ray spectroscopy shows that
very little gas is cooling below a temperature of about one-third of
the virial temperature of the cluster \citep{peterson:03}.

A further difficulty is that there is a fundamental mismatch between
the \emph{shape} of the dark matter halo mass function and that of the
observed mass function of cold baryons (cold gas and stars) in
galaxies \citep{sp:99,benson:03}. The galaxy mass function has a sharp
exponential cutoff above a mass of about a few times $10^{10} \msun$,
while the halo mass function has a shallower, power-law cutoff at much
higher mass ($\sim {\rm few} \times 10^{13}\msun$). There is also a
mismatch at the small-mass end, as the mass function of dark matter
halos is much steeper than that of galaxies. If we are to assume that
each dark matter halo hosts a galaxy, this then implies that the ratio
between the luminosity or stellar mass of a galaxy and the mass of its
dark matter halo varies strongly and non-monotonically with halo mass
\citep{kravtsov:04a,conroy:06,wang:06,moster:08}, such that ``galaxy
formation'' is much more inefficient in both small mass and large mass
halos, with a peak in efficiency close to the mass of our own Galaxy,
$\sim 10^{12} \msun$. On very small mass scales (below halo velocities
of $\sim 30-50 \kms$), the collapse and cooling of baryons may be
suppressed by the presence of a photoionizing background
\citep{efstathiou:92,thoul:96,quinn:96}. For larger mass halos (up to
$V_{\rm vir} \simeq 150-200 \kms$), the standard assumption is that
winds driven by massive stars and supernovae are able to heat and
expell gas, resulting in low baryon fractions in small mass halos
\citep{wr:78,dekel_silk:86,white_frenk:91}.  However, stellar feedback
probably cannot provide a viable solution to the overcooling problem
in massive halos \citep{benson:03}: stars do not produce enough energy
to expell gas from these large potential wells; and the massive, early
type galaxies in which the energy source is needed have predominantly
old stellar populations and little or no ongoing or recent star
formation. Other solutions, like thermal conduction, have been
explored, but probably do not provide a full solution
\citep{benson:03,voigt:04}.

The second problem is related to the correlation of galaxy structural
properties (morphology) and spectrophotometric properties (stellar
populations) with stellar mass. The presence of such a correlation has
long been known, in the sense that more massive galaxies tend to be
predominantly spheroid-dominated, with red colors, old stellar
populations, low gas fractions, and little recent star formation,
while low-mass galaxies tend to be disk-dominated and gas-rich, with
blue colors and ongoing star formation
\citep[e.g.][]{roberts:94}. More recently, with the advent of large
galaxy surveys such as the Sloan Digital Sky Survey (SDSS), we have
learned that the galaxy color distribution (and that of other related
properties) is strongly \emph{bimodal} \citep[e.g.][]{baldry:04}, and
that the transition in galaxy properties from star-forming disks to
``dead'' spheroids occurs rather sharply, at a characteristic stellar
mass of $\sim 3 \times 10^{10} \msun$
\citep{kauffmann:03a,brinchmann:04}. In contrast, the ``standard''
\LCDM-based galaxy formation models predict that massive halos have
been assembled relatively recently, and should contain an ample supply
of new fuel for star formation. These models predict an
\emph{inverted} color-mass and morphology-mass relation (massive
galaxies tend to be blue and disk dominated, rather than red and
spheroid dominated) and no sharp transition or strong
bimodality. Thus, the standard paradigm of galaxy formation does not
provide a physical explanation for the ``special'' mass scale (a halo
mass of $\sim 10^{12} \msun$, or a stellar mass of $\sim 3\times
10^{10} \msun$) which marks both the peak of galaxy formation
efficiency and the transition in galaxy properties seen in
observations.

Several pieces of observational evidence provide clues to the solution
to these problems. It is now widely believed that every
spheroid-dominated galaxy hosts a nuclear supermassive black hole
(SMBH), and that the mass of the SMBH is tightly correlated with the
luminosity, mass, and velocity dispersion of the stellar spheroid
\citep{kormendy_richstone:95,magorrian:98,ferrarese:00,gebhardt:00,marconi:03,haering:04}.
These correlations may be seen as a kind of ``fossil'' evidence that
black holes were responsible for regulating the growth of galaxies or
vice versa. This also implies that the most massive galaxies, where
quenching is observed to be the most efficient, host the largest black
holes, and therefore the available energy budget is greatest in
precisely the systems where it is needed, in contrast to the case of
stellar feedback. The integrated energy released over the lifetime of
a SMBH ($\simeq 10^{60}-10^{62}$ erg) is clearly very significant
compared with galaxy binding energies \citep{silk_rees:98}. In view of
these facts, it seems almost inconceivable that AGN feedback is
\emph{not} important in shaping galaxy properties.

However, in order to build a complete, self-consistent machinery to
describe the formation and growth of black holes within the framework
of a cosmological galaxy formation model, and to attempt to treat the
impact of the energy feedback from black holes in this context, we
need to address several basic questions: 1) When, where, and with what
masses do seed black holes form? 2) What triggers black hole
accretion, what determines the efficiency of this accretion, and what
shuts it off?  3) In what form is the energy produced by the black
hole released, and how does this energy couple with the host galaxy
and its surroundings? In order to address some of these questions, we
first identify two modes of AGN activity which have different
observational manifestations, probably correspond to different
accretion mechanisms, and have different physical channels of
interaction with galaxies.

\subsection{The Bright Mode of Black Hole Growth}
\label{sec:bright}

Classical luminous quasars and their less powerful cousins,
optical or X-ray bright AGN, radiate at a significant fraction of
their Eddington limit \citep[$L \sim (0.1-1) L_{\rm
Edd}$;][]{vestergaard:04,kollmeier:06}, and are believed to be fed by
optically thick, geometrically thin accretion disks
\citep{shakura:73}. We will refer to this mode of accretion as the
``bright mode'' because of its relatively high radiative efficiency
(with a fraction $\etabh \sim 0.1-0.3$ of the accreted mass converted
to radiation). The observed space density of these quasars and AGN is
low compared to that of galaxies, implying that if most galaxies
indeed host a SMBH, this ``bright mode'' of accretion is only ``on'' a
relatively small fraction of the time. Constraints from quasar
clustering and variability imply that quasar lifetimes must be
$\lesssim 10^{8.5}$ yr \citep{martini:01,martini:03}. These short
timescales combined with the large observed luminosities immediately
imply that fueling these objects requires funneling a quantity of gas
comparable to the entire supply of a large galaxy ($\sim
10^{9}-10^{10} \msun$) into the central regions on a timescale of
order the dynamical time ($\sim {\rm few} \times 10^7-10^8$ yr).

These considerations alone lead one to consider galaxy-galaxy mergers
as a promising mechanism for triggering this efficient accretion onto
nuclear black holes. The observational association of mergers with
enhanced star formation, particularly with the most violent observed
episodes of star formation exhibited by Ultra Luminous Infrared
Galaxies (ULIRGS), is well-established
\citep{sanders:96,farrah:01,colina:01,barton:00,woods:06,woods:07,barton:07,lin:07,li:07}.
Moreover, numerical simulations have shown that tidal torques during
galaxy mergers can drive the rapid inflows of gas that are needed to
fuel both the intense starbursts and rapid black hole accretion
associated with ULIRGS and quasars
\citep{hernquist:89,barnes:92,barnes:96,mihos:94,mihos:96,springel:05a,dimatteo:05}.
As well, it seems that if one can probe sufficiently deep to study the
SED beneath the glare of the quasar, one always uncovers evidence of
young stellar populations indicative of a recent starburst
\citep{brotherton:99,canalizo:01,kauffmann:03b,jahnke:04,sanchez:04,vandenberk:06}.
Near-equal mass (major) mergers also have the attractive feature that
they scramble stars from circular to random orbits, leading to
morphological transformation from disk to spheroid
\citep{toomre:72,barnes:88,barnes:92,hernquist:92,hernquist:93}. If
spheroids and black holes both arise from violent mergers, this
provides a possible explanation for why black hole properties always
seem to be closely associated with the spheroidal components of
galaxies.

What impact does the energy associated with this rapid, bright mode
growth have on the galaxy and on the growth of the black hole itself?
Long thought to be associated only with a small subset of objects
(e.g. Broad Absorption Line (BAL) quasars), high-velocity winds have
been detected in a variety of different types of quasar systems
\citep{dekool:01,pounds:03,chartas:03,pounds:06}, and are now believed
to be quite ubiquitous \citep{ganguly:08}. However, their impact on
the host galaxy remains unclear, as the mass outflow rates of these
winds are difficult to constrain \citep[though
see][]{steenbrugge:05,chartas:07,krongold:07}.  Recently, numerical
simulations of galaxy mergers including black hole growth found that
depositing even a small fraction ($\sim 5$ \%) of the energy radiated
by the BH into the ISM can not only halt the accretion onto the BH,
but can drive large-scale winds \citep{dimatteo:05}. These winds sweep
the galaxy nearly clean of cold gas and halt further star formation,
leaving behind a rapidly reddening, spheroidal remnant
\citep{springel:05b}.

To study how the interplay between feedback from supermassive black
hole accretion and supernovae, galaxy structure, orbital
configuration, and gas dissipation combine to determine the properties
of spheroidal galaxies formed through mergers, hundreds of
hydrodynamical simulations were performed by Robertson et
al. (\citeyear{robertson:06a,robertson:06b,robertson:06c}) and Cox et
al. (\citeyear{cox:06a,cox:06c}) using the methodology presented by
\citet{dimatteo:05} and \citet{springel:05a}. Robertson et al. and Cox
et al. analyzed the merger remnants to study the redshift evolution of
the BH mass-$\sigma$ relation, the Fundamental Plane, phase-space
density, and kinematic properties. This extensive suite has been
supplemented by additional simulations of minor mergers from
\citet{cox:08}. Throughout the rest of this paper, when we refer to
``the merger simulations'', we refer to this suite.

Based on their analysis of these simulations,
\citet{hopkins_cosmo1:07} have outlined an evolutionary sequence from
galaxy-galaxy merger, to dust-enshrouded starburst and buried AGN,
blow-out of the dust and ISM by the quasar- and starburst-driven
winds, to classical (unobscured) quasar, post-starburst galaxy, and
finally ``dead'' elliptical.
\citet{hopkins_bhfpth:07} find that in the merger simulations, the accretion
onto the BH is eventually halted by a pressure-driven outflow. Because
the depth of the spheroid's potential well determines the amount of
momentum necessary to entrain the infalling gas,
\citet{hopkins_bhfpth:07} find that this leads to a ``Black Hole
Fundamental Plane'', a correlation between the final black hole mass
and sets of spheroid structural/dynamical properties (mass, size,
velocity dispersion) similar to the one seen in observations
\citep{hopkins_bhfpobs:07,marconi:03}.

Furthermore, Hopkins et
al. (\citeyear{hopkins:05a,hopkins:05b,hopkins_qsolifetime:05}) have
shown that the self-regulated nature of black hole growth in these
simulations leads to a characteristic form for quasar lightcurves. As
the galaxies near their final coalescence, the accretion rises to
approximately the Eddington rate. After the critical black hole mass
is reached and the outflow phase begins, the accretion rate enters a
power-law decline phase. Although most of their growth occurs in the
near-Eddington phase, quasars spend much of their time in the decline
phase, and this implies that many observed low-luminosity quasars are
actually relatively massive black holes in the last stages of their
slow decline. \citet{hopkins_rss:06} found that when these lightcurves
are convolved with the observed mass function of merging galaxies, the
predicted AGN luminosity function is consistent with
observations. Moreover, Hopkins et al.
(\citeyear{hopkins_qsolifetime:05,hopkins_optx:05,hopkins_unified:06,hopkins_redgal:06})
have shown that this picture reproduces many quasar and galaxy
observables that are difficult to account for with more simplified
assumptions about QSO lightcurves, such as differences in the quasar
luminosity function in different bands and redshifts, Eddington ratio
and column density distributions, the X-ray background spectrum, and
relic red, early type galaxy population colors and distributions.

\subsection{The Radio Mode}

The second mode of AGN activity is much more common, and in general
less dramatic. A fairly large fraction of massive galaxies
(particularly galaxies near the centers of groups and clusters) are
detected at radio wavelengths \citep{best:05, best:07}. Most of these
radio sources do not have emission lines characteristic of classical
optical or X-ray bright quasars \citep{best:05,kauffmann:07}, and
their accretion rates are believed to be a small fraction of the
Eddington rate \citep{rafferty:06}. They are extremely radiatively
inefficient \citep{birzan:04}, and thought to be fuelled by optically
thin, geometrically thick accretion as expected in ADAF and ADIOS
models such as those proposed by \citet{narayan:94} and
\citet{blandford:99}. Because these objects are generally identified
via their radio emission, we refer to this mode of accretion and BH
growth as the ``radio mode'' \citep[following][]{croton:06}.

Although these black holes seem to be inefficient at producing
radiation, they can apparently be quite efficient at producing kinetic
energy in the form of relativistic jets. Intriguingly, the majority of
cooling flow clusters host these active radio galaxies at their
centers \citep{dunn:06,dunn:08}, and X-ray maps reveal that the radio
lobes are often spatially coincident with cavities, thought to be
bubbles filled with relativistic plasma and inflated by the jets
\citep[][and references therein]{mcnamara:07}. The observations of
these bubbles can be used to estimate the work required to inflate
them against the pressure of the hot medium
\citep{birzan:04,rafferty:06,allen:06}, and hence obtain lower limits
on the jet power.

While the idea that radio jets provide a heat source that could
counteract cooling flows has been discussed for many years
\citep[e.g.][]{binney_tabor:95,churazov:02,fabian:03,omma:04a,binney:04},
these observations now make it possible to investigate more
quantitatively whether the heating rates are sufficient to offset the
cooling rates in groups and clusters. Several studies conclude that in
the majority of the systems studied, the AGN heating traced by the
power in the X-ray cavities alone is comparable to or in excess of the
energy being radiated by the cooling gas
\citep{mcnamara:06,rafferty:06,fabian:06,best:06,mcnamara:07,dunn:08}.
Moreover, the net cooling rate is correlated with the observed star
formation rate in the central cD galaxy, indicating that there may be
a self-regulating cycle of heating and cooling \citep{rafferty:06}.

Several other physical processes that could suppress cooling in large
mass halos have been suggested and explored, such as thermal
conduction \citep{benson:03,voigt:04}, multi-phase cooling
\citep{maller_bullock:04}, or heating by sub-structure or clumpy
accretion \citep{khochfar:07,naab:07,dekel_birnboim:08}. While some or
all of these processes may well be important, in this paper we will
investigate whether it is plausible that ``radio mode'' heating alone
can do the job.

\subsection{A Unified Model for Black Hole Activity and AGN Feedback}

All of this begs the question: what determines whether a black hole
accretes in the ``bright mode'' or ``radio mode'' state? An interesting
possible answer comes from an analogy with X-ray binaries
\citep{jester:05,koerding:06}. Observers can watch X-ray binaries in
real time as they transition between two states: the ``low/hard''
state, in which a steady radio jet is present and a hard X-ray
spectrum is observed, and the ``high/soft'' state, in which the jet
dissapears and the X-ray spectrum has a soft, thermal component
\citep{maccarone:03,fender:04}. The transition between the two states
is thought to be connected to the accretion rate itself: the
``high/soft'' state is associated with accretion rates of $\gtrsim
(0.01$--0.02) $\dot{m}_{\rm Edd}$ and the existence of a classical
thin accretion disk, while the ``low/hard'' state is associated with
lower accretion rates and radiatively inefficient ADAF/ADIOS accretion
\citep{fender:04}. 

Recently, \citet{sijacki:07} have applied this idea in cosmological
hydrodynamic simulations, by assuming that when the accretion rate
exceeds a critical value, ``bright mode'' feedback occurs (AGN-driven
winds), while when the accretion rate is lower, ``radio mode''
(mechanical bubble feedback) is implemented. The results of their
simulations appear promising --- they produced black hole and stellar
mass densities in broad agreement with observations. In addition, they
found that their implementation of AGN feedback was able to suppress
strong cooling flows and produce shallower entropy profiles in
clusters, and to quench star formation in massive galaxies. However,
the very large dynamic range required to treat the growth of black
holes and galaxies in a cosmological context --- from the sub-pc
scales of the BH accretion disk to the super-Mpc scales of large scale
structure --- means that numerical techniques such as these will
likely need to be supplemented by semi-analytic or sub-grid methods
for some time to come.

Our approach is in many respects very similar in spirit to that of
\citet{sijacki:07}, although of course we are forced to implement both
modes of AGN feedback in an even more schematic manner because we are
using a semi-analytic model rather than a numerical simulation. We
adopt fairly standard semi-analytic treatments of the growth of dark
matter halos via accretion and mergers, radiative cooling of gas, star
formation, supernova feedback, and chemical evolution. We then adopt
the picture of self-regulated black hole growth and bright mode
feedback in mergers discussed in \S\ref{sec:bright}, and implement
these processes in our model using the results extracted from the
merger simulations described above. We assume that the radio mode is
fueled instead by hot gas in quasi-hydrostatic halos, and that the
accretion rate is described by Bondi accretion from an isothermal
cooling flow as proposed by \citet[][NF00]{nulsen_fabian:00}. We
calibrate the heating efficiency of the associated radio jets against
direct observations of bubble energetics in clusters.

A number of authors have previously explored the formation of black
holes and AGN in the context of CDM-based semi-analytic models of
varying complexity
\citep{efstathiou_rees:88,kh:00,wyithe_loeb:02,bromley:04,scannapieco:04,volonteri:03,volonteri:05},
and recently several studies have also investigated the impact of AGN
feedback on galaxy formation using such models
\citep{cattaneo:06,croton:06,bower:06,menci:06,schawinski:06,kang:06,monaco:07}. The
models that we present here differ from previous studies of which we
are aware, in two main respects: 1) we implement detailed modelling of
self-regulated black hole growth and bright mode feedback based on an
extensive suite of numerical simulations of galaxy mergers and 2) we
adopt a simple but physical model for radio mode accretion and
heating, and calibrate our model against direct observations of
accretion rates and radio jet heating efficiencies. We present a
broader and more detailed comparison with observations than previous
works, and highlight some remaining problems that have not previously
been emphasized. As well, unlike most previous studies, we calibrate
our models and make our comparisons in terms of ``physical'' galaxy
properties such as stellar mass and star formation rate, which can be
estimated from observations, rather than casting our results in terms
of observable properties such as luminosities and colors. Our results
are therefore less sensitive to the details of dust and stellar
population modelling, and easier to interpret in physical terms.

The goals of this paper are to present our new models in detail,
and to test and document the extent to which they reproduce basic
galaxy observations at $z=0$ and the global cosmic histories of the
main baryonic components of the Universe.  The structure of the rest
of this paper is as follows. In \S\ref{sec:model}, we describe the
ingredients of our models and provide a table of all of the model
parameters. In \S\ref{sec:results}, we present predictions for key
properties of galaxies at $z\sim0$ and for the global history of the
main baryonic components of the Universe: star formation, evolved
stars, cold gas, metals, and black holes. We conclude in
\S\ref{sec:conclude}.

\section{Model}
\label{sec:model}

\begin{table*}
\centering
\caption{Summary of Cosmological Parameters}
\begin{tabular}{llcc}
\hline \hline
parameter  & description & Concordance \LCDM & WMAP3 \\
\hline \hline
\textbf{cosmological parameters} &&& \\
$\Omega_m$ & present day matter density & 0.30 & 0.2383 \\
$\Omega_{\Lambda}$ & cosmological constant & 0.70 & 0.7617 \\
$H_0$ & Hubble Parameter [km/s/Mpc] & 70.0  & 73.2\\
$f_b$ & cosmic baryon fraction & 0.14 & 0.1746\\
$\sigma_8$ & power spectrum normalization & 0.9 & 0.761 \\
$n_s$ & slope of primordial power spectrum & 1.0 & 0.958 \\
\hline \hline
\end{tabular}
\label{tab:cosmo}
\end{table*}

\begin{table*}
\centering
\caption{Summary of the galaxy formation parameters in our
  ``fiducial'' model. We also specify the section in the paper where a
  more detailed definition of each set of parameters can be found, and
  whether the parameter is considered to be fixed based on direct
  observations or numerical simulations (F), or adjusted to match
  observations (A). }
\begin{tabular}{llcc}
\hline \hline
parameter  & description & fiducial value & fixed/adjusted \\
\hline \hline
\textbf{photoionization squelching (\S\ref{sec:model:squelch})} &&&\\
$z_{\rm overlap}$, $z_{\rm reionize}$  & redshift of overlap/reionization & $11$, $10$ & F\\
\hline
\textbf{quiescent star formation (\S\ref{sec:model:qsf})} &&&\\
$A_{\rm Kenn}$ & normalization of Kennicutt Law [\msun yr$^{-1}$ kpc$^{-2}$] & $8.33 \times 10^{-5}$ & A\\
$N_{\rm K}$ & power law index in Kennicutt Law  & 1.4 & F\\
$\chi_{\rm gas}$ & scale radius of gas disk, relative to stellar disk & 1.5 & A\\
$\Sigma_{\rm crit}$ & critical surface density for star formation [\msun pc$^{-2}$] & 
6.0  & A\\
\hline
\textbf{burst star formation (\S\ref{sec:model:bursts})} &&&\\
$\mu_{\rm crit}$ & critical mass ratio for burst activity & 0.1 & F\\ 
$e_{\rm burst,0}$ & burst efficiency for 1:1 merger & eqn.~\ref{eqn:eburst} & F\\
$\gamma_{\rm burst}$ & dependence of burst efficiency on mass ratio & eqn.~\ref{eqn:gamma_burst} & F\\
$\tau_{\rm burst}$ & burst timescale & eqn.~\ref{eqn:tburst} & F\\ 
\hline
\textbf{merger remnants \& morphology (\S\ref{sec:model:remnants})} &&&\\
$f_{\rm sph}$ & fraction of stars in spheroidal remnant & eqn.~\ref{eqn:fsph} & A\\
$f_{\rm scatter}$ & fraction of scattered satellite stars & 0.4 & A\\
\hline 
\textbf{supernova feedback (\S\ref{sec:model:snfb})} &&&\\
$\epsilon_{\rm SN}^0$ & normalization of reheating function & 1.3 & A\\
$\alpha_{\rm rh}$ & power law slope of reheating function & 2.0 & A\\
$V_{\rm eject}$ & velocity scale for ejection of reheated gas [km/s] & 120 & A\\
$\chi_{\rm reinfall}$ & timescale for re-infall of ejected gas & 0.1 & A\\ 
\hline
\textbf{chemical evolution (\S\ref{sec:model:chemev})} &&&\\
$y$ & chemical yield (solar units) & 1.5 & A\\
$R$ & recycled fraction & 0.43 & F\\
\hline
\textbf{black hole growth (\S\ref{sec:model:smbh})} &&&\\
$\etabh$ & efficiency of conversion of rest mass to radiation & 0.1 & F\\
$M_{\rm seed}$ & mass of seed BH [\msun] & 100 & F\\
$f_{\rm BH, final}$ & scaling factor for mass of BH at end of merger & 2.0 & A\\
$f_{\rm BH, crit}$ & scaling factor for ``critical mass'' of BH & 0.4 & F\\
\hline
\textbf{AGN-driven winds (\S\ref{sec:model:agnwinds})} &&&\\
$\epsilon_{\rm wind}$ & effective coupling factor for AGN driven winds & 0.5 & F\\
\hline
\textbf{radio mode feedback (\S\ref{sec:model:radio})} &&&\\
$\kappa_{\rm radio}$ & normalization of ``radio mode'' BH accretion rate & $3.5\times 10^{-3}$ & A\\
$\kappa_{\rm heat}$ & coupling efficiency of radio jets with hot gas & 1.0 & F\\
\hline \hline
\end{tabular}
\label{tab:param}
\end{table*}

Our model is based on the semi-analytic galaxy formation code
described in \citet[][SP99]{sp:99} and \citet[][SPF01]{spf:01}, with
several major updates and important new ingredients, which we describe
in detail here. Unless specified otherwise, we adopt the
``Concordance'' \LCDM\ model (C-\LCDM), with the parameters given
in Table~\ref{tab:cosmo}, which has been used in many recent
semi-analytic studies of galaxy formation. In
\S\ref{sec:results:global}, we also consider a model that uses the set
of parameters obtained from the three year results of the Wilkinson
Microwave Anisotropy Probe by \citet{spergel:07}, which are also
specified in Table~\ref{tab:cosmo}. We refer to this as the ``WMAP3''
model. We assume a universal Chabrier stellar initial mass function
\citep[IMF;][]{chabrier:03}, and where necessary we convert all
observations used in our comparisons to be consistent with this IMF.

\subsection{Dark Matter Halos, Merger Trees and Substructure}
\label{sec:model:trees}
We compute the number of ``root'' dark matter (DM) halos as a function
of mass at a desired output redshift using the model of
\citet{sheth_tormen:99}, which has been shown to agree well with
numerical simulations. Then, for each ``root'' halo of a given mass
$M_0$ and at a given output redshift, we construct a realization of
the merger history based on the method described in
\citet[][SK99]{sk:99}. We have introduced a modification to the SK99
algorithm, which we find leads to better agreement with N-body
simulations. We choose the timestep $\Delta t$ by requiring that the
\emph{average} number of progenitors $\bar{N}_p$ be close to two, by
inverting the equation for $N_{\rm prog}(M_0, \Delta t)$ (see
SK99). We then select progenitors as described in SK99, but do not
allow the number of progenitors to exceed
$\bar{N}_p+\sqrt{\bar{N}_p}+1$.  We follow halo merging histories down
to a minimum progenitor mass of $10^{10} \msun$, and our smallest
``root'' halos have a mass of $10^{11} \msun$. We have also
implemented our models within N-body based merger trees, and do not
find any significant changes to our results.

We assign two basic properties to every dark matter halo in each of
our merger trees: the angular momentum or spin parameter, and the
concentration parameter, which describes the matter density
profile. We express the angular momentum in terms of the dimensionless
spin parameter $\lambda \equiv J_h\vert E_h\vert ^{1/2} G^{-1}
\mvir^{-5/2}$ \citep{peebles:69}, where $E_h$ is the total energy of
the halo and $\mvir$ is the virial mass. Numerical N-body simulations
have demonstrated that $\lambda$ is uncorrelated with the halo's mass
and concentration \citep{bullock:01b,maccio:07} and does not evolve
with redshift. The distribution of $\lambda$ is log-normal, with mean
$\bar{\lambda}=0.05$ and width $\sigma_{\lambda}=0.5$
\citep{bullock:01b}. We assign each top-level halo a value of
$\lambda$ by selecting values randomly from this distribution,
assuming that it is not correlated with any other halo properties or
with redshift. The halo at the next stage of the merger tree inherits
the spin parameter of its largest progenitor.

We assume that the initial density profile of each halo is described
by the Navarro-Frenk-White (NFW) form \citep{navarro:97}, and compute
the characteristic concentration parameter \cnfw\ for the appropriate
mass and redshift using a fitting formula based on numerical
simulations \citep{bullock:01a}. We adopt the updated normalization of
\cnfw(\mvir) from \citet{maccio:07}. We neglect the scatter in \cnfw\
at fixed mass, as well as the known correlation between \cnfw\ and
halo merger history \citep{wechsler:02}.

At each stage in the merging hierarchy, one or more halos merge
together to form a new, virialized dark matter halo. The merged halos
(hereafter referred to as ``sub-halos'') and their galaxies, however,
can survive and continue to orbit within the potential well of the
parent DM halo for some time. The time it takes for the satellite to
lose all of its angular momentum due to dynamical friction and merge
with the central galaxy is typically modelled with some variant of the
Chandrasekhar formula \citep[see e.g. \S2.8 of][]{sp:99}. Here, we use
an updated version of this formula from \citet{boylan_kolchin:08},
which accounts for the tidal mass loss of sub-halos as they orbit
within the host halo, as well as the dependence on the energy and
angular momentum of the orbit. Because the merger time is proportional
to $M_{\rm host}/M_{\rm sat}$, accounting for this mass loss increases
the time it takes for small mass satellites to merge. 

These satellites may eventually lose so much of their mass that they
become tidally disrupted. Based on the results of
\citet{taylor_babul:04} and \citet{zentner_bullock:03}, we assume that
satellites lose $\sim 30-40$ percent of their mass per orbital period,
and that when the mass has been stripped down to the mass within the
NFW scale radius $r_s \equiv \rvir/\cnfw$, we consider the satellite
to be tidally destroyed. Sub-halos that survive until they reach the
center of the parent halo are assumed to merge with the central
object. Subhalos that are tidally destroyed before they can merge are
assumed to contribute their stars to a ``diffuse stellar component''
(DSC), which may be associated with the stellar halo or the
Intra-Cluster Light. We have verified that our model reproduces the
conditional multiplicity function of sub-halos over the relevant range
of host halo masses. Details and tests of our new algorithm for the
treatment of sub-structure will be presented in Maulbetsch et al. (in
prep).

\subsection{Cooling}
\label{sec:model:cooling}

The rate of gas condensation via atomic cooling is computed based on
the model originally proposed by \citet{white_frenk:91}, and utilized
in various forms in virtually all semi-analytic models. Here we use a
slightly different implementation of the cooling model than that used
in SP99 and subsequent papers, which we find is numerically better
behaved. We first compute the ``cooling time'', which is the time
required for the gas to radiate away all of its energy, assuming that
it all starts out at the virial temperature:

\begin{equation}
t_{\rm cool} = \frac{\frac{3}{2} \mu m_p kT}{\rho_g(r) \Lambda(T, Z_h)} \, .
\label{eqn:tcool}
\end{equation}
Here, $\mu m_p$ is the mean molecular mass, $T$ is the virial
temperature $T_{\rm vir} = 35.9 (V_{\rm vir}/({\rm km/s}))^2$ K,
$\rho_g(r)$ is the radial density profile of the gas, $\Lambda(T,
Z_h)$ is the temperature and metallicity dependent cooling function
\citep{sutherland:93}, and $Z_h$ is the metallicity of the hot halo
gas. We assume that the gas density profile is described by that of a
singular isothermal sphere: $\rho_g(r) = m_{\rm hot}/(4 \pi r_{\rm
vir} r^2)$. Substituting this expression for $\rho_g(r)$, we can solve
for the cooling radius $r_{\rm cool}$, which is the radius within
which all of the gas can cool within a time $\tcool$. Writing the
expression for the mass within $\rcool$, and differentiating, we
obtain the rate at which gas can cool:
\begin{equation}
\frac{dm_{\rm cool}}{dt} = \frac{1}{2} m_{\rm hot} \frac{r_{\rm
    cool}}{r_{\rm vir}} \frac{1}{t_{\rm cool}}.
\label{eqn:mcooldot}
\end{equation}

There are various possible choices for the cooling time $\tcool$. Some
early works used the Hubble time, $t_{\rm cool}=t_H$
\citep[e.g.][]{kwg:93}. In our earlier models (e.g. SP99), we used the
time since the last halo major merger $t_{\rm mrg}$, defined as a
merger in which the halo grows in mass by at least a factor of
two. Here, we follow \citet{springel:01} and \citet{croton:06} and
assume that the cooling time is equal to the halo dynamical time,
$t_{\rm cool} = t_{\rm dyn} = \rvir/\vvir$. Note that because in
general $t_{\rm dyn} < t_{\rm mrg} < t_H$, and the cooling rate
$dm_{\rm cool}/dt \propto \tcool^{-1/2}$, the choice $t_{\rm cool} =
t_{\rm dyn}$ results in higher cooling rates than assuming $t_{\rm
cool} = t_{\rm H}$, while using $t_{\rm cool} = t_{\rm mrg}$ produces
intermediate results.

It can occur that $\rcool > \rvir$, indicating that the cooling time is
shorter than the dynamical time. In this case, we assume that the
cooling rate is given by the rate at which gas can fall into the halo,
which is governed by the mass accretion history. 

We note that there are several rather arbitrary choices that must be
made in any semi-analytic cooling model --- for example, the profile
of the hot gas and whether it is ``reset'', and the time to which the
cooling time is compared (see above) --- and different groups tend to
make slightly different choices for these ingredients. Changing these
ingredients in reasonable ways leads to overall variations in the
cooling rates (changes in the redshift and halo mass dependence tend
to be small) of at most a factor of two to three. These differences
are then typically compensated by adjusting the supernova feedback
and/or AGN feedback parameters.

We have adopted choices similar to those of \citet{croton:06}, in part
to facilitate comparison with their results, and also because it has
been shown that this recipe produces good agreement with the cooling
rates and accumulation of gas in fully 3-D hydrodynamic simulations
\citep{yoshida:02} without star formation, SN feedback, or chemical
enrichment. We have also compared our results with the cooling rates
presented by \citet{keres:05}, and find good agreement.

Recently, studies based on 1D and 3D hydrodynamic simulations
\citep{birnboim_dekel:03,keres:05} have highlighted a distinction
between gas which is accreted in a ``cold flow'' mode, in which the
gas particles are never heated much above $\sim 10^4$ K, and a ``hot
flow'' mode in which gas is first shock heated to close to the virial
temperature of the halo, forming a quasi-hydrostatic halo, and then
cools in a manner similar to a classical cooling flow. The possible
importance of distinguishing between gas flows occuring in the regime
$t_{\rm cool} < t_{\rm ff}$ vs. $t_{\rm cool} > t_{\rm ff}$ (where
$t_{\rm ff}$ is the free-fall time) has been highlighted many times in
the literature
\citep{silk:77,binney:77,rees_ostriker:77,white_frenk:91}. Although
other criteria have been proposed \citep[see][]{croton:06}, we will
identify gas cooling which occurs in timesteps in which $\rcool >
\rvir$ as ``cold mode'' and the reverse ($\rcool < \rvir$) as ``hot
mode''. This distinction will be relevant later, when we begin to
consider the impact of heating by AGN-driven radio jets.

As in most semi-analytic models, we assume that all new cold gas is
accreted by the central galaxy in the halo. Because of this, satellite
galaxies tend to consume their gas and become red, non-starforming,
and gas poor. Realistically, satellite galaxies can probably retain
their hot gas halos, and thus receive new cold gas, for some time
after they merge with another halo. This aspect of the modelling
should be improved; however, for the moment, we simply keep this
problem in mind, and in some cases restrict our analysis to central
galaxies.

\subsection{Photo-ionization Squelching}
\label{sec:model:squelch}

Photoionization heating may ``squelch'' or suppress the collapse of
gas into small mass halos
\citep{efstathiou:92,thoul:96,quinn:96}. This mechanism may play an
important role in reconciling the (large) number of small-mass
satellite halos predicted by CDM with the observed number of satellite
galaxies in the Local Group
\citep{squelch,benson:02}. \citet[][G00]{gnedin:00} showed that the
fraction of baryons that can collapse into halos of a given mass in
the presence of a photo-ionizing background can be described in terms
of the ``filtering mass'' $M_F$. Halos less massive than $M_F$ contain
fewer baryons than the universal average. G00 parameterized the
collapsed baryon fraction as a function of redshift and halo mass with
the expression:
\begin{equation}
f_{\rm b, coll}(z, M_{\rm vir}) = \frac{f_{b}}{[1 + 0.26 M_F(z)/M_{\rm
vir}]^3}\, ,
\label{eqn:fbar_squelch}
\end{equation}
where $f_{b}$ is the universal baryon fraction and $\mvir$ is the
halo virial mass.

The filtering mass is a function of redshift, and this function
depends on the reionization history of the
universe. \citet{kravtsov:04b} provide fitting formulae for the
filtering mass in the simulations of G00, parameterized according to
the redshift at which the first $H_{\rm II}$ regions begin to overlap
($z_{\rm overlap}$) and the redshift at which most of the medium is
reionized ($z_{\rm reion}$). In the simulations of G00, reionization
occurs fairly late ($z_{\rm overlap}=8$, $z_{\rm reion}=7$). Recent
results from the WMAP satellite, however, suggest an earlier epoch of
reionization, $z_{\rm reion} \ga 10$ \citep{spergel:07}. We
make use of the fitting functions (B2) and (B3) from Appendix B of
\citet{kravtsov:04b} to compute the initial fraction of baryons that
can collapse as a function of halo mass and redshift, with $z_{\rm
overlap}=11$ and $z_{\rm overlap}=10$.

As shown by \citet{squelch} using a similar treatment of
photo-ionization squelching, we find that our model reproduces the
luminosity function of satellite galaxies in the Local Group (Macci\`o
et al. in prep).

\subsection{Disk Sizes}
\label{sec:model:disk_sizes}
When gas cools, it is assumed to initially settle into a thin
exponential disk, supported by its angular momentum. We assume that
the gas has aquired angular momentum before its collapse, along with
the dark matter, via tidal torques \citep{peebles:69}. Given the
halo's concentration parameter \cnfw, spin parameter $\lambda$ and the
fraction of baryons in the disk $f_{\rm disk}$, we can use angular
momentum conservation arguments to compute the scale radius of the
exponential disk after collapse. We include the ``adiabatic
contraction'' of the halo due to the gravitational force of the
collapsing baryons. Our approach is based on work by
\citet{blumenthal:86}, \citet{flores:93}, and \citet[][MMW98]{mo:98},
and is described in detail in \citet[][S08]{somerville:08}. In S08, we
showed that this model produces good agreement with the observed
radial sizes of disks as a function of stellar mass both locally and
out to $z\sim 2$.

\subsection{Star Formation}
\label{sec:model:sf}

\subsubsection{Quiescent Star Formation}
\label{sec:model:qsf}
During the `quiescent' phase of galaxy evolution (i.e. in undisturbed
disks) we adopt a star formation recipe based on the empirical
Schmidt-Kennicutt law \citep{kennicutt:89,kennicutt:98}. The star
formation rate density (per unit area) is given by:
\begin{equation}
\dot{\Sigma}_{\rm SFR} = A_{\rm Kenn} \, {\Sigma_{\rm gas}}^{N_K} ,
\end{equation}

where $A_{\rm Kenn} = 1.67 \times 10^{-4}$, $N_K=1.4$, $\Sigma_{\rm
gas}$ is the surface density of cold gas in the disk (in units of
\msun pc$^{-2}$), and $\Sigma_{\rm SFR}$ has units of \msun yr$^{-1}$
kpc$^{-2}$. The normalization quoted above is appropriate for a
Chabrier IMF, and has been converted from the value given in
\citet{kennicutt:98}, which was based on a Salpeter IMF.

We assume that the gas profile is also an exponential disk, with a
scale length proportional to the scale-length of the stellar disk:
$r_{\rm gas} = \chi_{\rm gas}\, r_{\rm disk}$, where the stellar
scalelength $r_{\rm disk}$ is determined as described in
\S\ref{sec:model:disk_sizes}. We adopt $\chi_{\rm gas} =1.5$, which
yields average gas surface densities in the range $\sim 4-60$ \msun
pc$^{-2}$, consistent with the observations of
\citet{kennicutt:98}. This value is also consistent with observations
of the radial extent of \HI\ gas in spiral galaxies
\citep{broeils_rhee:97}.

We further adopt a critical surface density threshold $\Sigma_{\rm
crit}$, and assume that only gas lying at surface densities above this
value is available for star formation. We can then compute the radius
within which the gas density exceeds the critical value: 
\begin{equation}
r_{\rm crit} = -\ln \left[\frac{\Sigma_{\rm crit}}{\Sigma_0} \right] r_{\rm gas}\end{equation}
where $\Sigma_0 \equiv m_{\rm cold}/(2\pi r^2_{\rm gas})$. The
fraction of the total gas supply that is eligible for star formation
is then:
\begin{equation}
f_{\rm gas}(r<r_{\rm crit}) = 1 - (1+r_{\rm crit}/r_{\rm gas}) \frac{\Sigma_{\rm crit}}{\Sigma_0} \,
\end{equation}
and the total star formation rate is: 
\begin{eqnarray*}
\dot{m}_{*}  & = & \int_0^{r_{\rm crit}} \dot{\Sigma}_{\rm SFR} \, 2\pi r \, {\rm d}r \\
 & = &\frac{2 \pi A_{\rm K} {\Sigma_0}^{N_K} r^2_{\rm gas}}{N^2_K} \\
 &&     \times  \left [1 - \left(1+\frac{N_K r_{\rm crit}}{r_{\rm gas}}\right)
	  \exp(-N_K r_{\rm crit}/r_{\rm gas}) \right]
\end{eqnarray*}  

\citet{schaye:04} investigated star formation thresholds in models of
isolated self-gravitating disks embedded in dark matter halos,
containing metals and dust, and exposed to a UV background. They found
that the gas was able to form a cold interstellar phase only above a
critical surface density threshold of $\Sigma_{\rm crit} \sim 3-10$
\msun pc$^{-2}$, which is consistent with observations of SF
thresholds in spiral galaxies \citep{martin_kenn:01}. We find that
adopting a value of $\Sigma_{\rm crit} = 6$ \msun\, pc$^{-2}$ produces
good agreement with the observations of global SFR vs. gas density of
\citet{kennicutt:98}, and also with observed gas fractions as a
function of stellar mass.

We allow the normalization of the star formation law $A_{\rm Kenn}$ to
be adjusted as a free parameter. We find that using a value of $A_{\rm
Kenn}=8.33 \times 10^{-5}$, a factor of two lower than the one
measured by \citet{kennicutt:98} gives good agreement with observed
star formation rates and gas fractions as a function of stellar
mass. We adopt the observed value for the slope of the SFR law,
$N_{K}=1.4$.

We account for the mass loss from stars (recycled gas) using the
instantaneous recyling approximation. Thus, for an instantaneous star
formation rate $\dot{m}$, we form a mass $dm_* = (1-R)\, \dot{m}\, dt$
of long-lived stars in a timestep $dt$. We adopt a recycled fraction
$R=0.43$, appropriate for a Chabrier IMF \citep{bruzual:03}.

\subsubsection{Merger-Driven Starbursts}
\label{sec:model:bursts}

As in SPF01, we parameterize the efficiency of star formation in a
merger-triggered ``burst'' mode as a function of the mass ratio of the
merging pair. This is supported both by observations of star formation
enhancement in galaxy pairs \citep{woods:07} and by numerical
simulations of galaxy mergers \citep{cox:08}. However, first, an
important question arises: which quantity should we use for the mass
ratio? Many previous works have either used the ratio of the virial
masses of the two dark matter halos, or else the baryonic masses of
the two galaxies. Because the ratio of baryons to dark matter can vary
by several orders of magnitude across halos of different masses, and
moreover is a systematic function of halo mass (see
\S\ref{sec:results:galprop}), we find that our results can depend quite
sensitively on this choice. Moreover, the baryonic mass ratio is
sensitive to the modelling of supernova and AGN feedback.

When we consider that the simulation results clearly indicate that the
efficiency of the starburst is mainly determined by the strength of the
torques during the later stages of the merger, it is clear that what
should be relevant is the \emph{total} mass (baryons and dark matter)
in the \emph{central} parts of the galaxies. Therefore we define
$m_{\rm core} = M_{\rm DM}(r<2r_s)$, i.e., the dark matter mass within
twice the characteristic NFW scale radius $r_s \equiv r_{\rm
vir}/\cnfw$, assuming that the dark matter follows an NFW profile. For
a Milky Way sized halo ($M_{\rm vir} \sim 2 \times 10^{12}\msun$),
$r_s \sim 27$ kpc and so $2 r_s$ corresponds to about 60 kpc, close to
the scale that we expect to be relevant. We then define the mass ratio
$\mu \equiv (m_{\rm core, 1} + m_{\rm bar, 1})/(m_{\rm core, 2} +
m_{\rm bar, 2})$, i.e. as the ratio of the dark matter ``core'' plus
the total baryonic mass (stars plus cold gas) of the smaller to the
larger galaxy.

Now defining $e_{\rm burst}$ as the fraction of the total cold gas
reservoir in the galaxy that is consumed by the burst, we parameterize
the burst efficiency via:
\begin{equation}
e_{\rm burst} = e_{\rm burst, 0} \, \mu^{\gamma_{\rm
burst}} .
\end{equation}
This functional form has been shown to describe well the scaling of
burst efficiency with merger mass ratio in hydrodynamic simulations of
galaxy mergers \citep[][C08]{cox:08}. Again based on C08, we assume
that mergers with mass ratios below 1:10 do not produce bursts,
i.e. $e_{\rm burst}=0$ for $\mu<0.1$.

Numerical studies \citep{mihos:94,cox:08} have furthermore shown that
the burst efficiency in \emph{minor} mergers ($\mu \lesssim 0.25$)
depends on the bulge-to-total ratio of the progenitor galaxies,
because the presence of a bulge stablizes the galaxy and reduces the
efficiency of the burst. To reflect the joint dependence on merger
mass ratio and bulge fraction, we adopt the results of C08:
\begin{equation}
\gamma_{\rm burst} = 
\begin{cases}
0.61 & {B/T \leq 0.085} \\
0.74 & {0.085 < B/T \leq 0.25} \\
1.02 & {B/T > 0.25} 
\end{cases}
\label{eqn:gamma_burst}
\end{equation}
where $B/T$ is the ratio of the stellar mass in the spheroidal
component to the total stellar mass (disk plus spheroid) in the larger
progenitor galaxy at the beginning of the merger.

We have studied the burst efficiency $e_{\rm burst, 0}$ and burst
timescale $\tau_{\rm burst}$ in equal mass mergers in a large suite of
numerical simulations containing stellar feedback as well as feedback
from energy released by accretion onto a central black hole
\citep{robertson:06a}. We find that the burst efficiency can be fit by:
\begin{multline}
e_{\rm burst,0} = 0.60 (V_{\rm vir}/({\rm km/s}))^{0.07}
                (1+q_{EOS})^{-0.17}\\ 
\times (1+f_g)^{0.07}(1+z)^{0.04}
\label{eqn:eburst}
\end{multline}
with a scatter of 4.9 percent, and the burst timescale (assuming a
double exponential form for the star formation rate peak) is fit by:
\begin{multline}
\tau_{\rm burst} = 191 \, {\rm Gyr} \, (V_{\rm vir}/({\rm
km/s}))^{-1.88}(1+q_{EOS})^{2.58}\\ 
 (1+f_g)^{-0.74} (1+z)^{-0.16}
\label{eqn:tburst}
\end{multline}
with a logarithmic scatter of 0.36. Here, \vvir\ is the virial
velocity of the progenitor galaxies, $q_{\rm EOS}$ is the effective
equation of state of the gas \citep[see][]{robertson:06a}, $f_g \equiv
m_{\rm cold}/(m_{\rm cold}+m_{\rm star})$ is the cold gas fraction in
the disk, and $z$ is the redshift for which the progenitor disk models
were constructed. It is important to note that the simulations used to
obtain these fitting functions span the range $\vvir = 60$ -- 500
km/s, $q_{EOS}=0.25$--1, $f_g$=0.01--0.8, and $z=0$--6. The results of
the fitting formulae should be used with caution outside of this range
of values for the input parameters.

The parameter $q_{EOS}$ can be thought of as parameterizing the
multi-phase nature of the Interstellar Medium (ISM), such that
$q_{EOS}=0$ corresponds to an isothermal gas, and $q_{EOS}=1$
corresponds to the fully pressurized multiphase ISM. Increasing
$q_{EOS}$ (or adopting a ``stiffer'' equation of state) increases the
dynamical stability of the gas, and suppresses the starburst. Thus
larger values of $q_{EOS}$ give smaller values of $e_{\rm burst,0}$
and larger values of $\tau_{\rm burst}$ (because the burst is more
extended). In this work we adopt a value corresponding to a stiff
equation of state, $q_{EOS} = 1$.

For the burst efficiency, we can see that the only significant
 dependence on these parameters is on the equation of state
 $q_{EOS}$. The burst timescale is more sensitive to other parameters
 (as also found by C08), and in particular has quite a strong
 dependence on \vvir. For our adopted fiducial value of $q_{EOS} = 1$,
 the typical value of the burst efficiency is $e_{\rm burst} \sim
 0.8$, and the burst timescale (exponential decline time) for a Milky
 Way sized galaxy ($\vvir \sim 130$ km/s) is $\tau_{\rm burst} \sim
 100$ Myr.

We now parameterize the ``burst'' mode of star formation as $\dot{m}_*
= m_{\rm burst}/\tau_{\rm burst}$. At the beginning of the merger, we
allocate a reservoir of ``burst fuel'' $m_{\rm burst} = e_{\rm burst}
m_{\rm cold}$, where $m_{\rm cold}$ is the combined cold gas from both
of the progenitor galaxies. The burst continues until this fuel is
exhausted, and in the absence of new sources of fuel, the burst SFR
will decline exponentially, with exponential decline time $\tau_{\rm
burst}$. However, particularly in the early universe, it can
frequently happen that a new merger occurs while a burst from an
earlier merger is still going on. In this case, we add the new burst
fuel to the reservoir, and assign a new burst timescale based on the
updated galaxy properties. Note that the ``quiescent'' mode of star
formation still goes on as before (the burst efficiencies computed
from the simulations have the quiescent star formation subtracted
out).

\subsection{Merger Remnants and Morphology}
\label{sec:model:remnants}

\subsubsection{Spheroid Formation}
\label{sec:model:spheroid}

Numerical simulations of mergers of galaxy disks have also shown that
major mergers $\mu > 0.25$ leave behind a spheroidal remnant, while
smaller mass ratio minor mergers ($\mu < 0.25$) tend to just thicken
the disk, perhaps driving minor growth of a spheroid via bar
instabilities. Most previous semi-analytic models have assumed a sharp
threshold in merger mass ratio (e.g. $\mu > f_{\rm ellip} \simeq
0.25-0.3$) for determining whether the stars after a merger are placed
in a ``spheroidal'' component or not. However, in reality there will
be a continuum, whereby larger mass ratios result in more heating and
a transfer of more material to a dynamically hot spheroidal
component. To represent this continuum, we define the function
\begin{equation}
f_{\rm sph} = 1 - \left [1+\left(\frac{\mu}{f_{\rm ellip}}\right)^8 \right]^{-1}
\label{eqn:fsph}
\end{equation}
which determines the fraction of the disk stars that is transferred to
the ``spheroid'' or bulge component following a merger (as with
bursts, we assume that mergers with mass ratio $\mu < 0.1$ have no
effect).

Thus, consider a merger of two galaxies with bulge masses $B_1$ and
$B_2$, and disk masses $D_1$ and $D_2$. The mass of the new bulge will
be $B_{\rm new} = B_1 + B_2 + f_{\rm sph} (D_1+D_2)$, and the mass of
the surviving disk will be $D_{\rm new} = (1-f_{\rm sph})
(D_1+D_2)$. All new stars formed in the burst mode are also deposited
in the spheroid component.

In this paper, we assume that all spheroid growth is connected with
mergers. That is, we do not consider formation of spheroids via disk
instabilities. We have experimented with including spheroid formation
via disk instabilities, and find that it has only a minor impact on
the results presented here.

\subsubsection{Formation of Diffuse Stellar Halos}
\label{sec:model:dsh}

There is now considerable observational evidence for spatially
extended stellar components surrounding brightest group and cluster
galaxies \citep[e.g.][]{zibetti:05,gonzalez:05}. It is now thought
that these ``diffuse stellar halos'' (DSH) originated from tidally
disrupted merging satellites and/or scattering of stars during major
mergers
\citep{murante:04,monaco:06,conroy:07,murante:07,purcell:07}. In our
models, the stars from all satellites that are deemed to be tidally
destroyed before they merge (according to the criteria described in
\S\ref{sec:model:trees}), are deposited in a DSH component. In
addition, when two galaxies merge, we assume that a fraction $f_{\rm
scatter}$ of the stars from the satellite may be scattered into the
DSH (thus the galaxy's mass following a merger increases by $(1-f_{\rm
scatter})m_{\rm star, sat}$, where $m_{\rm star, sat}$ is the stellar
mass of the merging satellite).

\subsection{Supernova Feedback}
\label{sec:model:snfb}

Cold gas may be ejected from the galaxy by winds driven by supernova
feedback. The rate of reheating of cold gas is given by:
\begin{equation}
\dot{m}_{\rm rh} = \epsilon^{SN}_0 \left( \frac{V_{\rm disk}}{200\, {\rm km/s}} \right)^{\alpha_{\rm rh}} \, \dot{m}_*
\label{eqn:snfb}
\end{equation}
where $\epsilon^{SN}_0$ and $\alpha_{\rm rh}$ are free parameters
\citep[we expect $\alpha_{\rm rh}\simeq 2$ for ``energy driven''
winds; see e.g.][]{kwg:93}. We take the circular velocity of the disk
$V_{\rm disk}$ to be equal to the maximum rotation velocity of the DM
halo, $V_{\rm max}$.

The heated gas is either trapped within the potential well of the dark
matter halo, so deposited in the ``hot gas'' reservoir, or is ejected
from the halo into the ``diffuse'' Intergalactic Medium (IGM). The
fraction of reheated gas that is ejected from the halo is given by:
\begin{equation}
f_{\rm eject}(\vvir) =  \left[1.0+(\vvir/V_{\rm eject})^{\alpha_{\rm eject}}
  \right]^{-1} ,
\end{equation}
where $\alpha_{\rm eject}=6$ and $V_{\rm eject}$ is a free parameter
in the range $\simeq 100$ --150 km/s.

We keep track of this ejected gas in a ``diffuse gas reservoir'',
which recollapses into the halo in later timesteps and once again
becomes available for cooling. Following \citet{springel:01} and
\citet{delucia:04}, we model the rate of reinfall of ejected gas by:
\begin{equation}
\dot{m}_{\rm reinfall} = \chi_{\rm reinfall} 
\left(\frac{m_{\rm eject}}{t_{\rm dyn}}\right)
\end{equation} 
where $\chi_{\rm reinfall}$ is a free parameter, $m_{\rm eject}$ is
the mass of ejected gas in the ``diffuse reservoir'', and $t_{\rm
dyn}=\rvir/\vvir$ is the dynamical time of the halo.

Varying $\chi_{\rm reinfall}$ is degenerate with variations in the
other supernova feedback parameters, $\epsilon^{SN}_0$, $\alpha_{\rm
  rh}$, and $V_{\rm eject}$. For larger values of $\chi_{\rm
  reinfall}$, the ejected gas is reincorporated in the halo more
quickly and tends to cool rapidly, so the supernova feedback must be
made more efficient in order to retain good agreement with the
abundance of low-mass galaxies. On the other hand, if $\chi_{\rm
  reinfall} =0$ (ejected gas is never reaccreted), then the baryon
fractions in clusters are too low. We have chosen to adopt the minimal
value of $\chi_{\rm reinfall}$ that allows us to fit the cluster
baryon fractions and the mass function of low-mass galaxies
simultaneously.

\subsection{Chemical Evolution}
\label{sec:model:chemev}

We track the production of metals using a simple approach that is
commonly adopted in semi-analytic models \cite[see
  e.g.][]{sp:99,cole:00,delucia:04}. In a given timestep, where we
create a parcel of new stars ${\rm d}m_*$, we also create a mass of
metals ${\rm d}M_Z = y \, {\rm d}m_*$, which we assume to be
instantaneously mixed with the cold gas in the disk. The yield $y$ is
assumed to be constant, and is treated as a free parameter in our
model. We track the mean metallicity of the cold gas $Z_{\rm cold}$,
and when we create a new parcel of stars they are assumed to have the
same metallicity as the mean metallicity of the cold gas in that
timestep.  Supernova feedback ejects metals from the disk, along with
cold gas. These metals are either mixed with the hot gas in the halo,
or ejected from the halo into the ``diffuse'' Intergalactic Medium
(IGM), in the same proportion as the reheated cold gas. The ejected
metals in the ``diffuse gas'' reservoir are also reaccreted into the
halo in the same manner as the gas (see \S\ref{sec:model:snfb}).

Throughout this paper, the yield $y$ and all metallicities are given
in solar units, which we take to be $Z_{\odot} = 0.02$. Although this
formally represents the total metallicity, we note that as we track
only the enrichment associated with Type II supernovae, our
metallicity estimates probably correspond more closely with
$\alpha$-type elements.

\subsection{The Growth of Supermassive Black Holes}
\label{sec:model:smbh}

We assume that every top-level halo in our merger tree contains a seed
black hole with mass $M_{\rm seed}$. Typically, we assume $M_{\rm
seed} \simeq 100 \msun$, however, we have checked that the results
presented here are not sensitive to this choice for a range of values
$M_{\rm seed} \sim 100-10^4 \msun$. Black holes of approximately this
mass could be left behind as remnants of massive Pop III stars
\citep[e.g.][]{abel:02}, or could form via direct core-collapse. In
our models, all ``bright mode'' accretion onto supermassive black
holes is triggered by galaxy-galaxy mergers, and we assume that this
mode of BH accretion is regulated, and eventually halted, by feedback
from the BH itself. Our treatment of BH growth and AGN activity is
closely based on an analysis of a large suite of numerical
hydrodynamic simulations including BH growth and feedback
\citep{robertson:06a,robertson:06b,robertson:06c,cox:06a,hopkins_bhfpth:07},
which utilize the methodology developed in \citet{dimatteo:05} and
\citet{springel:05a}. We now briefly summarize the results of those
simulations and the manner in which we implement them in our
semi-analytic model.

In the merger simulations, as the galaxies near their final
coalescence, the accretion onto the BH rises to approximately the
Eddington rate. This rapid accretion continues until the energy being
deposited into the ISM in the central region of the galaxy is
sufficient to significantly offset and eventually halt accretion via a
pressure-driven outflow. \citet{dimatteo:05} and \citet{robertson:06b}
found that the merger simulations naturally produced black holes and
spheroidal remnants that obeyed the observed BH mass vs. spheroid mass
relationship. The normalization of the relationship depends on the
fraction of the AGN's energy that is coupled to the ISM, and was
chosen to reproduce the normalization of the observed relation. Based
on further analysis of these simulations, \citet{hopkins_bhfpth:07}
suggested that the BH mass is largely determined by the depth of the
potential well in the central regions of the galaxy. As shown by
\citet{robertson:06a} and \citet{cox:06a}, mergers of progenitor
galaxies with higher gas fractions suffer more dissipation, and
produce more compact remnants than those with less gas. Therefore,
mergers with high gas fractions will produce a remnant with a deeper
potential well and a larger BH mass to spheroid mass ratio than
gas-poor mergers. This picture predicts that there should be a ``Black
Hole Fundamental Plane'', whereby galaxies with smaller effective
radius for their mass host larger mass black holes; there is
observational evidence for the existence of such a BH fundamental
plane in nearby dormant BH host galaxies
\citep{marconi:03,hopkins_bhfpobs:07}.

\citet{hopkins_bhfpth:07} find that the relationship between progenitor gas
fraction and the final BH mass to spheroid stellar mass ratio at the
end of the merger obtained in their simulations can be parameterized as:
\begin{equation}
\log (\mbh/M_{\rm sph}) = -3.27 + 0.36 \, {\rm erf}[(f_{\rm gas}-0.4)/0.28]
\label{eqn:mbh_msph}
\end{equation}
with a scatter around this relationship of $\sim$ 0.2-0.3 dex,
corresponding to the expected range of orbital parameters.

In our semi-analytic model, at the beginning of each merger above a
critical mass ratio ($\mu_{\rm crit} \sim 0.1$), we compute the
expected mass of the spheroid that will be left behind at the end of
the merger, where we assume that all of the new stars formed in the
burst mode will end up in the spheroid, along with the `heated' disk
stars specified by eqn.~\ref{eqn:fsph}. We then use
eqn.~\ref{eqn:mbh_msph}, above, to compute $m_{\rm BH, final}$, the BH
mass at the end of the merger, based on the initial ``effective'' gas
fraction $f_{\rm gas, eff} = (m_{\rm cold, 1} + m_{\rm cold,
2})/(m_{\rm bar, 1} + m_{\rm bar, 2})$ (i.e., the sum of the cold gas
masses in both galaxies, divided by the sum of their baryonic
masses). We allow the value of $m_{\rm BH, final}$ given by
Eqn.~\ref{eqn:mbh_msph} to be scaled by an adjustable free parameter
$f_{\rm BH, final}$.

We assume that the BH in the two progenitor galaxies merge rapidly to
form a new BH, and that mass is conserved in the BH merger.  We allow
the BH to grow at the Eddington rate until it reaches a mass $M_{\rm
BH, crit}$, whereupon it enters the ``blowout'' phase and begins a
power-law decline in the accretion rate, according to the family of
lightcurves defined by \citet{hopkins_faintev:06}. From the
simulations, $M_{\rm BH, crit} = f_{\rm BH, crit}\, 1.07
\left(M_{\rm BH, final}/10^9 \msun \right)^{1.1}$, where we
introduce the adjustable parameter $f_{\rm BH, peak}$, which
determines how much of the BH growth occurs in the Eddington-limited
vs. power-law decline (``blow-out'') phases. When the BH reaches the
mass $M_{\rm BH, final}$, ``bright mode'' accretion is switched
off. If the pre-existing BH is more massive than $M_{\rm BH, crit}$,
it goes straight into the ``blowout'' mode until it reaches $M_{\rm
BH, final}$. If the pre-existing BH is more massive than $M_{\rm BH,
final}$, the BH does not grow at all, and there is no AGN activity.

\subsection{AGN-driven galactic scale winds}
\label{sec:model:agnwinds}

\begin{figure} 
\begin{center}
\plotone{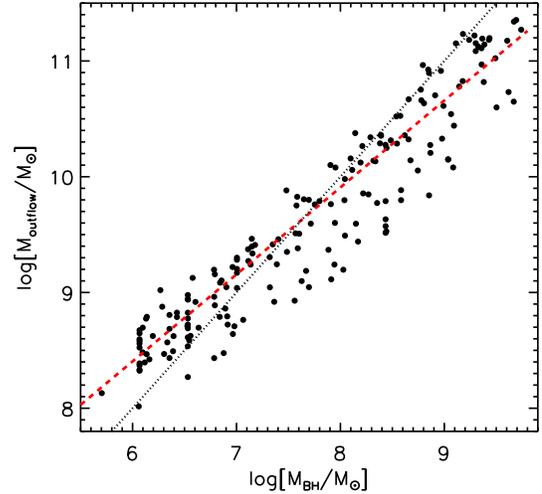}
\end{center}
\caption{Mass ejected in an outflow as a function of the final BH
mass, from the numerical merger simulations. The dashed line shows the
scaling predicted by the momentum conservation argument
(Eqn.~\protect\ref{eqn:agnwind}), with $\epsilon_{\rm wind}=0.5$. The
dotted line shows a simple scaling of ejected mass with BH mass, which
does not fit the simulation results as well.
\label{fig:moutflow}}
\end{figure}

In the numerical merger simulations, the energy being released during
the rapid growth of the BH also drives powerful galactic-scale winds
\citep{dimatteo:05,springel:05b}. We again make use of the simulations
to parameterize this process in our semi-analytic model. We start by
equating the momentum associated with the radiative energy from the
accreting BH with the momentum of the outflowing wind:
\begin{equation}
\frac{\epsilon_{\rm wind} E_{\rm BH}}{c} = M_{\rm outflow} V_{\rm esc},
\end{equation}
where $\epsilon_{\rm wind}$ is the effective coupling efficiency,
$E_{\rm BH} = \etabh m_{\rm acc} c^2$, $M_{\rm outflow}$ is the
mass of the ejected gas, and $V_{\rm esc}$ is the escape velocity of
the galaxy. We then obtain the following expression for the mass
outflow rate due to the AGN driven wind:
\begin{equation}
\frac{dM_{\rm out}}{dt} = \epsilon_{\rm wind}\, \etabh \frac{c}{V_{\rm
esc}} \dot{m}_{\rm acc} \, .
\label{eqn:agnwind}
\end{equation}

We find that this simple formula provides quite a good description of
the outflow rates in the simulations, as shown in
Fig.~\ref{fig:moutflow}. We see that Eqn.~\ref{eqn:agnwind} provides a
much better description of the simulation results than simply assuming
$\dot{M}_{\rm outflow} \propto \dot{m}_{\rm BH}$.

\subsection{Radio Mode Feedback}
\label{sec:model:radio}

\begin{figure} 
\begin{center}
\plotone{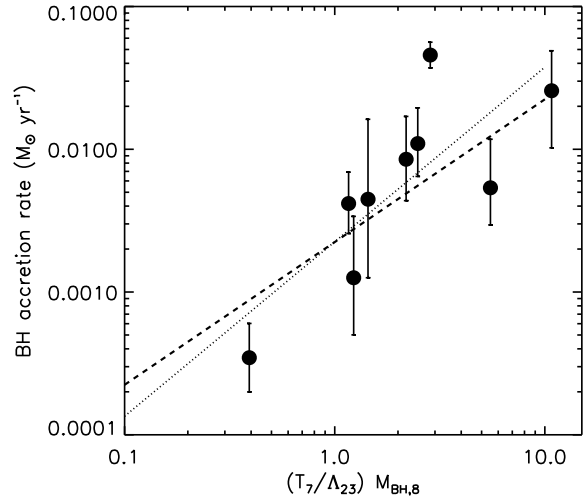}
\end{center}
\caption{\small Black hole accretion rate as a function of $\zeta
\equiv (T_7/\Lambda_{23}) (m_{\rm BH,8})$, from the observational analysis
of Allen et al. (2006) (filled circles). The dashed line shows the
scaling predicted by the NF00 ``isothermal cooling flow'' model (see
text), while the dotted line shows a linear fit to the data points.
\label{fig:bondi}}
\end{figure}

In addition to the rapid growth of BH in the merger-fueled,
radiatively efficient ``bright mode'', we assume that BH also
experience a low-Eddington-ratio, radiatively inefficient mode of
growth associated with efficient production of radio jets that can
heat gas in a quasi-hydrostatic hot halo. We base our fiducial model
on the assumption that the ``radio mode'' is fueled by Bondi-Hoyle
accretion \citep{bondi:52}:
\begin{equation}
\dot{m}_{\rm Bondi} = \pi (G \mbh)^2 \rho_{0}  c_{s}^{-3},
\label{eqn:bondi}
\end{equation} 
where $\rho_0 \equiv \rho(r_A)$ is the density of the gas at the
accretion radius $r_A$, $c_s$ is the sound speed of the gas, and we
have assumed an adiabatic index of $\gamma_1=5/3$ for the gas. We
adopt the isothermal cooling flow solution of
\citet[][NF00]{nulsen_fabian:00}, in which thermal instabilities act
to maintain the density such that the sound crossing time is of order
the local cooling time:
\begin{equation}
\frac{r_A}{c_s} = K\, \frac{3}{2} \frac{\mu m_p kT}{\rho(r_A) \Lambda(T, Z_h)}.  
\end{equation}
Here, $r_A \equiv 2G\mbh/c_{s}^2$ is the Bondi accretion radius, K is
a dimensionless constant which depends on the details of the flow,
$kT$ is the temperature of the gas, and $\Lambda(T, Z_h)$ is the
cooling function. Solving for the density $\rho_0$ and substituting
into eqn.~\ref{eqn:bondi}, we obtain:
\begin{equation}
\dot{m}_{\rm radio} = \kappa_{\rm radio} \left(\frac{kT}{\Lambda(T, Z_h)}\right)
\left(\frac{\mbh}{10^8 \msun}\right)
\end{equation}
where we have subsumed all constants into the factor $\kappa_{\rm
radio}$. A similar model has also been considered by
\citet{churazov:05} and \citet{croton:06}.

We can test the validity of this model using recent observations of
the central density and temperature of hot X-ray emitting gas in nine
nearby elliptical galaxies by \citet[][A06]{allen:06}. Deep Chandra
observations allowed A06 to obtain measurements or reliable
extrapolations of the gas properties within one order of magnitude of
the Bondi radius for eight of the systems. Each system also has a
measured velocity dispersion, which allows an estimate of the BH mass
using the relation of \citet{tremaine:02}. In Fig.~\ref{fig:bondi}, we
compare the Bondi accretion rates with the quantity $\zeta \equiv
(T_7/\Lambda_{23}) (m_{\rm BH,8})$, where we define $T_7 \equiv T/10^7
K$, $\Lambda_{23} \equiv \Lambda(T)/(10^{-23}\, {\rm erg \, cm^3\,
s^{-1}})$, and $m_{\rm BH,8} \equiv \mbh/10^8 \msun$. We use the
published values of gas density and temperature and the BH mass
estimates from A06, and assume that the hot gas has a metallicity of
one-third solar. A formal fit gives a slope of 1.23 in $\zeta$, but we
see that the NF00 isothermal cooling flow model is quite consistent
with the data. In terms of the scaled quantities $T_7$,
$\Lambda_{23}$, and $m_{\rm BH,8}$, $\kappa_{\rm radio}=2.25 \times
10^{-3}$ provides the best fit to the data.

In our fiducial semi-analytic model, we assume that whenever ``hot
mode'' gas is present in the halo, the central BH accretes at the rate
given by eqn.~\ref{eqn:bondi}, but we allow $\kappa_{\rm radio}$ to be
adjusted as a free parameter. We then assume that the energy that
effectively couples to and heats the hot gas is given by $L_{\rm heat}
= \kappa_{\rm heat} \etabh \dot{m}_{\rm radio} c^2$. Assuming that all
the hot gas is at the virial temperature of the halo $T_{\rm vir}$,
the mass of gas that can be heated per unit time is then
\begin{equation}
\dot{m}_{\rm
heat} = \frac{L_{\rm heat}}{\frac{3}{2} kT/(\mu m_p)} = \frac{L_{\rm
heat}}{\frac{3}{4} V^2_{\rm vir}},
\end{equation}
using $kT/(\mu m_p)=\frac{1}{2} V^2_{\rm vir}$.  The net cooling rate
is then the usual cooling rate $\dot{m}_{\rm cool}$ minus this heating
rate $\dot{m}_{\rm heat}$. If the heating rate exceeds the cooling
rate, the cooling rate is set to zero.

We apply this heating term \emph{only} for timesteps in which the
halos are cooling in the ``hot mode'' (see
\S\ref{sec:model:cooling}). That is, if $r_{\rm cool}>r_{\rm vir}$ in
a given timestep, we assume that the gas is not susceptible to the
heating by radio jets, so it cools at the normal rate.

\section{Results}
\label{sec:results}

\begin{figure*} 
\begin{center}
\plottwo{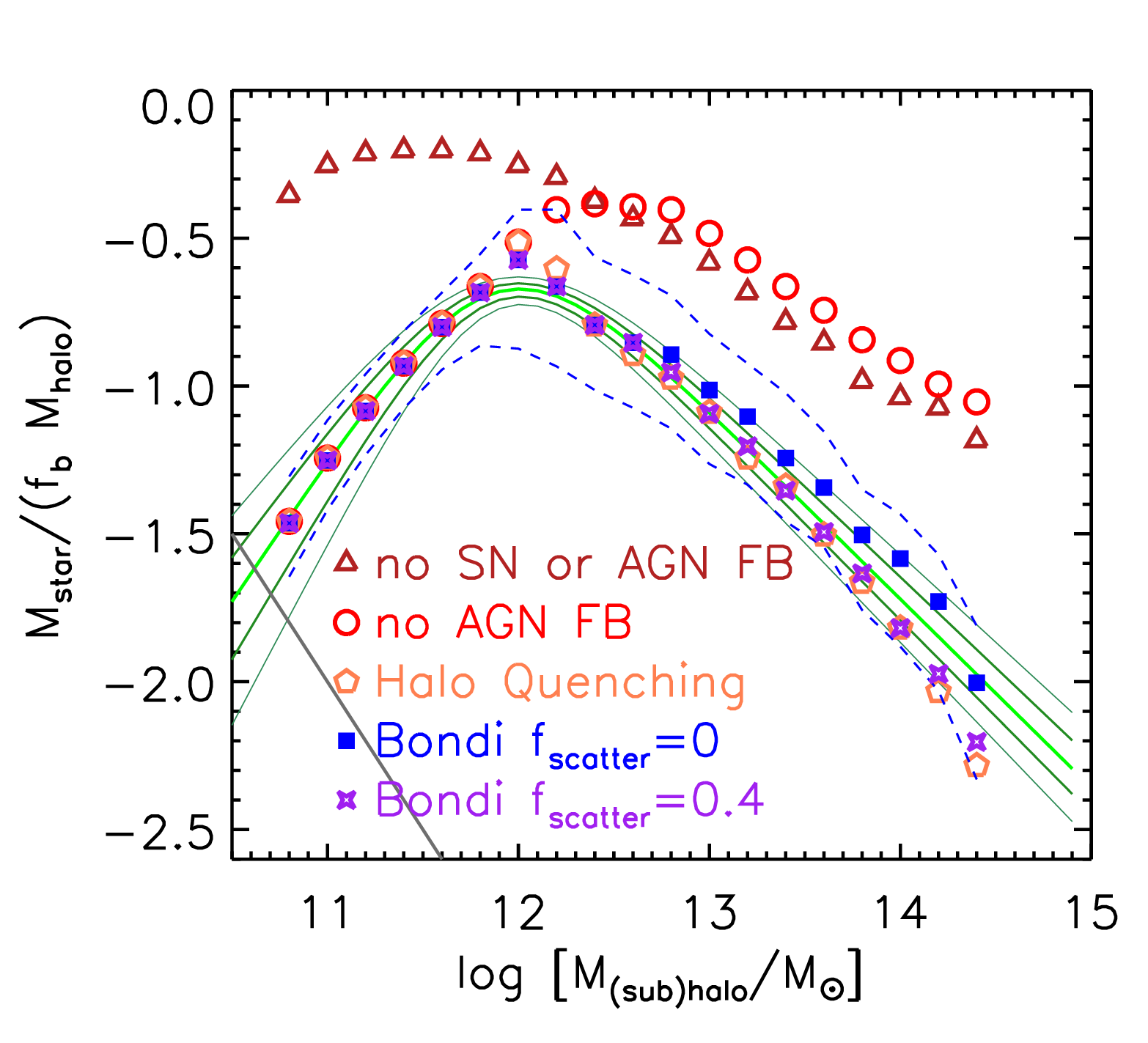}{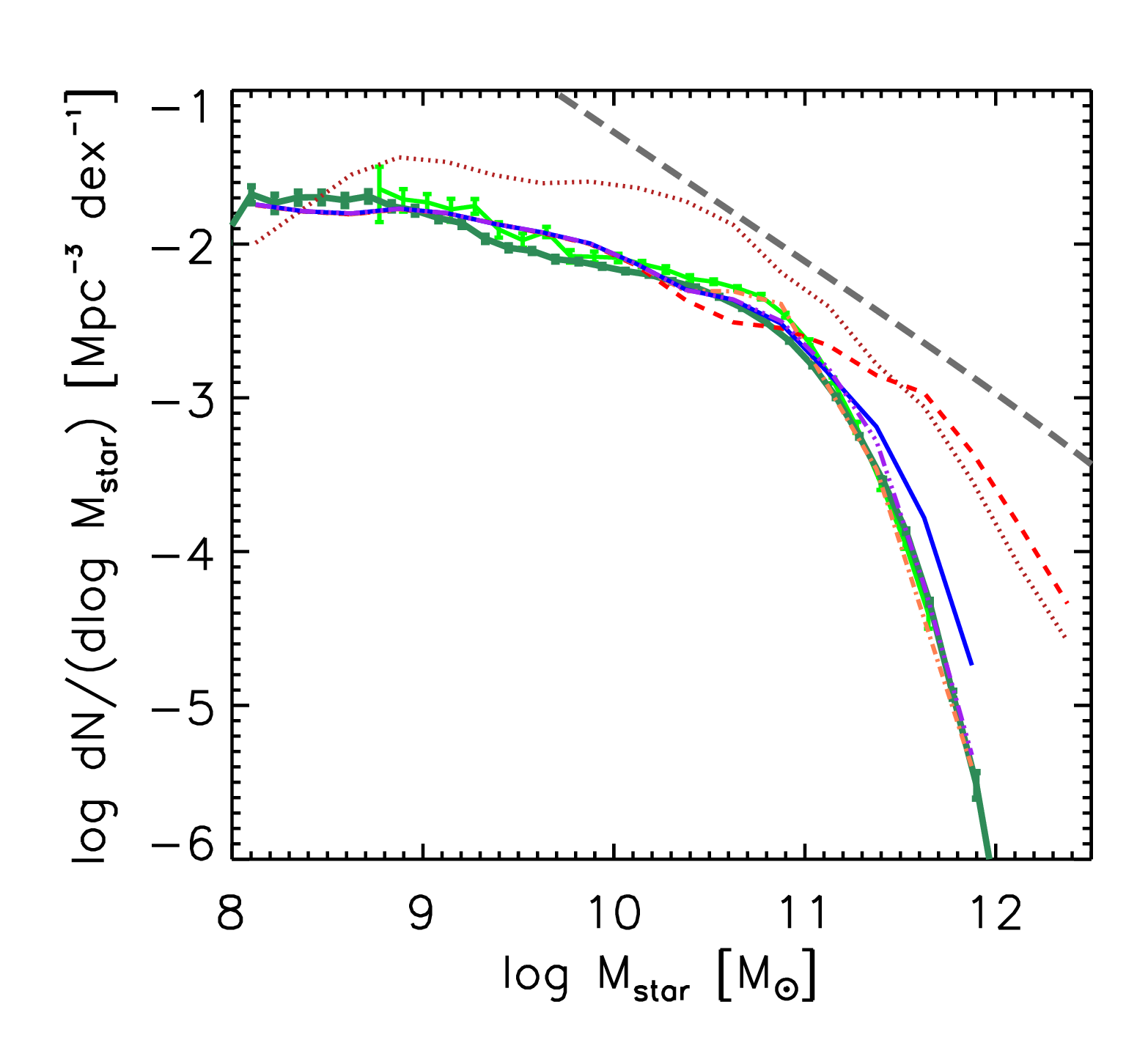}
\end{center}
\caption{\small Left panel: Fraction of baryons in the form of stars
  as a function of halo mass (for central galaxies) or sub-halo mass
  (for satellite galaxies). The solid green lines show the empirical
  relation (with 1- and 2-$\sigma$ errors) obtained by Moster et
  al. (2008; see text). Triangles (brown) show the models with no SN
  or AGN FB; open dots (red) show the model without AGN feedback;
  pentagons (orange) show the Halo Quenching model; solid (blue)
  squares show the fiducial (isothermal Bondi) model with $f_{\rm
    scatter}=0$, and crosses (purple) show $f_{\rm scatter}=0.4$. The
  dashed lines show the sixteen and eighty-fourth percentiles for the
  fiducial model.  The diagonal gray line in the bottom left corner
  shows the stellar mass corresponding to the smallest galaxies that
  we can accurately resolve, $\sim 10^{9} \msun$. Right panel: Galaxy
  stellar mass functions for the same models (dotted (brown) no SN or
  AGN FB; short dashed (red) no AGN FB; dot-dashed (orange) Halo
  Quenching; solid (blue) fiducial isothermal Bondi ($f_{\rm
    scatter}=0$); triple dot-dashed (purple) ($f_{\rm
    scatter}=0.4$). Green lines with error bars show the observed
  Galaxy Stellar Mass functions derived from SDSS by
  \protect\citet[][light green]{bell:03b} and \protect\citet[][dark
    green]{panter:07}. The long-dashed gray line shows the DM halo
  mass function with the masses shifted by a factor equal to the
  universal baryon fraction.
\label{fig:fstar}}
\end{figure*}

\subsection{Properties of nearby galaxies}
\label{sec:results:galprop}

If we adopt a specific set of values for the cosmological parameters,
it is relatively straightforward to answer the following question: how
must dark matter halo mass and galaxy mass (or luminosity) be related
in order to reconcile CDM with observations? Numerical N-body
simulations can now accurately predict the multiplicity function of
dark matter halos and sub-halos (i.e., the number density of halos of
a given mass), and it is then straightforward to adopt a parametric or
non-parametric model relating halo properties to galaxy properties,
and to adjust the model to fit the observed stellar mass function or
luminosity function of galaxies. This exercise has been carried out in
terms of luminosity by \citet[e.g.][]{kravtsov:04a}, and in terms of
stellar mass by \citet{wang:06} and \citet[][M08]{moster:08}. It has
been shown that if galaxies and halos are related in this way, one
then also reproduces the observed correlation functions of galaxies as
a function of stellar mass \citep{wang:06,moster:08}. We show the
function $f_{\rm star}(M_{\rm halo})$ derived by M08 in
Fig.~\ref{fig:fstar}. This quantity is defined as $f_{\rm star}(M_{\rm
  halo}) \equiv m_{\rm star}/(f_b M_{\rm halo})$, or the galaxy's
stellar mass divided by the universal baryon fraction times the halo
mass. For central galaxies, $M_{\rm halo}$ is the virial mass of the
halo. For non-central galaxies, $M_{\rm halo}$ is the virial mass of
the halo just before it became subsumed in a larger halo.

Just by comparing the halo mass function with the observed stellar
mass function (see Fig.~\ref{fig:fstar} right panel), we see that in
order to reconcile the DM halo mass function predicted by CDM with the
observed galaxy stellar mass function, star formation must not only be
inefficient overall ($f_{\rm star} \sim 0.2$--0.3 at its peak), but
the function must be a strong function of halo mass. Apparently, the
conversion of baryons into stars is highly inefficient both in small
mass halos and in large ones, and this efficiency peaks in halos with
mass $\sim 10^{12} \msun$. The interesting question that then arises,
of course, is which physical processes are responsible for shaping
this highly variable efficiency, and for setting the characteristic
halo mass scale $\sim 10^{12} \msun$?

The semi-analytic models can give us some insights into this
question. We note that our adopted halo mass resolution ($10^{11}
\msun$ for host halos, $10^{10} \msun$ for sub-halos) means that our
simulations should be reliable and complete for galaxies with stellar
masses greater than $\sim 10^9 \msun$. Below this mass, we cannot
accurately resolve a galaxy's formation history. If we switch off both
AGN feedback and SN feedback\footnote{In this model, we fix the
metallicity of the hot gas to be one-third of solar for purposes of
computing the cooling rates.}, $f_{\rm star}$ is far too high and too
flat below $10^{12}\msun$ (the mild decline at the low-mass end is due
to photo-ionization squelching). We adjust the parameters of our model
for SN-driven winds in order to match the empirical values of $f_{\rm
star}$ below $M_{\rm halo} \sim 10^{12}\msun$, and find that we
require $\epsilon_{SN}^0 \sim 1.3$, $\alpha_{\rm} \sim 2$, and $V_{\rm
eject} \sim 120$ km/s. Consulting Eqn.~\ref{eqn:snfb}, we see that
this implies that in large galaxies ($V_{\rm disk} \sim 200$ km/s),
the SN-driven mass outflow rate is comparable to the star formation
rate, and the outflow rate increases fairly strongly with decreasing
disk circular velocity $V_{\rm disk}$. This normalization is in good
agreement with the observational results of e.g. \citet{martin:99},
although it is unclear that the strong scaling with circular velocity
is supported by these observations. Also, these winds can escape the
potential well of the dark matter halo in halos with $V_{\rm vir}
\lesssim 120$ km/s, which is again consistent with the observations of
\citet{martin:99}.

In models with no feedback from AGN, we can see from
Fig.~\ref{fig:fstar} that $f_{\rm star}$ does turn over at large halo
masses: this is because large mass objects have formed more recently,
and have had less time to cool. In some of the earliest explorations
of galaxy formation in the CDM paradigm, it was suggested that this
cooling time argument could explain the characteristic mass scale of
galaxies \citep{bfpr:84,wr:78}. However, one can see that the turnover
occurs at too high a mass, and too much gas cools and forms stars in
large mass halos. This result is obtained not only in semi-analytic
models by many different groups
\citep[e.g.][]{benson:03,croton:06,cattaneo:06}, but also in numerical
hydrodynamic simulations \citep{balogh:01,borgani:06,cattaneo:07}.

For comparison, Fig.~\ref{fig:fstar} also shows the predictions for
$f_{\rm star}(M_{\rm halo})$ in a very simple implementation of the
concept of AGN heating, what we shall call the ``Halo Quenching'' (HQ)
model. In this model, we simply shut off cooling flows when the host
halo exceeds a mass of $M_Q = 1.3 \times 10^{12} \msun$. This model is
based on the idea that halos around $10^{12} \msun$ lie near the
transition between the ``cold flow mode'' (gas cooling more rapidly
than the free-fall time) and the formation of quasi-hydrostatic hot
halos (``hot flow mode''), and that radio jets from SMBH can easily
keep gas hot if it is in a quasi-hydrostatic hot halo, but not if it
is cooling in the cold flow mode
\citep{dekel_birnboim:06,binney:04}. This idea has been previously
implemented in a full semi-analytic model by \citet{cattaneo:06}, and
was found to very successfully reproduce both the luminosity function
and magnitude dependent color distributions of galaxies. We also find
that this model reproduces $f_{\rm star}(M_{\rm halo})$, and hence the
galaxy stellar mass function, extremely well.

We can then compare this with the prediction of our fiducial model, in
which accretion onto a central SMBH is modelled assuming Bondi
accretion and the isothermal cooling flow model of NF00 (see
\S\ref{sec:model:radio}). We adjust the scaling factor $\kappa_{\rm
  radio}$ in order to reproduce $f_{\rm star}(M_{\rm halo})$ as well
as possible over its whole range. Of course, in this model, the
scaling of the heating rate as a function of halo mass, and hence the
shape of $f_{\rm star}(M_{\rm halo})$, is determined by the isothermal
Bondi accretion model. Indeed, we can see that our fiducial model
(with $f_{\rm scatter}=0$) slightly overpredicts $f_{\rm star}$ for
large mass halos $M_{\rm halo} \gtrsim 10^{13} \msun$. This results in
a small excess of large mass galaxies ($M_{\rm star} \gtrsim 10^{11}
\msun$) in the predicted stellar mass function (Fig.~\ref{fig:fstar},
right panel). Note that at halo masses $M_{\rm halo} \lesssim
10^{12}$, the results for $f_{\rm star}$ are the same as in the model
without AGN feedback. The radio heating mode is ineffective in small
mass halos for multiple reasons: 1) low mass halos cool mainly in the
``cold flow'' mode, and we have assumed that the ``radio mode'' is
fueled by hot gas and that radio jets can only heat gas that is in a
quasi-hydrostatic hot halo 2) low mass halos tend to host
disk-dominated galaxies, which do not contain massive black holes.

How concerned should we be about the galaxies in high-mass halos being
too heavy in our fiducial model? The discrepancy amounts to about
0.15--0.2 dex in stellar mass\footnote{Although we do not present any
results in terms of luminosity in this paper, we see a similar
discrepancy in the predicted luminosity functions in all
bands. Therefore we probably cannot ascribe the problem to the stellar
mass estimates.}. A number of recent observational studies have
suggested that the luminosities and therefore stellar masses of the
central galaxies in clusters may be underestimated by a significant
factor (as much as 1.5 mags) in surveys such as SDSS and 2MASS
\citep{lauer:07,desroches:07,vonderlinden:07}, upon which our local
galaxy stellar mass function and luminosity function estimates are
based. Other studies have shown that a significant fraction of the
stars in these galaxies are distributed in a very extended ``halo'' or
envelope \citep{gonzalez:05,zibetti:05}.  One possible origin of this
extended ``diffuse stellar halo'' (DSH) is stars that are scattered to
large radii in mergers \citep{murante:04,murante:07}. In order to
explore this idea, we run a model in which a fraction $f_{\rm
scatter}$ of the stars in merged satellite galaxies is added to such a
diffuse component, which is tracked separately from the main stellar
body of the galaxy. We show the results of such a model with $f_{\rm
scatter}=0.4$ (probably an upper limit on the physically plausible
value of this parameter) in Fig.~\ref{fig:fstar}, and find that in
this model, the stellar mass function is reproduced extremely
accurately.

\begin{figure*} 
\begin{center}
\includegraphics[width=6.5in]{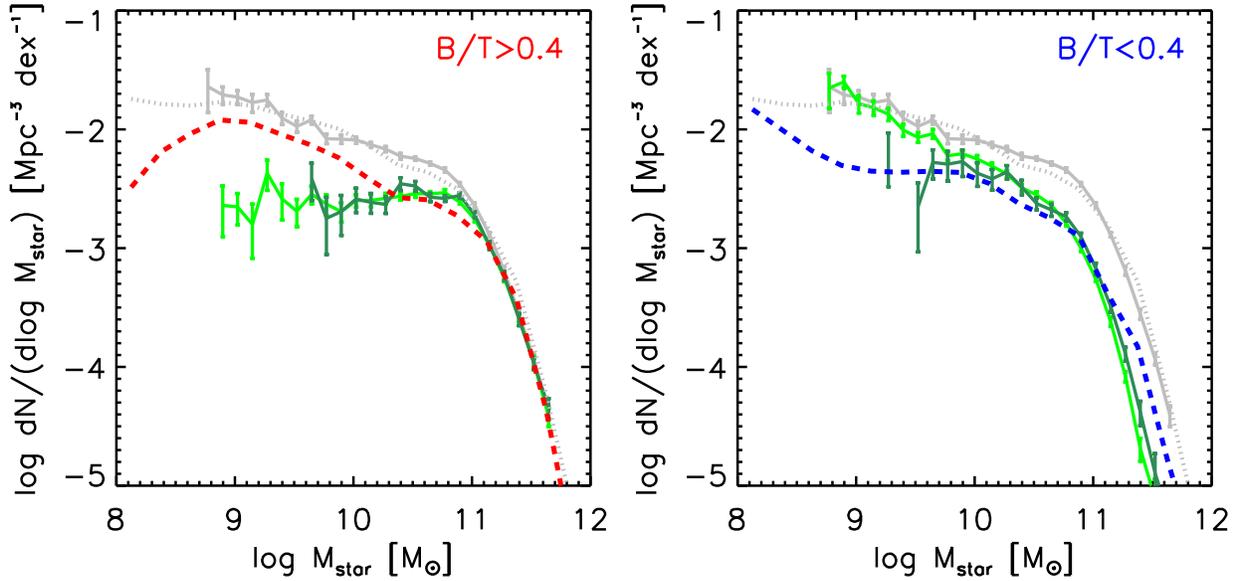}
\end{center}
\caption{\small The galaxy stellar mass function divided by
  morphological type. Solid lines with error bars show observational
  estimates from SDSS (light green) and 2MASS (dark green; Bell et
  al. 2003) for early type galaxies (left) and late type galaxies
  (right). Light (gray) lines show the observed mass function for
  galaxies of all morphological types. Dashed lines show the model
  predictions for bulge dominated ($B/T>0.4$; left) and disk-dominated
  ($B/T<0.4$; right) galaxies, for the fiducial isothermal Bondi model
  with $f_{\rm scatter}=0.4$.
\label{fig:mfstar_type}}
\end{figure*}

As discussed in \S\ref{sec:model:spheroid}, in each galaxy, we track
separately the stars that have survived in an undisturbed disk and
stars that have been ``heated'' by mergers to form a spheroid. Thus we
can assign a crude morphological type based on the ratio of the mass
in the ``bulge'' to that in the ``disk'', $B/T$. In
Fig.~\ref{fig:mfstar_type}, we show the stellar mass functions divided
into spheroid-dominated and disk-dominated galaxies, compared with
observational estimates similarly divided in terms of morphological
type \citep{bell:03b}, for the fiducial model with $f_{\rm
scatter}=0.4$.  There is a small excess of massive disk-dominated
galaxies, which may indicate that there is still a small degree of
overcooling in our most massive halos. There is also quite a large
excess of low-mass spheroid-dominated galaxies, and a deficit of
low-mass disk-dominated galaxies. These problems persist even if we
exclude satellite galaxies from our analysis, and cannot be eliminated
by simply adjusting the parameter $f_{\rm sph}$ without ruining the
agreement for massive galaxies.

\begin{figure*} 
\begin{center}
\plottwo{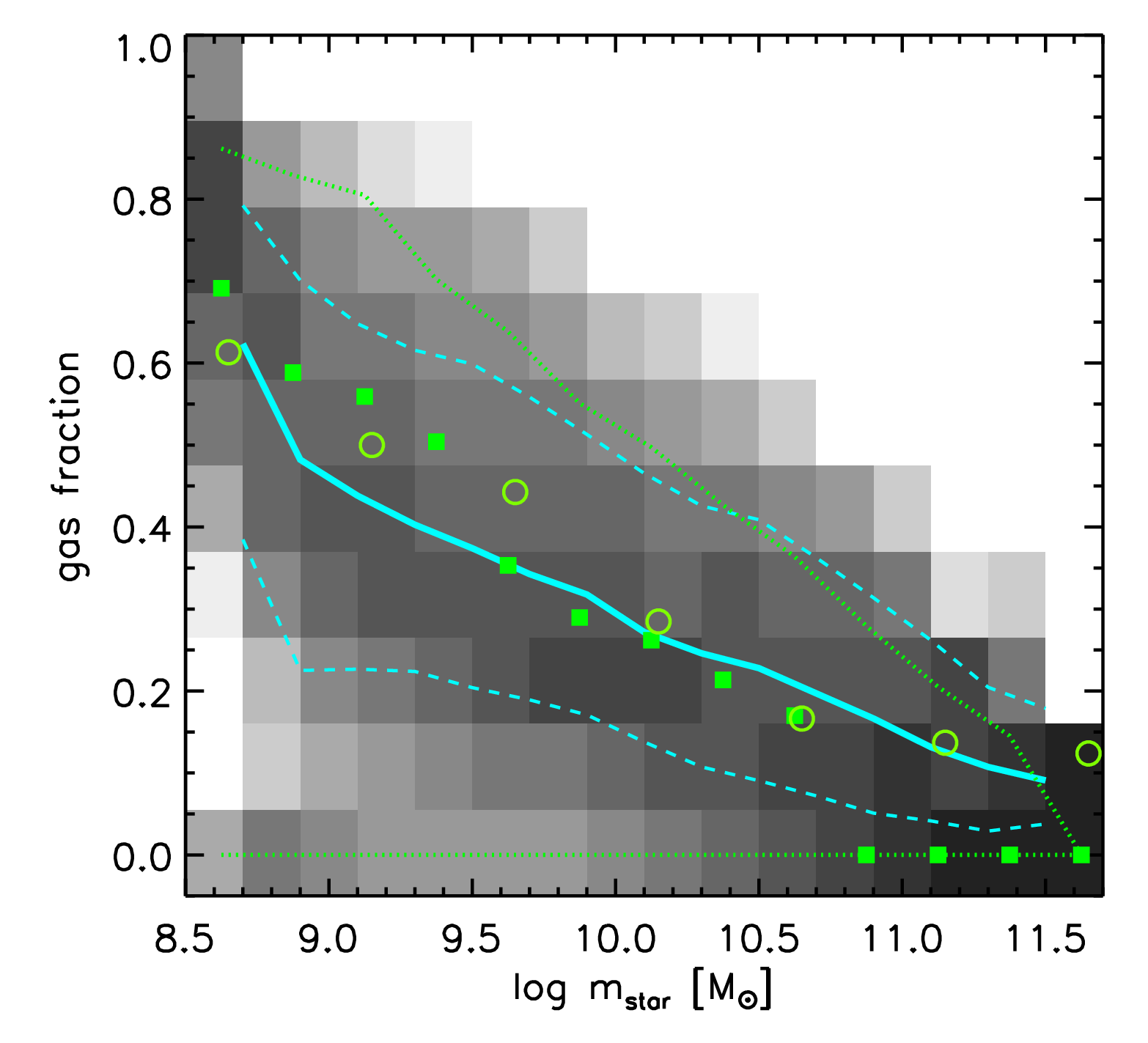}{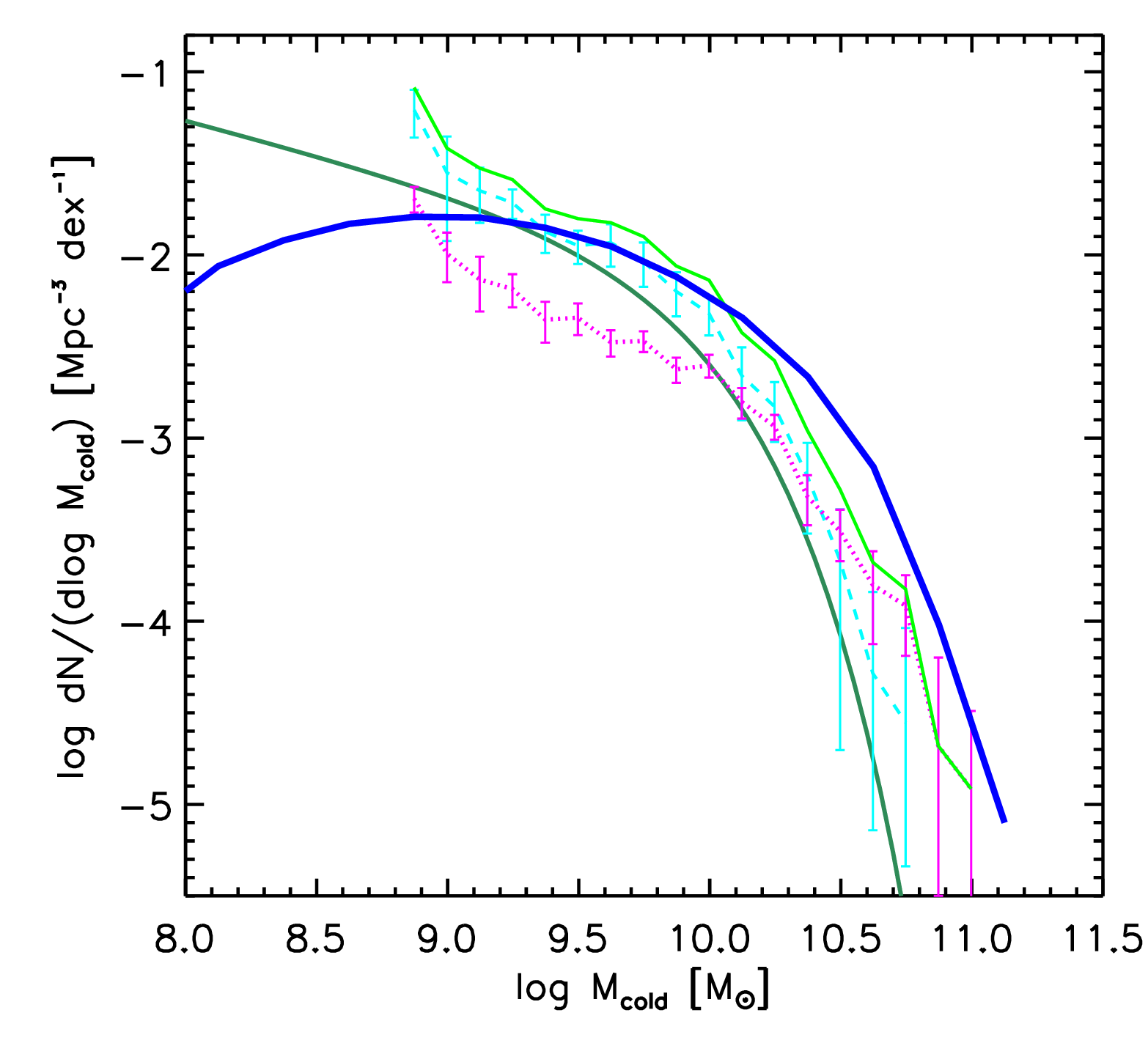}
\end{center}
\caption{\small Left: the cold gas fraction as a function of stellar
  mass.  The (green) squares show observational estimates for
  morphologically late-type galaxies derived from the data of
  \protect\citet{bell:03a}, and open circles show the observational
  estimates for blue galaxies from \protect\citet{kannappan:04}. The
  shaded area shows the conditional probability distribution $P(f_{\rm
    gas}|m_{\rm star})$ for central disk-dominated galaxies predicted
  by our fiducial (isothermal Bondi) model. The (light blue) solid
  line shows the median of this distribution, and dashed (light blue)
  lines show the 16 and 84th percentiles. Right: the galactic cold gas
  mass function. The thick (dark blue) line shows the prediction of
  our fiducial model. The thick solid (dark green) curve shows the
  observed \HI\ mass function of \protect\citet{zwaan:05}, the dashed
  (light blue) line the \HI\ mass function of
  \protect\citet{rosenberg:02}, the dotted (magenta) line the
  \Htwo\ mass function of \citet{keres:03}, and the solid (light
  green) line shows the sum of the \protect\citet{rosenberg:02}
  \HI\ mass function and the \Htwo\ mass function.
\label{fig:coldgas}}
\end{figure*}

Another of the important free parameters in our model is the
normalization of the star formation recipe $A_{\rm Kenn}$.  The
strongest constraint on this parameter is the ratio of cold gas to
stars in galactic disks. Increasing $A_{\rm Kenn}$ causes gas to be
converted into stars more rapidly and leads to lower gas fractions. We
compare the predicted cold gas fractions $f_{\rm gas} \equiv m_{\rm
cold}/(m_{\rm cold}+m_{\rm star})$ as a function of stellar mass in
our fiducial model with observational estimates in
Fig.~\ref{fig:coldgas}. For the models, we consider disk-dominated
galaxies ($B/T<0.4$) which are the central galaxies in their halo (we
suspect that the cold gas fractions of satellite galaxies may be too
low because we assume that all cooling gas is accreted onto the
central galaxy). We compare with the observational estimates of
\citet{bell:03a} for morphologically late-type galaxies and with
galaxies on the blue sequence from \citet{kannappan:04}.  The
agreement is good at stellar masses greater than $\sim 10^{9.5}$, but
the gas fractions are a bit low for lower mass galaxies.  Note that if
we had not adopted a critical density in our star formation law (see
\S\ref{sec:model:sf}), i.e. if we tune $\Sigma_{\rm crit} \rightarrow
0$, then the gas fractions in low mass galaxies come out far too low
and we do not reproduce the observed trend that gas fractions are
higher in low mass galaxies. A further check on the cold gas content
of our galaxies comes from observations of the H$_{\rm I}$ and \Htwo\
mass function. We show the prediction of our fiducial model compared
with these observations in Fig.~\ref{fig:coldgas}, and find reasonable
agreement, but with a hint of a deficit of low-gas-mass galaxies, and
a small excess on the high-gas-mass end.

\begin{figure*} 
\begin{center}
\plottwo{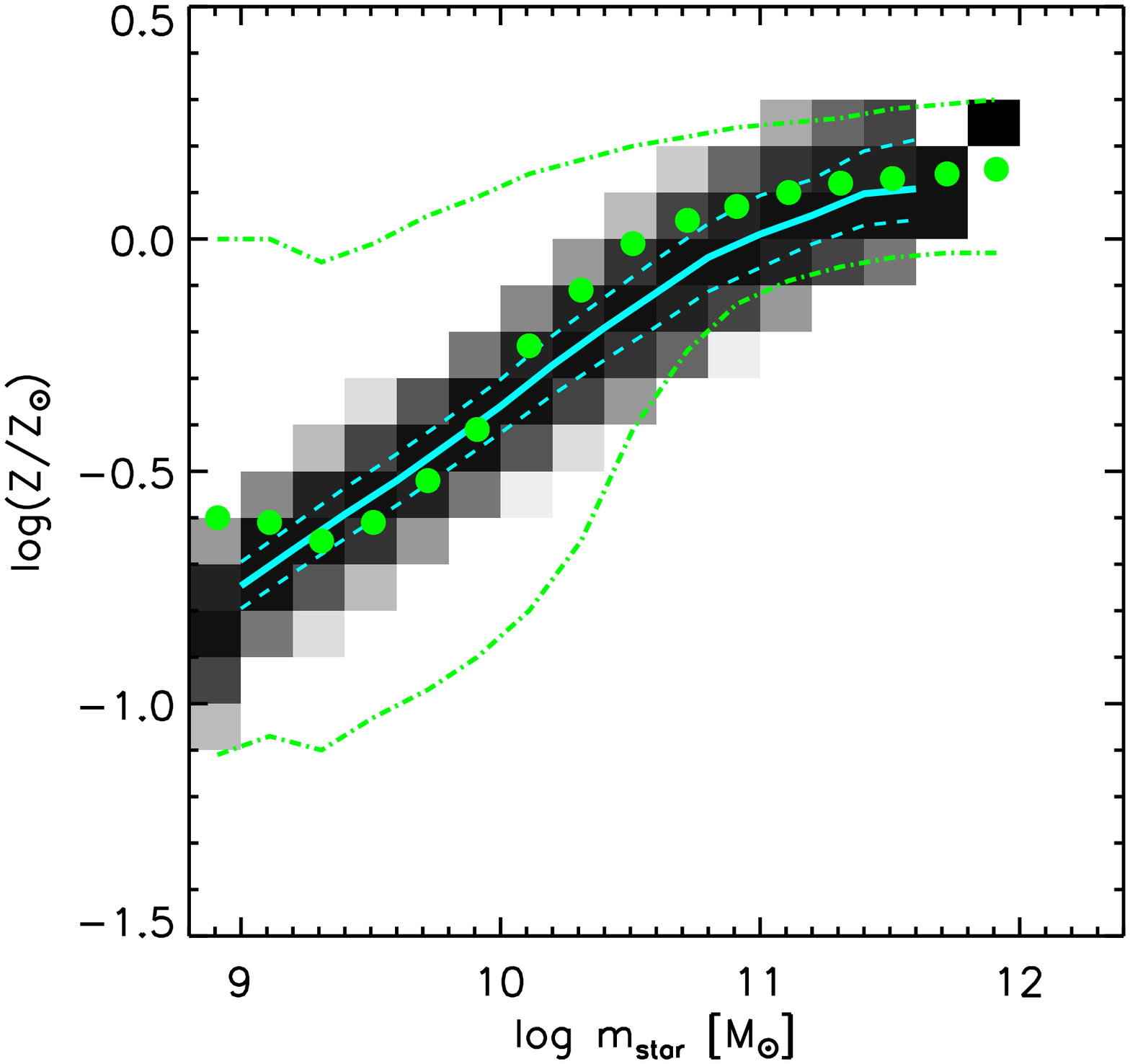}{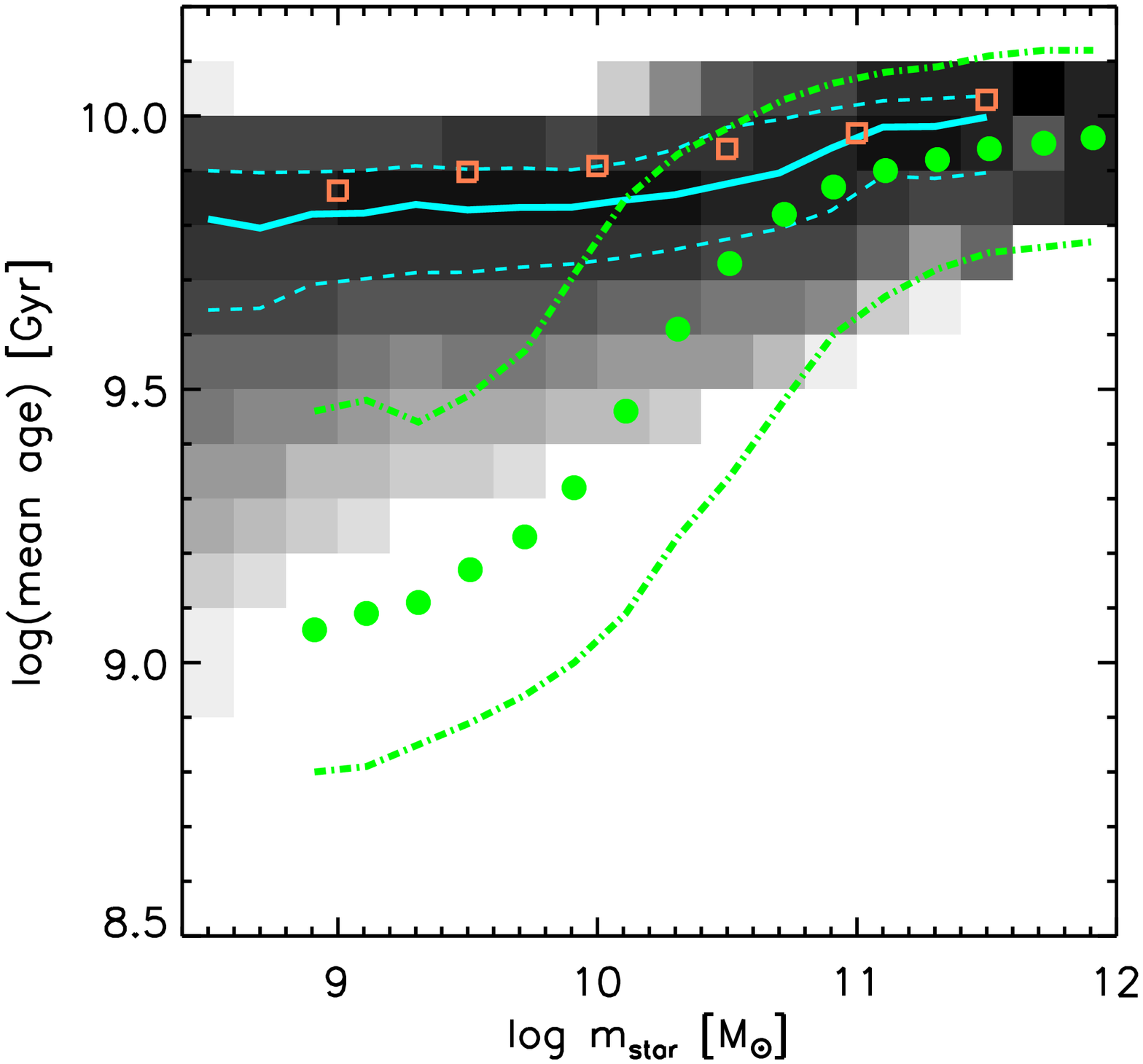}
\end{center}
\caption{\small Left: Stellar mass vs. stellar metallicity. Grey
  shading shows the conditional probability $P(Z_*|m_*)$ for our
  fiducial model ($f_{\rm scatter}=0.4$), and the (light blue) solid
  and dashed lines show the 50th, 16th and 84th percentiles. The
  (green) dots and dashed lines show the observational estimates from
  \protect\citet{gallazzi:05}. Right: Stellar mass vs. stellar
  mass-weighted mean stellar age. Shading, lines, and symbols are as
  in the left panel. Open square symbols show the predictions of the
  semi-analytic model of \protect\citet{croton:06} for comparison.
\label{fig:massmet}}
\end{figure*}

A complementary tracer of star formation is the heavy elements locked
up in stars. We show the stellar-mass weighted mean stellar
metallicity as a function of stellar mass for galaxies in our fiducial
model ($f_{\rm scatter}=0.4$) in Fig.~\ref{fig:massmet}. Our
predictions may be compared with observational estimates based on SDSS
spectra from \citet{gallazzi:05}. It is worth noting that the
estimates of \citet{gallazzi:05} effectively measure a combination of
$\alpha$-process elements and Fe, while our modelling includes only
enrichment due to Type II supernova, and therefore our
``metallicities'' correspond more closely to $\alpha$-type
elements. Also, the \citet{gallazzi:05} estimates are effectively
luminosity weighted, not stellar mass weighted, and may be
systematically biased towards higher values for supersolar
metallicities \citep[see the discussion
in][]{gallazzi:05}. Considering these potential biases, and the
relatively crude nature of our chemical evolution model, we find
fairly good agreement with the observed stellar mass vs. metallicity
relation. Note that although the normalization of this relation can be
adjusted by tuning the value of the stellar yield, $y$, (the results
shown here adopt $y=1.5$ in solar units), the shape of this relation
is a fairly complex product of various ingredients of the model. For
example, the low-mass slope is primarily determined by the mass
dependent star formation efficiency caused by our star formation
threshold and the strongly mass-dependent supernova feedback
efficiency that we have assumed. The turnover on the high mass end is
shaped by the quenching of star formation by AGN feedback and gas-poor
mergers.

In Fig.~\ref{fig:massmet} we also show our model predictions for the
stellar mass weighted mean stellar age of galaxies as a function of
stellar mass, compared with the observational estimates of
\citet{gallazzi:05}.  The models predict a weak trend of older ages in
more massive galaxies, with the ages of massive galaxies slightly
older than the observational estimates. However, the predicted trend
in our models is much weaker than the observed trend found by
\citet{gallazzi:05}, and low-mass galaxies in our models are much
older than the observations indicate. \citet{croton:06} showed that
models without AGN feedback predicted that massive galaxies have ages
as young as low-mass galaxies, and that introducing radio mode AGN
feedback produced an age-mass trend with the correct sense (more
massive galaxies are older). However, they did not compare directly
with observational estimates. We reproduce the predictions from the
AGN feedback model shown in Figure~10 of \citet{croton:06}, and see
that their results are very similar to ours (in fact low-mass galaxies
are slightly older in their models than in ours). We should keep in
mind that if a galaxy has a significant-by-mass older stellar
population with a small ``frosting'' of young stars, the ages derived
from stellar absorption lines (mainly Balmer lines) as in the
\citet{gallazzi:05} approach will be biased towards young ages. This
discrepancy is worth examining in more detail, but this is beyond the
scope of this paper.

\begin{figure} 
\begin{center}
\plotone{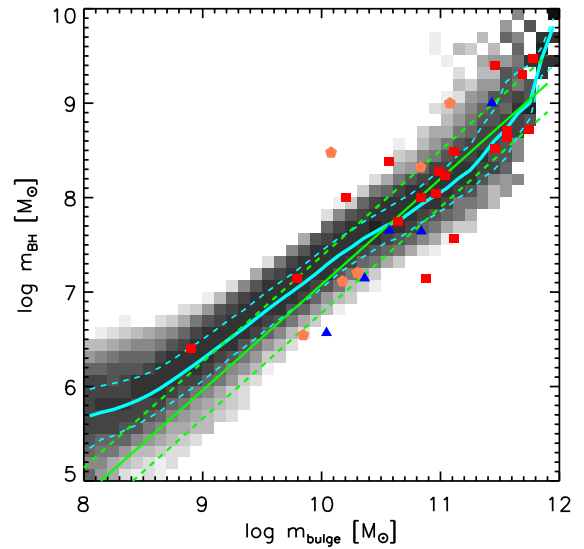}
\end{center}
\caption{\small Predicted relationship between bulge mass and black
hole mass (grey shading indicates the conditional probability
$P(m_{\rm bh}|m_{\rm bulge})$; light blue solid and dashed line shows
the median and 16th and 84th percentiles) compared with the observed
relation from \protect\citet[][green lines]{haering:04}. Symbols
show the measurements for individual galaxies from
\protect\citet{haering:04}. 
\label{fig:mbh}}
\end{figure}

The relationship between galaxy mass and BH mass is clearly a key
result that our model should reproduce. Recall that in our model, this
relationship is set by the depth of the potential well of the galaxy
at the time when the BH forms, which in turn is determined by the gas
fraction of the progenitor galaxies of the last merger (see
\S\ref{sec:model:smbh}). More gas-rich progenitors suffer more
dissipation when they merge, and produce more compact remnants with
deeper potential wells. A deeper potential well requires more energy,
and therefore a more massive BH in order to halt further accretion and
growth. Although we have seen that the predicted gas fractions of
disks at the present day agree reasonably well with observations, the
gas fractions of the progenitors of black hole hosts depend on many
factors, such as the masses of those progenitors at the time when the
BH is formed, the epoch of formation of BH of a given mass, and the
details of the star formation and feedback modelling. It is therefore
a non-trivial success of our model that we reproduce the observed
slope and scatter of the \mbh-$M_{\rm sph}$ (black hole mass
vs. spheroid mass) relationship, as seen in Fig.~\ref{fig:mbh}. It is
interesting that our model predicts a small upward curvature at the
high-mass end, which \citet{wyithe:06} argue is present in the
observed relation.  Our predicted relation also has a somewhat flatter
slope at low BH masses than the extrapolation of the
\citet{haering:04} results; however, there are currently very few
robust BH mass estimates at such low masses.

\subsection{Group and Cluster Properties}
\label{sec:results:groupcluster}

\begin{figure*} 
\begin{center}
\plottwo{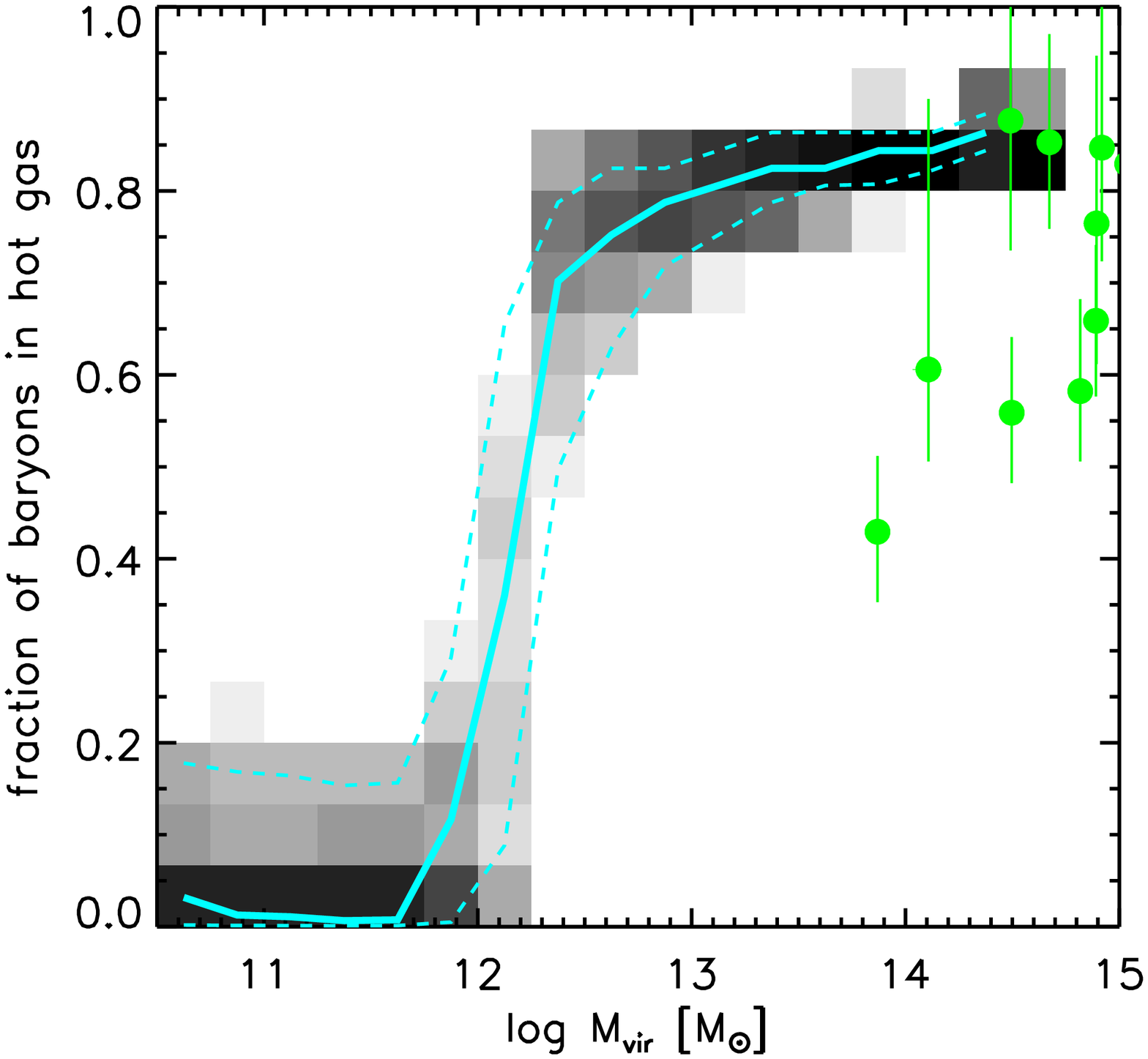}{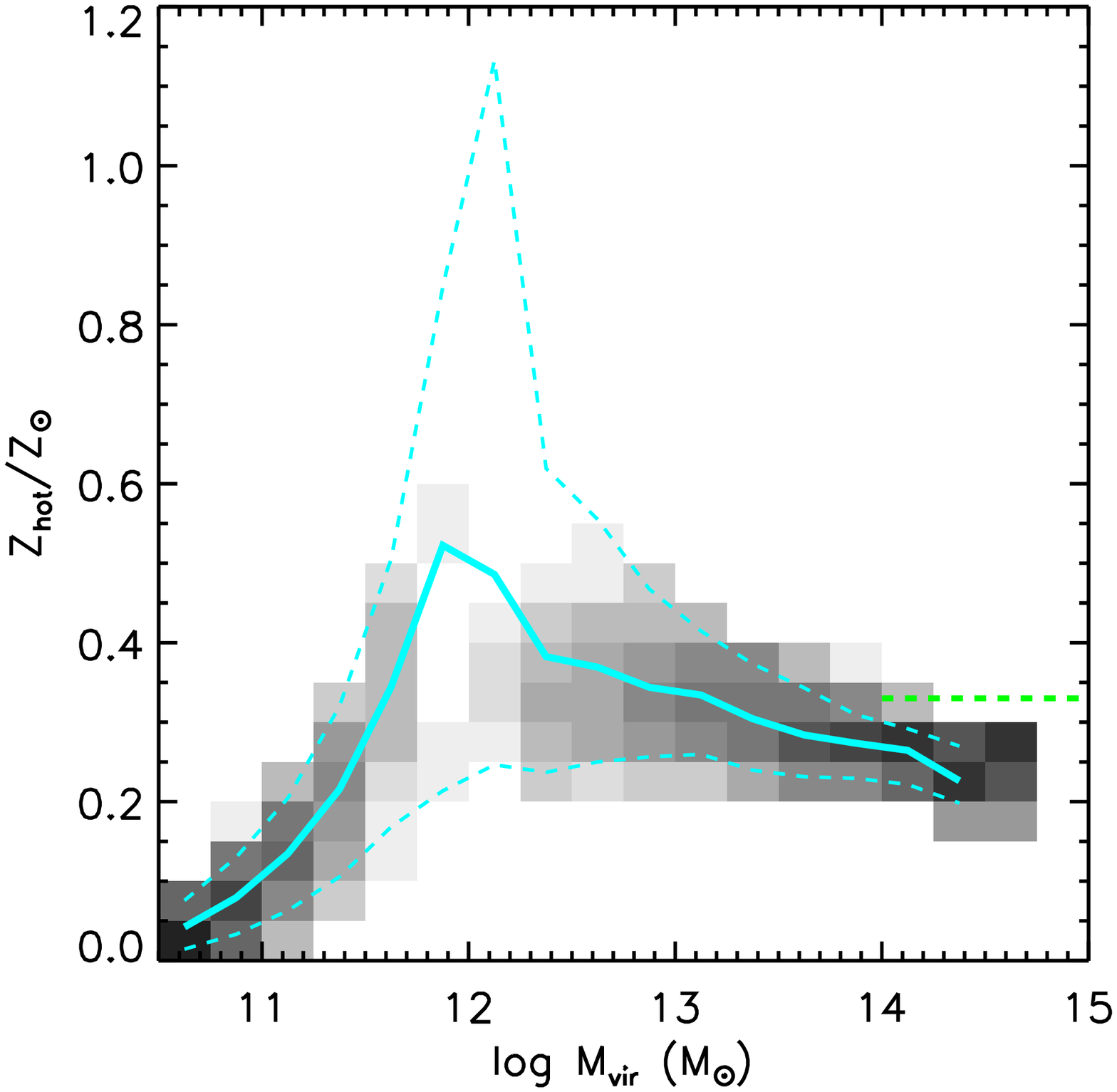}
\end{center}
\caption{\small Left: the fraction of halo baryons in the form of hot
gas as a function of halo virial mass. The gray shaded area shows the
conditional probability distribution $P(f_{\rm hot}|M_{\rm vir})$ for
our fiducial model, with the 16, 50, and 84th percentiles shown with
(light blue) curves.  The sharp discontinuity at $M_{\rm vir} \simeq
10^{12} \msun$ represents the transition from rapid cooling (cold
flows) to the formation of a hot halo.  The (green) solid circles show
the observational estimates of hot gas fraction from
\protect\citet{vikhlinin:06}. Right: the metallicity of hot cluster
gas as a function of halo mass in our fiducial model. The green dashed
line shows the observed value in clusters, approximately one-third of
the Solar value \protect\citep{arnaud:92}.
\label{fig:clusters}}
\end{figure*}

We have so far focussed on the properties of individual galaxies. We
now consider predictions for a few properties of groups and
clusters. We have selected these quantities because they help to
constrain some of the free parameters or uncertain ingredients in our
models. In Fig.~\ref{fig:clusters}, we show the hot gas fraction
($f_{\rm hot} \equiv m_{\rm hot}/M_{\rm vir}$), i.e., the mass of hot
gas contained in the dark matter halo divided by the total
virial mass of the halo. The hot gas fraction in our models has a
sharp ``step'' at $M_{\rm vir} \sim 10^{12}\msun$, because of the
rapid transition between halos in which the gas cools rapidly compared
with the dynamical time, so there is typically little hot gas present
in the halo, and halos in which the cooling time is longer compared
with the dynamical time, so the halo can build up a reservoir of hot
gas. In our fiducial model, this hot gas is then maintained by AGN
``radio mode'' heating. Our results are in reasonable agreement with
the hot gas fractions in clusters estimated from observations of their
X-ray emitting gas by \citet{vikhlinin:06}. These observations are
somewhat uncertain, because the X-ray emission typically cannot be
detected all the way out to the cluster virial radius, so it must be
extrapolated. However, these observations provide an important
constraint on the modelling of re-infall of gas that has been ejected
by supernovae (see \S\ref{sec:model:snfb}). If we do not allow this
ejected gas to be reaccreted at all, then the baryon fractions in
clusters are predicted to be significantly smaller than the universal
value, in conflict with observations.

In Fig.~\ref{fig:clusters} we also show the predicted metallicity of
the hot gas in halos. In our model, the hot gas is enriched by the
ejection of metals from the cold gas in galactic disks by
supernova-driven winds. We tuned the chemical yield $y$ to reproduce
the metallicities of stars in galaxies, so the metallicity of the hot
cluster gas is a cross-check on our chemical evolution and supernova
feedback modelling. We find that the hot gas in cluster-mass halos is
enriched to about 0.25 of the solar value, and is nearly constant
above about $M_{\rm vir} \sim 10^{13} \msun$. This is close to the
value of $\sim 0.3 Z_{\rm \odot}$ measured for hot gas in clusters
\citep{arnaud:92}. These measurements of hot gas metallicity are
primarily sensitive to Iron, while as we discussed earlier in this
section, our chemical evolution modelling traces only the metals
produced by Type II supernovae, so we do not expect perfect
agreement. Also, some additional metals may be driven out of the
galaxies by strong shocks during mergers \citep{cox:06c}.

\begin{figure} 
\begin{center}
\plotone{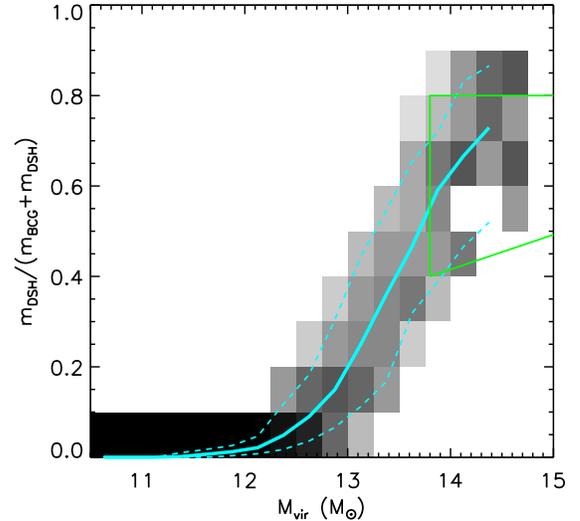}
\end{center}
\caption{\small Mass contained in a ``diffuse stellar halo'' (DSH)
relative to the mass of the central galaxy plus the DSH mass. The
shading and (light blue) curves show the predictions of our fiducial
model; the green box shows the approximate locus of observational
estimates from \protect\citet{gonzalez:05}.
\label{fig:fdsc}}
\end{figure}

We have argued that a significant fraction of stars may be scattered
into a ``diffuse stellar halo'' (DSH) by mergers. It is important to
check whether the predicted mass of stars in these DSH is in agreement
with direct observational measurements. In Fig.~\ref{fig:fdsc} we show
the mass of the DSH divided by the total mass of the DSH plus the main
galaxy ($f_{\rm DSH} \equiv m_{\rm DSH}/(m_{\rm DSH}+m_{\rm BCG})$),
as a function of the virial mass of the halo. The model predictions
are consistent with the range of observational estimates from
\citet{gonzalez:05}, which we have adopted from the results presented
in \citet{conroy:07}. In agreement with previous studies by
\citet{monaco:06} and \citet{conroy:07}, we find that the model with
$f_{\rm scatter} =0.4$ is able to reproduce the observed stellar mass
function of galaxies at $z\sim0$ as well as the fraction of stellar
mass in the DSH component \footnote{We note that the published value
  of $f_{\rm DSH} \simeq 0.33$ from \protect\citet{zibetti:05} is
  significantly lower than the values we have adopted here. While
  several factors may play a role, such as the different extent of the
  photometry and sample selection, much of the discrepancy with the
  results of \protect\citet{gonzalez:05} is apparently due to the
  details of the way the BCG and DSH (or intracluster light) are
  defined (A. Gonzalez and S. Zibetti, priv. comm.; see also
  \protect\citet{zibetti:iau}, Section 5.1). When Zibetti adopts the
  same decomposition method as Gonzalez, he finds much more consistent
  values $f_{\rm DSH} \simeq 0.67$ \protect\citep{zibetti:iau}.}.

\subsection{Radio Mode Heating}
\label{sec:results:radio}

\begin{figure} 
\begin{center}
\includegraphics[width=6.8cm]{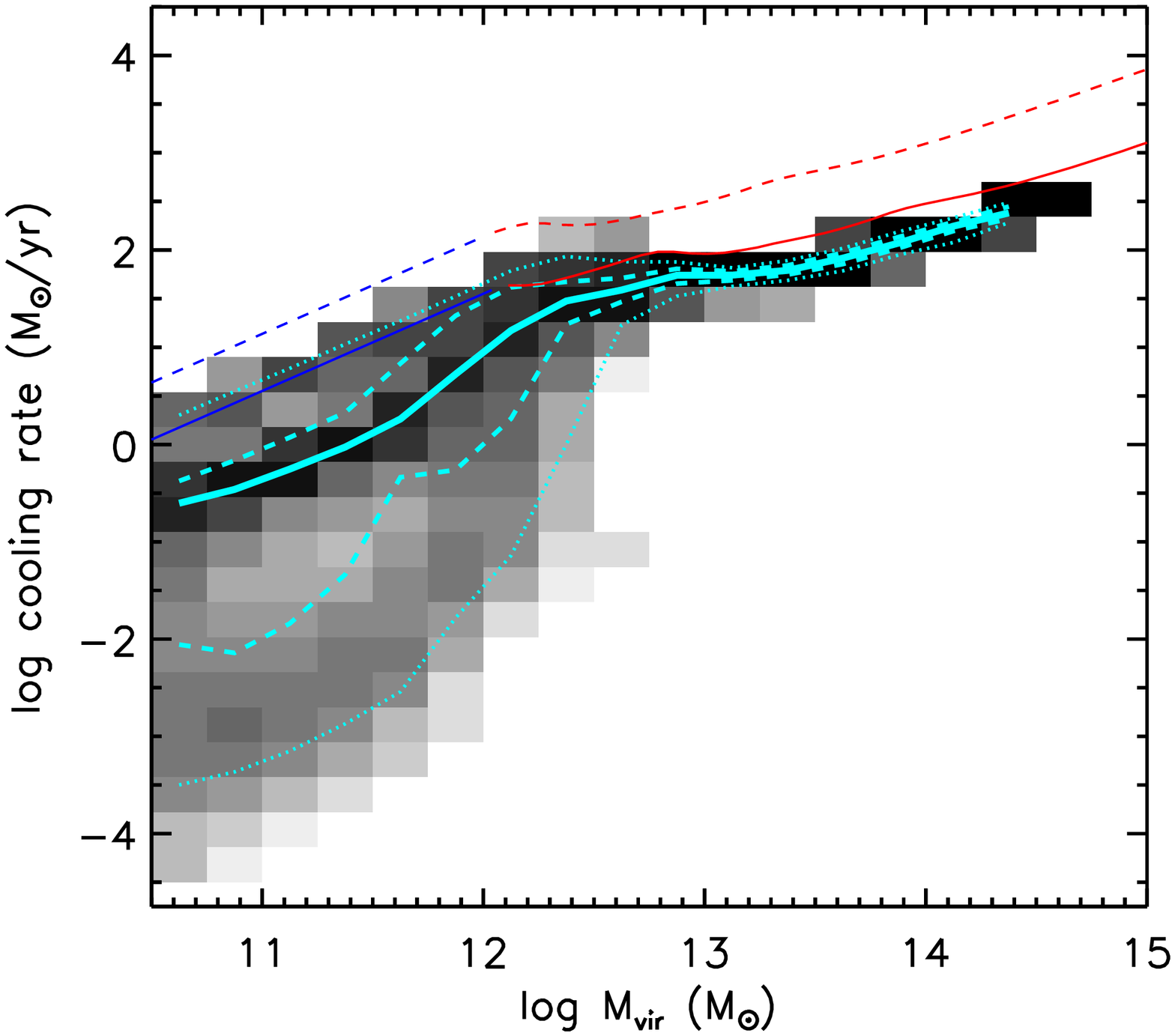}
\includegraphics[width=6.8cm]{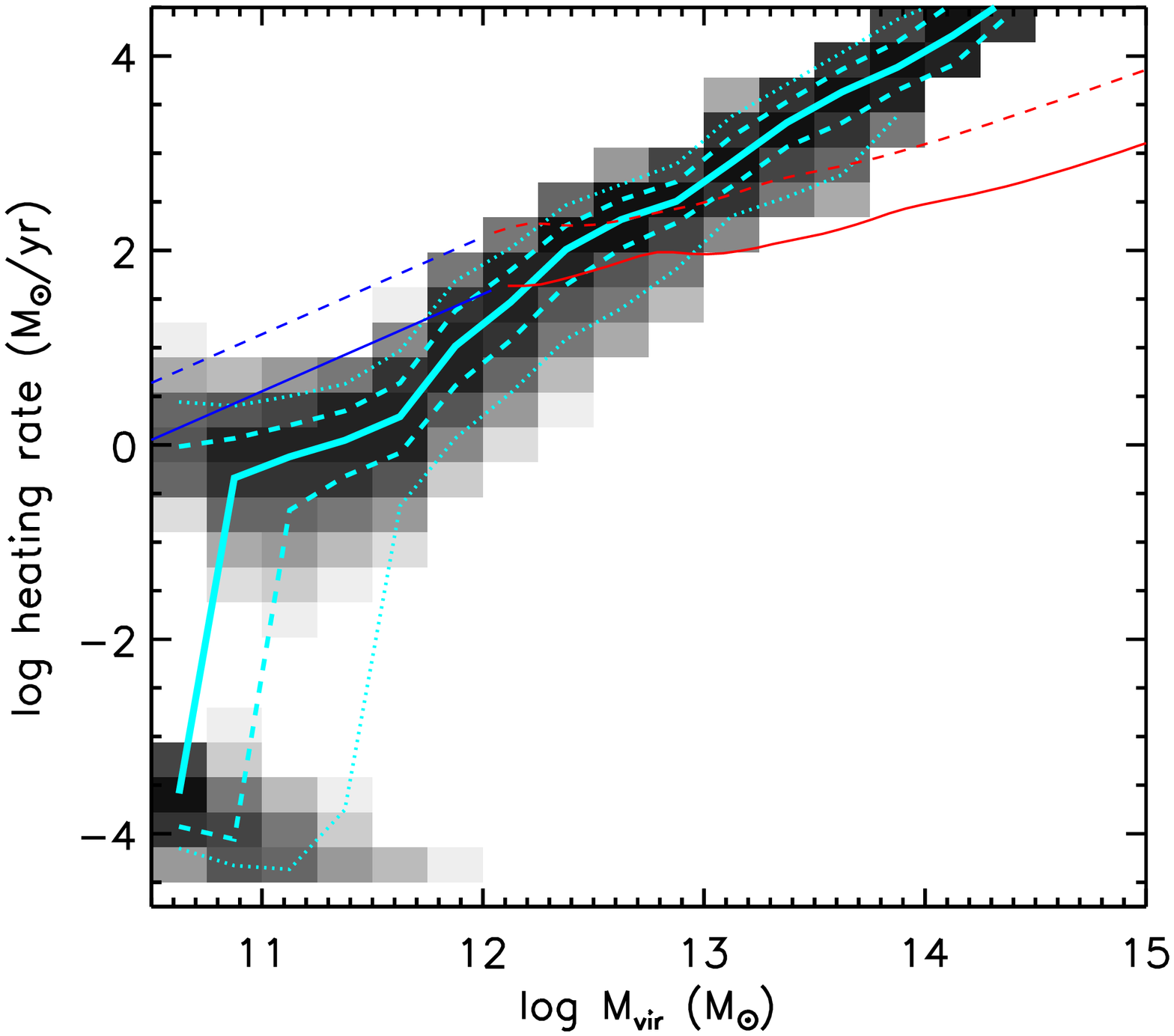}
\includegraphics[width=6.8cm]{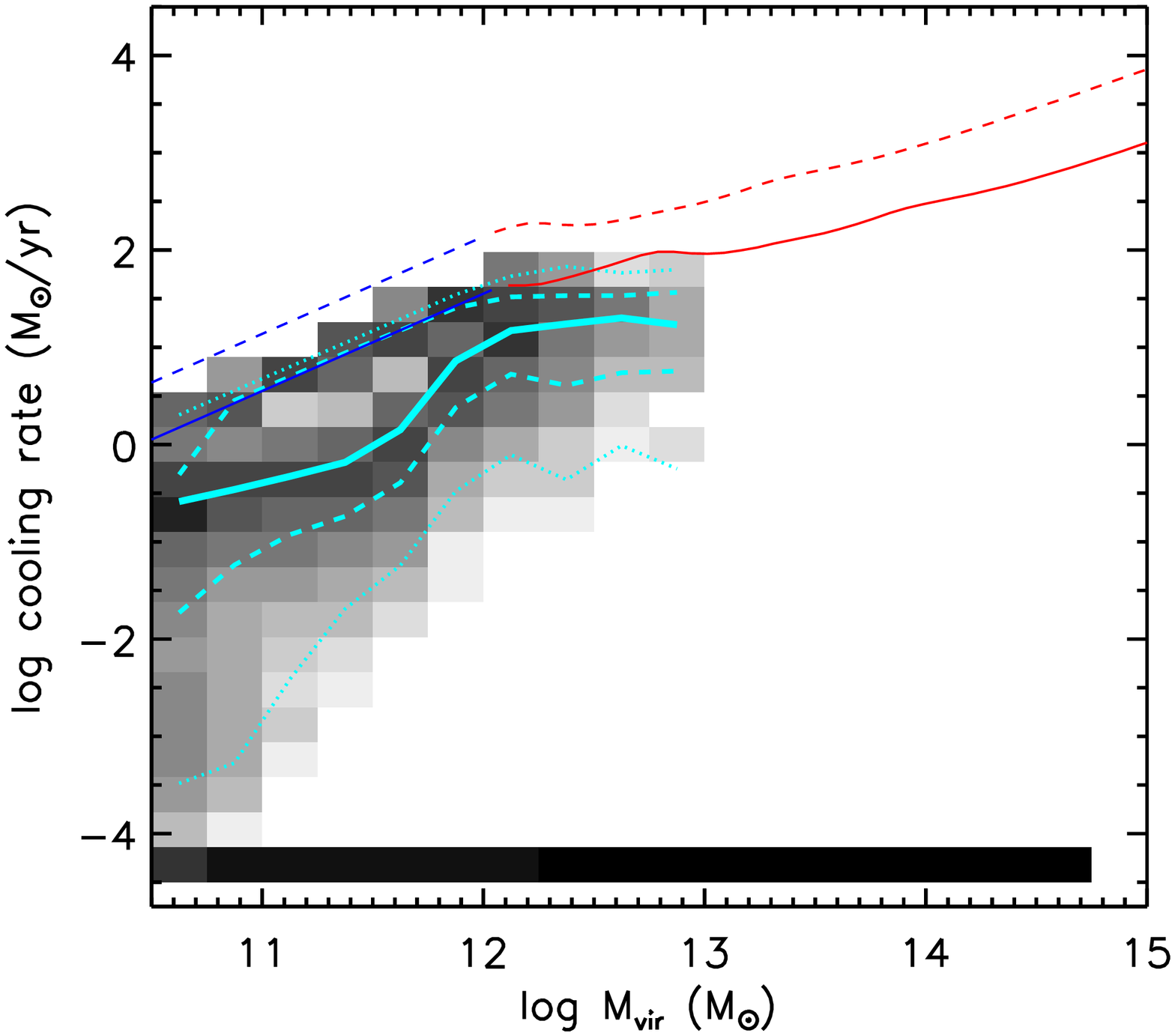}
\end{center}
\caption{\small Top: cooling rate as function of halo mass, for model
  with no AGN feedback. Grey shading indicates the conditional
  probability $P(\dot{m}_{\rm cool}|M_{h})$. Light blue lines indicate
  the 2.2, 16, 50, 84, and 98th percentiles. The smooth red and blue
  solid and dashed lines indicate the expected cooling rates in a
  ``static halo'' model (see text) for $z=0$ and $z=2$, respectively;
  the dark blue section of these lines indicates the approximate
  regime for ``cold mode'' infall, and the red lines for ``hot mode''
  (see text). Middle: heating rate by ``radio jets'' in our fiducial
  (isothermal Bondi) model. Bottom: net cooling rate in fiducial model
  with radio mode feedback. Halos with cooling rates below $10^{-4.5}
  \msun \rm{yr}^{-1}$ are plotted in the bottom-most bin.
\label{fig:cooling}}
\end{figure}

Several other groups have implemented heating by radio jets from AGN
in semi-analytic models, and shown that in this way they can solve the
overcooling problem and other related problems
\citep{croton:06,bower:06,monaco:07}. However, these works have not
addressed whether the \emph{amount of energy required} or the
\emph{scalings as a function of halo mass} adopted in these models are
consistent with constraints from observations of radio galaxies and
cooling flow clusters. We turn now to this
question. Fig.~\ref{fig:cooling} shows the predicted cooling rate as a
function of halo mass in a model without AGN heating. The shaded area
shows the cooling rates predicted by the full semi-analytic model,
while the smooth lines show the cooling rate given by
Eqn.~\ref{eqn:mcooldot}, assuming that $m_{\rm hot} = f_b M_{\rm vir}$
and $Z_{\rm hot}/Z_{\odot}=0.33$ (we refer to this as the ``static
halo'' cooling model). The lower of these lines is for redshift $z=0$,
and the the higher is for $z=2$. The divide between ``cold mode'' and
``hot mode'' halos at $\sim 10^{12} \msun$ is indicated by the color
of the lines (with blue indicating cold mode, red indicating hot
mode). Clearly, the static halo model prediction can differ from the
cooling rate in the full SAM because, as we saw in
Fig.~\ref{fig:clusters}, halos of a given mass have a range of values
of hot gas fraction and metallicity due to their different formation
histories. Here we see that, in the absence of AGN heating, cluster
mass halos would be expected to have cooling flows of hundreds up to
one thousand solar masses per year, which we know to be in conflict
with X-ray observations. In the middle panel we show the rate at which
gas is heated by the ``radio jets'' in our fiducial (isothermal Bondi)
model. Note that although we show non-zero heating rates below $\sim
10^{12} \msun$, actually, most of these halos are cooling in the
``cold flow'' mode and so their cold gas accretion rates are
unaffected by the AGN heating. We note that the heating and cooling
rates cross near the ``magic'' halo mass of $\sim 10^{12} \msun$, and
that the heating rate is a steeper function of halo mass than the
cooling rate at large masses, so that there is a lot of ``excess''
energy being deposited in the hot gas. At the moment, in our simple
modelling, this excess energy is not accounted for. The final panel in
this plot shows the net cooling rate including the AGN
heating. Cooling flows are quenched entirely in halos more massive
than $M_{\rm vir} \sim 10^{13} \msun$. However, this is not a sharp
cutoff. There is a transition region between $10^{12} \lesssim M_{\rm
  vir} \lesssim 10^{13}$ where some halos have had their cooling flows
quenched and some have only had them reduced.

\begin{figure*} 
\begin{center}
\plottwo{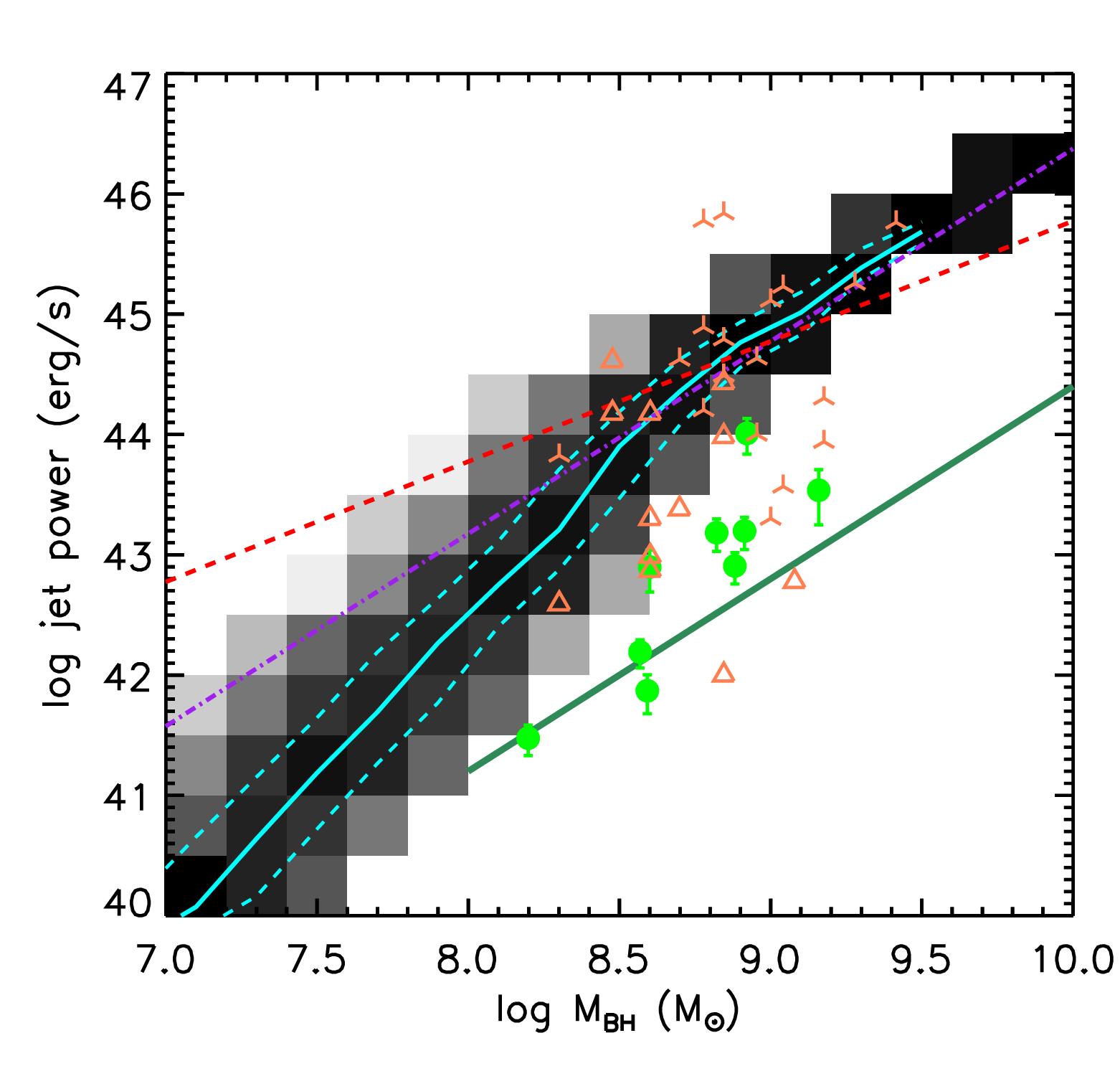}{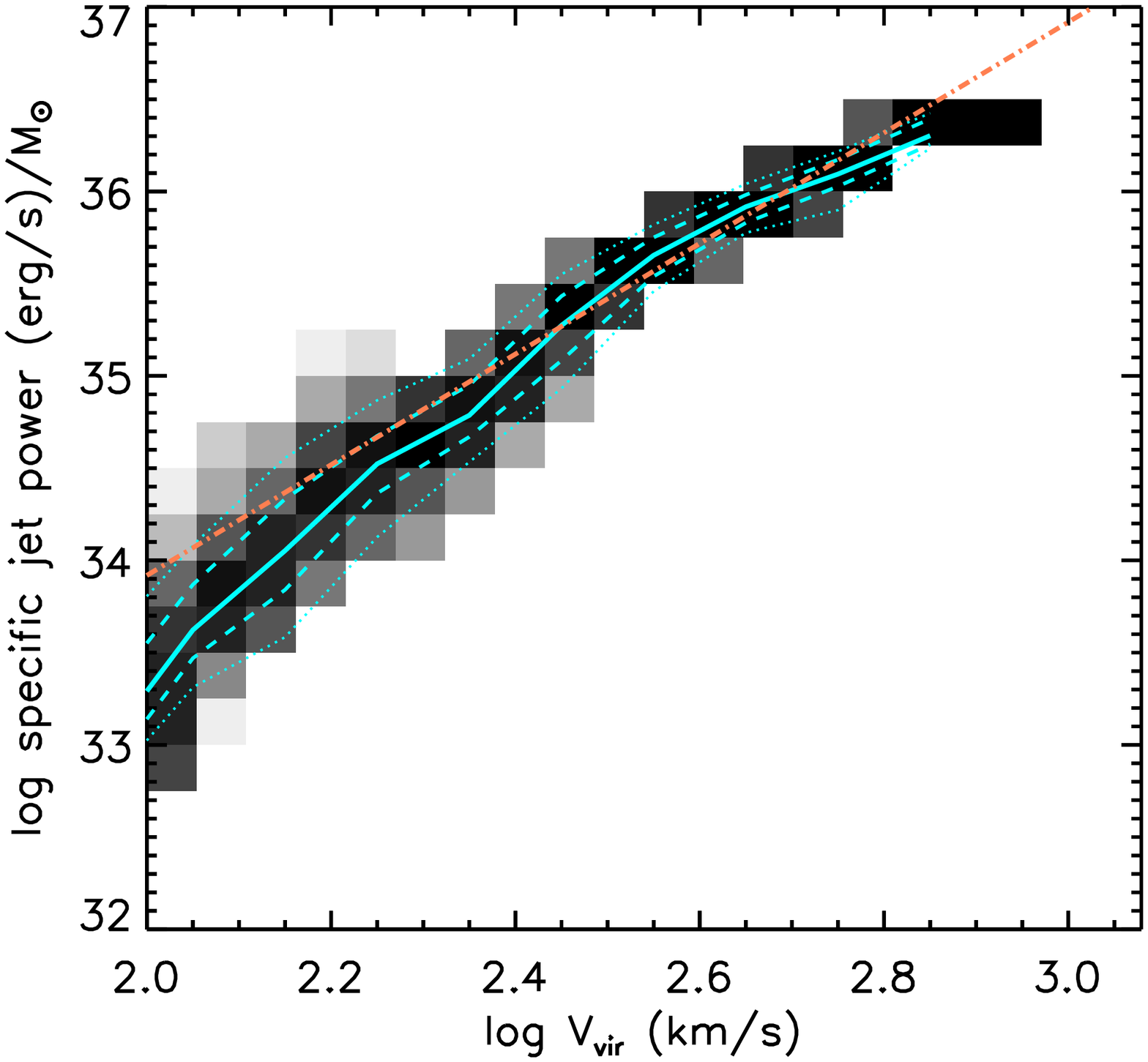}
\end{center}
\caption{\small Left: Rate of energy input by the ``radio mode'' (jet
  power) as a function of BH mass. Large, solid green circles show the
  observational estimates from \protect\citet{allen:06}. Open and
  skeletal triangles (orange) show the observational estimates from
  \protect\citet{rafferty:06}, for systems at redshift $z<0.05$ and
  $z>0.05$, respectively. The thick (dark green) solid line shows the
  time-averaged heating rate derived from observations by
  \protect\citet{best:06}. The gray shaded area shows the conditional
  probability distribution $P(P_{\rm jet}|m_{\rm BH})$, and the light
  blue curves show the median and 16 and 84th percentiles of this
  distribution. The (red) dashed line shows the heating rate that
  would result if all BH accreted at a fixed fraction ($f_R = 3.5
  \times 10^{-3}$ is shown here) of their Eddington rate, and the
  (purple) dot-dashed line shows $P_{\rm jet} \propto m_{\rm
    BH}^{1.6}$.  Right panel: the jet power divided by the black hole
  mass (specific jet power) as a function of halo virial velocity
  ($\propto T_{\rm vir}^{1/2}$). The gray shaded region shows the
  conditional probability distribution, and light blue lines show the
  2.2, 16, 50, 84, and 98th percentiles (from bottom to top), for our
  fiducial isothermal Bondi model. The dot-dashed (orange) line shows
  the scaling for the fiducial model of \protect\citet{croton:06}.
\label{fig:lheat}}
\end{figure*}

We can compare the heating rates needed in our model in order to solve
the overcooling problem and the galaxy mass problem with observations
of the hot bubbles associated with radio jets, seen in the X-ray gas
in groups and clusters. By estimating the amount of energy required to
inflate the bubbles, these systems can be used as ``calorimeters'',
giving an estimate of the power being injected by the
jets. \citet{allen:06} find a tight correlation between the Bondi
accretion rate, and the jet power. For systems that also have a black
hole mass estimate or its proxy, for example from a measured velocity
dispersion or bulge mass, we can then assess the fraction of the black
hole's rest mass that is being extracted as kinetic energy that can
heat the gas. In Fig.~\ref{fig:lheat}, we show observational estimates
of the rate of energy injection, or jet power, as a function of black
hole mass, from observations of elliptical galaxies with associated
hot gas bubbles by \citet{allen:06} and \citet{rafferty:06}.  The
\citet{rafferty:06} estimates of jet power overlap those of
\citet{allen:06}, but extend to considerably higher values for a given
BH mass. This may be because the Rafferty et al. sample, which extends
to higher redshift, contains more massive clusters than the very
nearby sample of \citet{allen:06}. These massive clusters may have
higher gas densities at the relevant radii, thus allowing more
efficient coupling of the radio jet with the ICM (S. Allen,
priv. comm.).  We also show the time averaged heating rate as a
function of BH mass estimated by \citet{best:06}, from observations of
the radio-loud fraction of SDSS galaxies.  The observed scaling of jet
power with BH mass is a bit steeper than it would be if the jet power
were proportional to the Eddington luminosity: the jet power scales
approximately as $\mbh^{1.6}$, or as $L_{\rm Edd} \mbh^{0.6}$.

These results may be compared directly with the ``jet power'' as a
function of BH mass incorporated in our fiducial model. Recall that in
our model, we assumed that the central density and temperature of the
gas was set by the isothermal cooling flow model of NF00 (see
\S\ref{sec:model:radio}) and the accretion rate onto the BH was then
set by the Bondi accretion rate. We allowed an overall scaling factor
$\kappa_{\rm radio}$, which we adjusted to the \emph{minimum} value
that produced a good fit to the empirical constraint on galaxy stellar
mass as a function of host halo mass discussed in
\S\ref{sec:results:galprop}. We find that the resulting accretion
rates are about an order of magnitude higher than the Bondi rates
estimated by \citet{allen:06}, and the jet power at a given BH mass is
also higher than the \citet{allen:06} estimates. However, the jet
powers are consistent with the higher values in the
\citet{rafferty:06} sample. It is also important to remember that
these observational estimates are lower limits. They include only the
energy associated with inflating the bubbles, while significant energy
can also be dissipated through sound waves, viscosity, and weak shocks
\citep{mcnamara:05,nulsen:05a,nulsen:05b,fabian:06,forman:07}.
\citet{binney:07} analyzed 3D adaptive grid simulations of heating of
cooling flows, and found that the bubbles reflected only $\sim 10$
percent of the total injected energy.

However, it is also possible that we are overestimating the energy
required, because we are insisting that AGN heating does the whole
job, while as we have discussed, there may be several other processes
that help to reduce the efficiency of cooling in group and
cluster-mass halos. Furthermore, the semi-analytic cooling model is
probably only accurate to a factor of two to three at best, and may be
overestimating the cooling rates. We intend to test and better
calibrate our models by comparing with the results of numerical
simulations, but this is not entirely straightforward, because
supernova feedback probably plays an important role in altering the
equation of state and metallicity of the gas, which affects the
cooling rates. Given these uncertainties, we conclude that the heating
due to AGN predicted by our simple model is not only very successful at
solving the overcooling problem, it is also reasonably consistent with
the direct observational constraints.

Fig.~\ref{fig:lheat} (right panel) also shows the predicted ``specific
jet energy'' (jet energy divided by BH mass) as a function of the halo
virial velocity. One can see that the isothermal Bondi model predicts
a fairly strong dependence of jet power on halo virial velocity (or
temperature), and in fact this scaling is important in fitting the
shape of the $f_{\rm star}(M_{\rm halo})$ function at intermediate
masses --- we find that if we assume that the accretion rate (and
hence the jet power) is just a function of BH mass, we obtain
qualitatively similar results, but we do not get as good a match to
the $f_{\rm star}(M_{\rm halo})$ function and hence the galaxy stellar
mass function. For comparison, we also show the empirical scaling
adopted in the fiducial Munich SAM \citep[e.g.][]{croton:06}. As also
noted by \citet{croton:06}, their adopted scaling $\dot{m}_{\rm
  acc}/m_{\rm BH} \propto V_{\rm vir}^3$ is very similar to that
predicted by the isothermal Bondi model, and it yields very similar
results when we adopt it in our SAM.

\begin{figure*} 
\begin{center} 
\plottwo{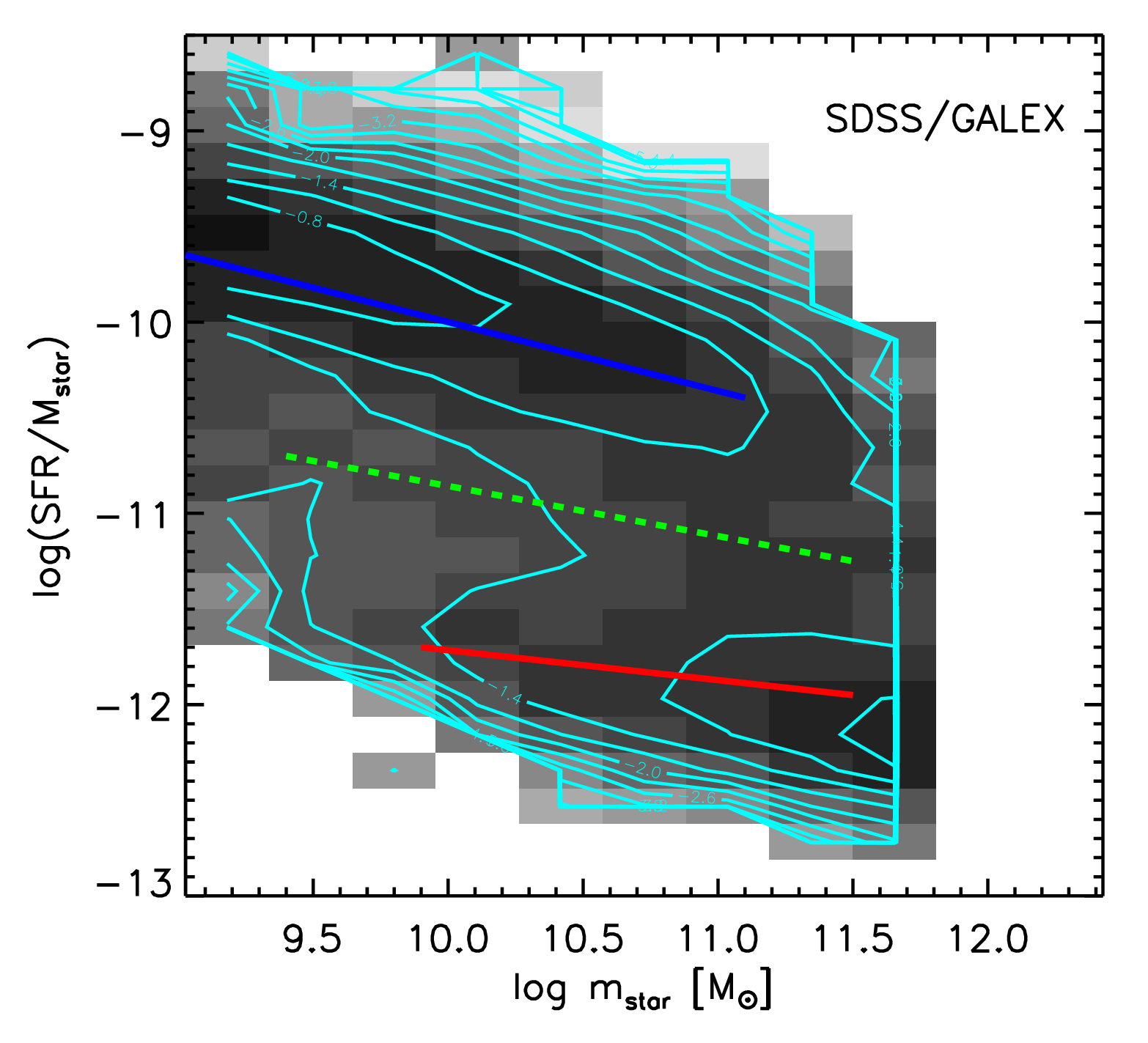}{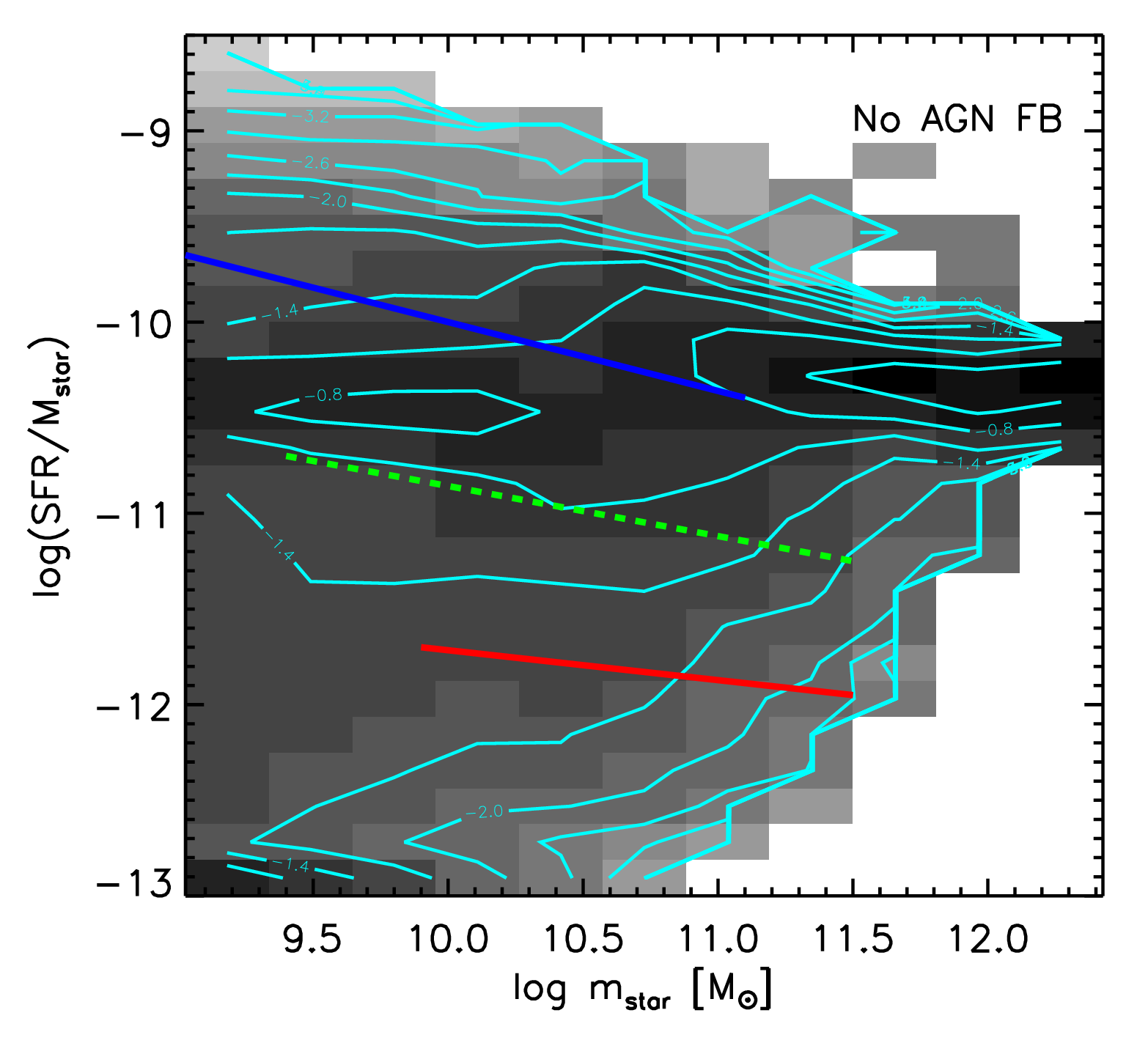}
\plottwo{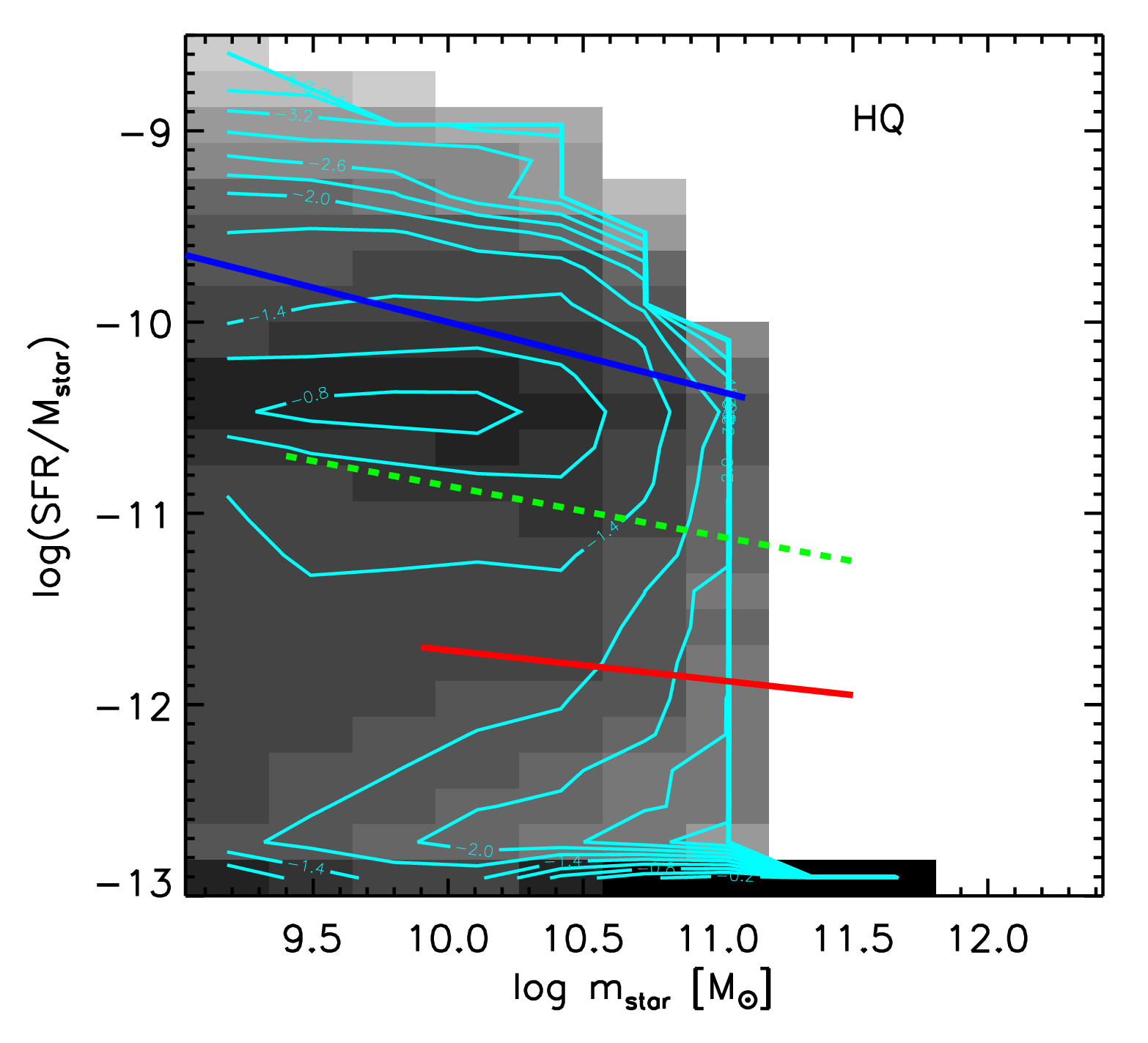}{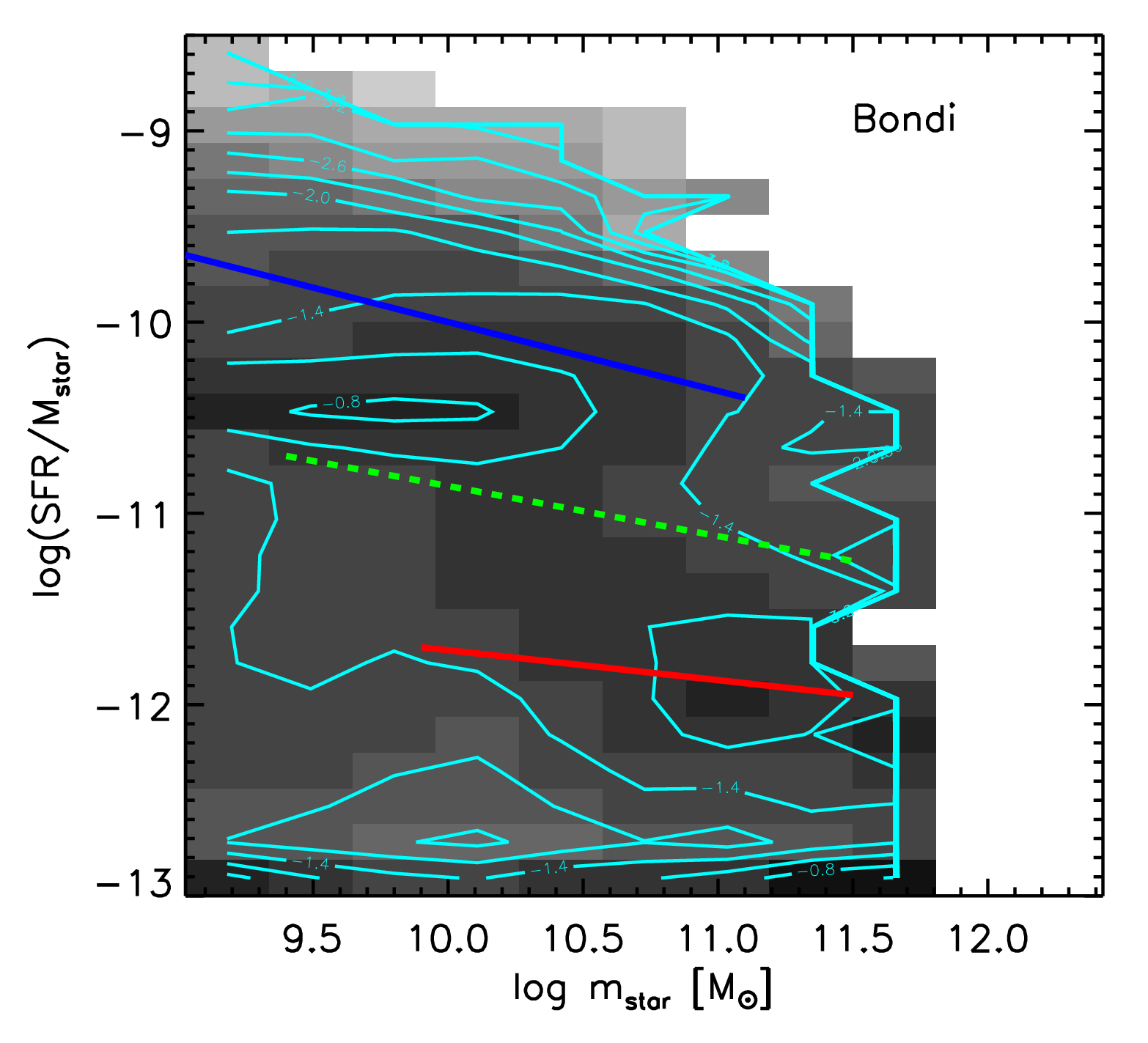}
\end{center}
\caption{\small Specific star formation rate (star formation rate
divided by stellar mass) vs. stellar mass. Gray shading and contours
indicate the conditional probability $P(\rm{SSFR}|m_{\rm star})$. The
diagonal (dark blue) solid line in the upper left part of the plot and
the (red) line in the lower right part of the plot indicate the ``star
forming sequence'' and ``quenched sequence'' from the observational
results of \protect\citet{salim:07}. The middle, dashed (green) line
indicates the dividing line between the star forming or active
galaxies and quenched galaxies (sometimes called the ``green
valley''). These active, valley, and quenched sequences based on the
observed distributions from GALEX are repeated on all four panels. Top
left: observed SSFR vs. mass distribution from GALEX
\protect\citep{schiminovich:07}. Top right: predicted distribution
from the model with no AGN feedback. Bottom left: predicted
distribution from the HQ model. Bottom right: predicted distribution
from the fiducial (isothermal Bondi) model.
\label{fig:ssfr}}
\end{figure*}

\subsection{Star formation Quenching}
\label{sec:results:quenching}

\begin{figure*} 
\begin{center}
\includegraphics[width=6.5in]{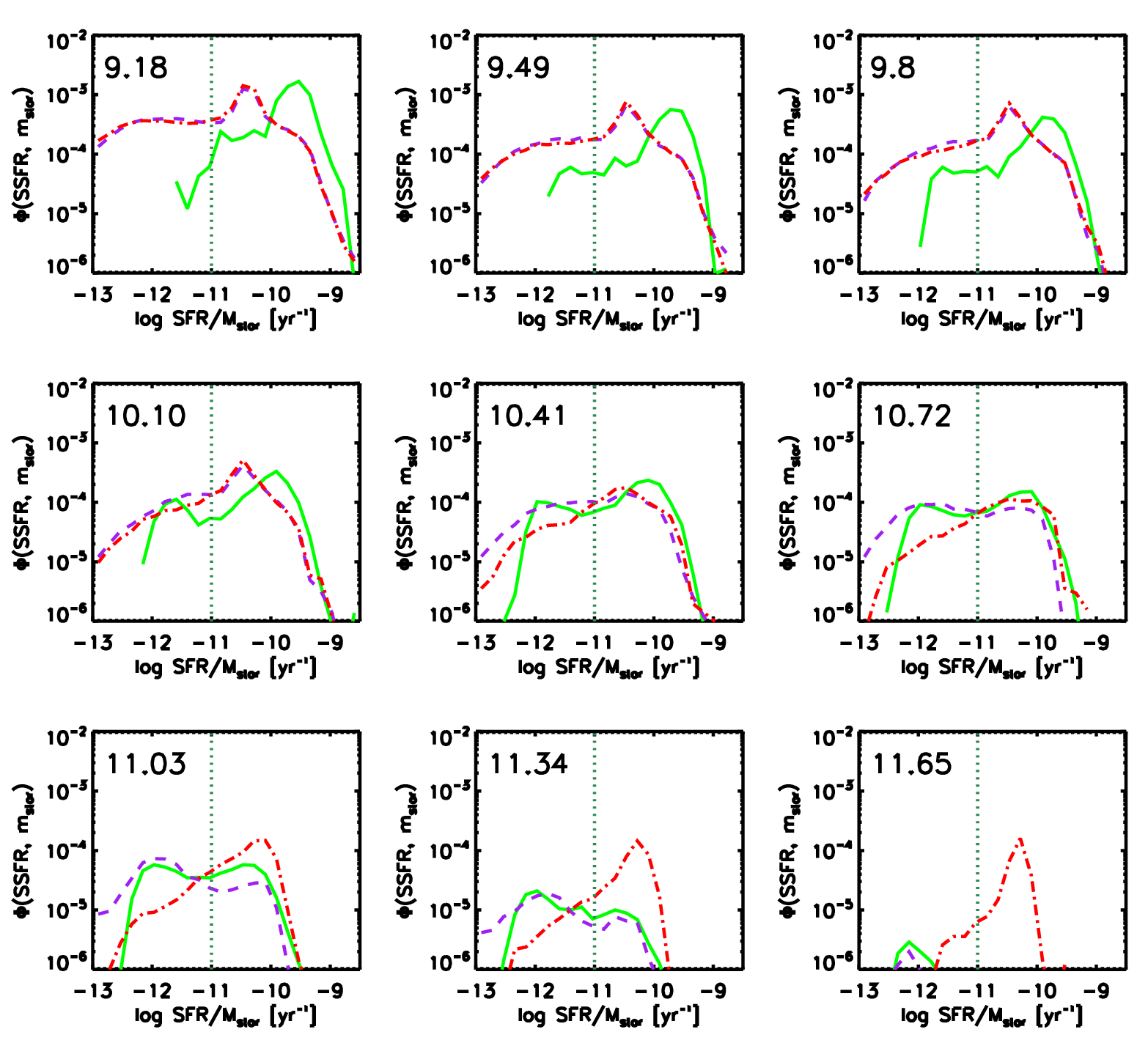}
\end{center}
\caption{\small Distribution of specific star formation rates (SSFR)
  in stellar mass bins (as indicated on the panels). Solid (green)
  lines show the observational results from GALEX
  \protect\citep{schiminovich:07}, dot-dashed (red) lines show the no
  AGN FB model, and dashed (purple) lines show the fiducial
  (isothermal Bondi) model. Vertical dotted lines show the rough
  location of the division between the ``active'' and ``quenched''
  populations.
\label{fig:ssfrdist}}
\end{figure*}

As we noted in the introduction, there are two puzzles in galaxy
formation that we wish to address in this paper. One is that real
galaxies do not grow as massive as we would have predicted in the
absence of AGN feedback. The other is that star formation in most
massive galaxies has been quenched, while most low-mass galaxies
continue to form stars. These two problems seem very likely to be
interconnected, but it is not obvious that a specific model which
solves one problem will necessarily solve the other. In order to
assess the quenching problem, it is common practice to compare model
predictions with observed optical or optical-NIR color-magnitude
distributions. However, these kinds of predictions are quite sensitive
to metallicity and dust, which complicates the
interpretation. Instead, we make use of the physical properties,
specific star formation rate (SSFR $\equiv \dot{m}_{\rm star}/m_{\rm
star})$ and stellar mass, derived from GALEX UV photometry plus SDSS
five-band optical photometry
\citep{salim:07,schiminovich:07}. \citet{yi:05} and \citet{kaviraj:07}
have shown that the NUV-optical colors are a highly effective way to
probe small amounts of recent star formation in galaxies with an
underlying old stellar population.

Fig.~\ref{fig:ssfr} shows the conditional probability distribution of
SSFR as a function of stellar mass $P(\rm{SSFR}|m_{\rm star})$. The
top left panel shows the observed distribution derived from the
GALEX+SDSS data by \citet{schiminovich:07}. The star forming sequence,
quenched sequence, and the dividing line (sometimes called the ``green
valley'') derived from GALEX+SDSS by \citet{salim:07} are also shown,
and are repeated on every panel. It is important to remember that the
GALEX-SDSS survey is incomplete in the bottom left region of the plot.
We also show the same distribution, as predicted by the no AGN FB
model (top right), the Halo Quenching model (bottom left), and the
fiducial isothermal Bondi model (bottom right). The star formation
rates shown for the models have been averaged over the past $10^8$
years.

All of the models shown have the same problem with low-mass
galaxies. The star forming sequence is nearly flat, rather than being
tilted such that less massive galaxies have higher SSFR, as in the
observations, and the specific star formation rates are too
low. Though we have tried extensive experiments with parameter
variation, we have not succeeded in solving this problem. The problem
also seems to be quite robust to the star formation recipe that we
adopt. If we remove the SF threshold, thereby effectively increasing
the star formation efficiency in low-mass galaxies, then the galaxies
consume more gas and have lower gas fractions at the present
day. Their SSFR are still low, because they have very little fuel for
star formation. If we increase the star formation threshold, which
makes star formation even more inefficient in low-mass galaxies, the
galaxies have higher gas fractions at the present day, but most of
that gas is not allowed to make stars, so the SSFR are again almost
the same as before.

In the model without AGN feedback, nearly all massive galaxies have
high SSFR and would also have blue colors, in drastic conflict with
the observations. In the HQ model, which did as well or better than
our fiducial model at matching $f_{\rm star}(M_{\rm halo})$ and the
galaxy stellar mass function, essentially \emph{all} massive galaxies
are completely quenched. This is not surprising, as we know that there
is a fairly tight relationship between stellar mass and halo mass, so
a sharp cutoff in halo mass produces a fairly sharp cutoff in stellar
mass. This model cannot account for the population of massive galaxies
with small but detectable amounts of recent star formation, seen in
the GALEX observations (Yi et al. \citeyear{yi:05} argue that the NUV
light in these galaxies is indeed due to star formation and not a UV
upturn or AGN). However, our fiducial isothermal Bondi model does
produce such a population of massive galaxies whose star formation has
been substantially, but not completely, quenched. The SSFR vs. $m_{\rm
  star}$ distribution predicted by this model looks qualitatively
quite similar to the observations (for massive galaxies).

We analyze the distribution of SSFR vs. stellar mass in more detail in
Fig.~\ref{fig:ssfrdist}. Here, we show histograms of SSFR in stellar
mass bins, for the observations and two of the models: the no AGN FB
model, and the fiducial model. From this comparison we can see that
our fiducial model is not producing quite enough massive, actively
star-forming galaxies ($m_{\rm star} \gtrsim 10.7$). It seems that
star formation is actually being quenched a bit too effectively in
massive galaxies. However, overall the agreement is quite good. 

\subsection{Global formation histories of stars, cold gas, 
and metals}
\label{sec:results:global}

\begin{figure*} 
\begin{center}
\plottwo{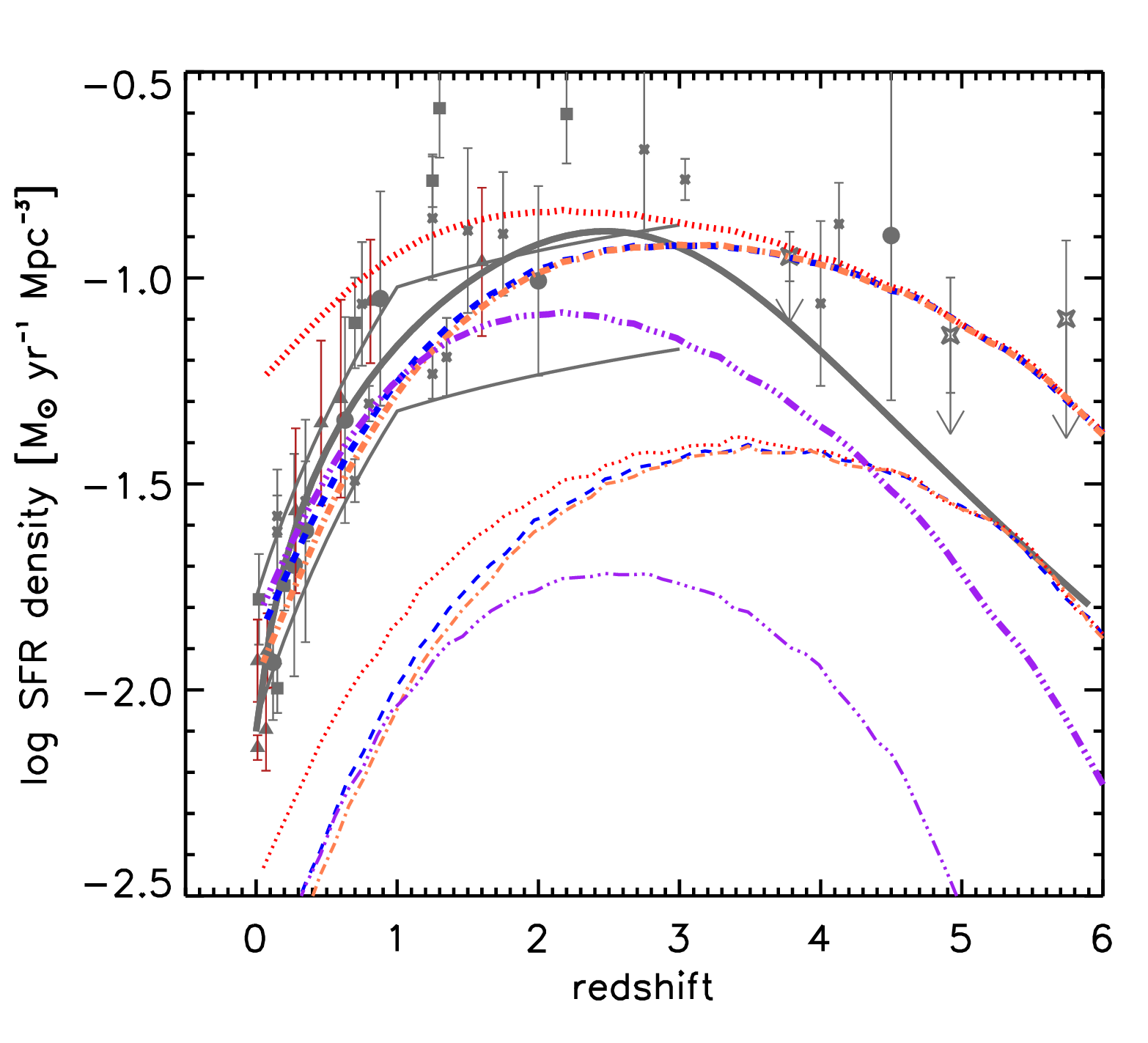}{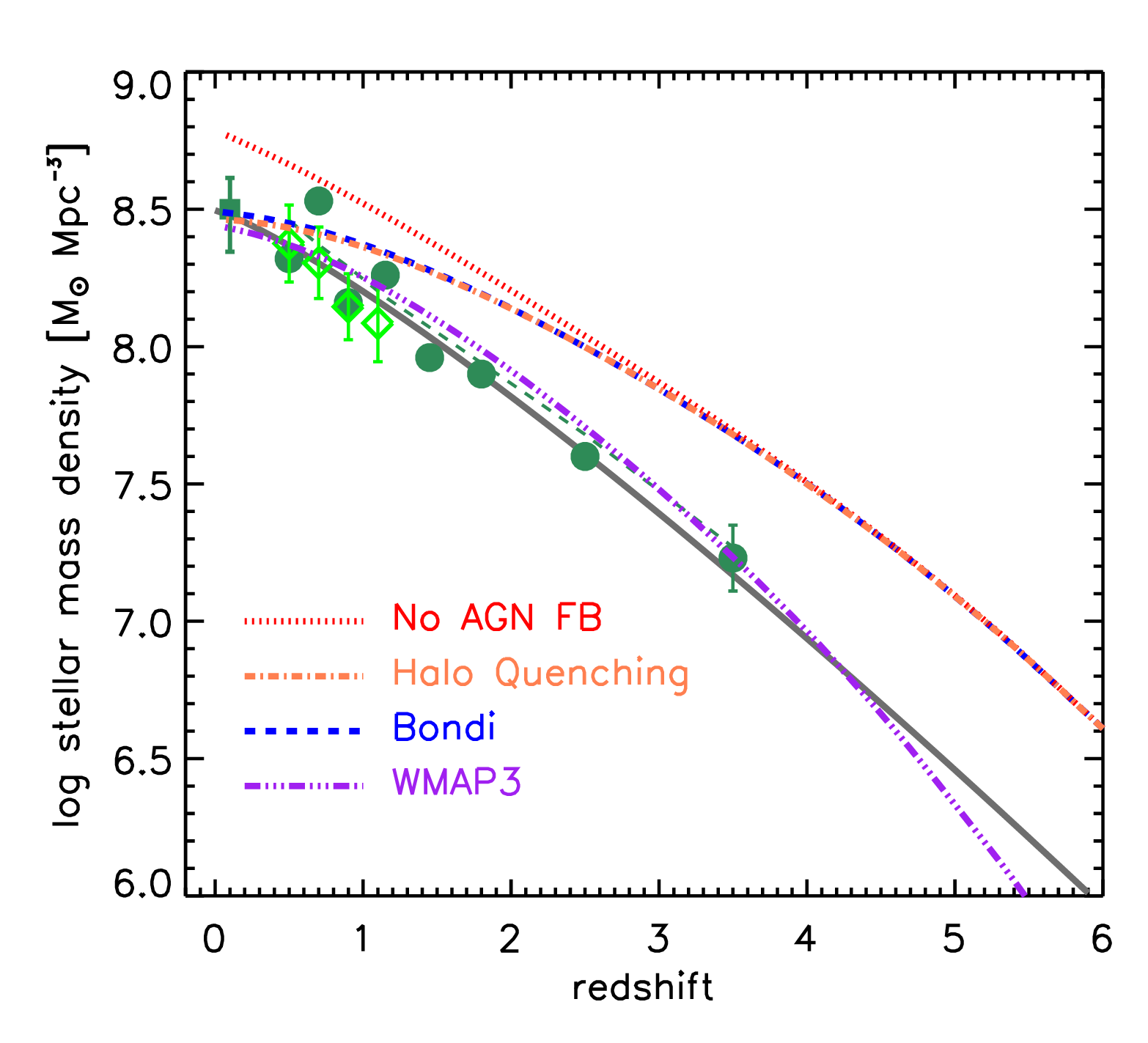}
\plottwo{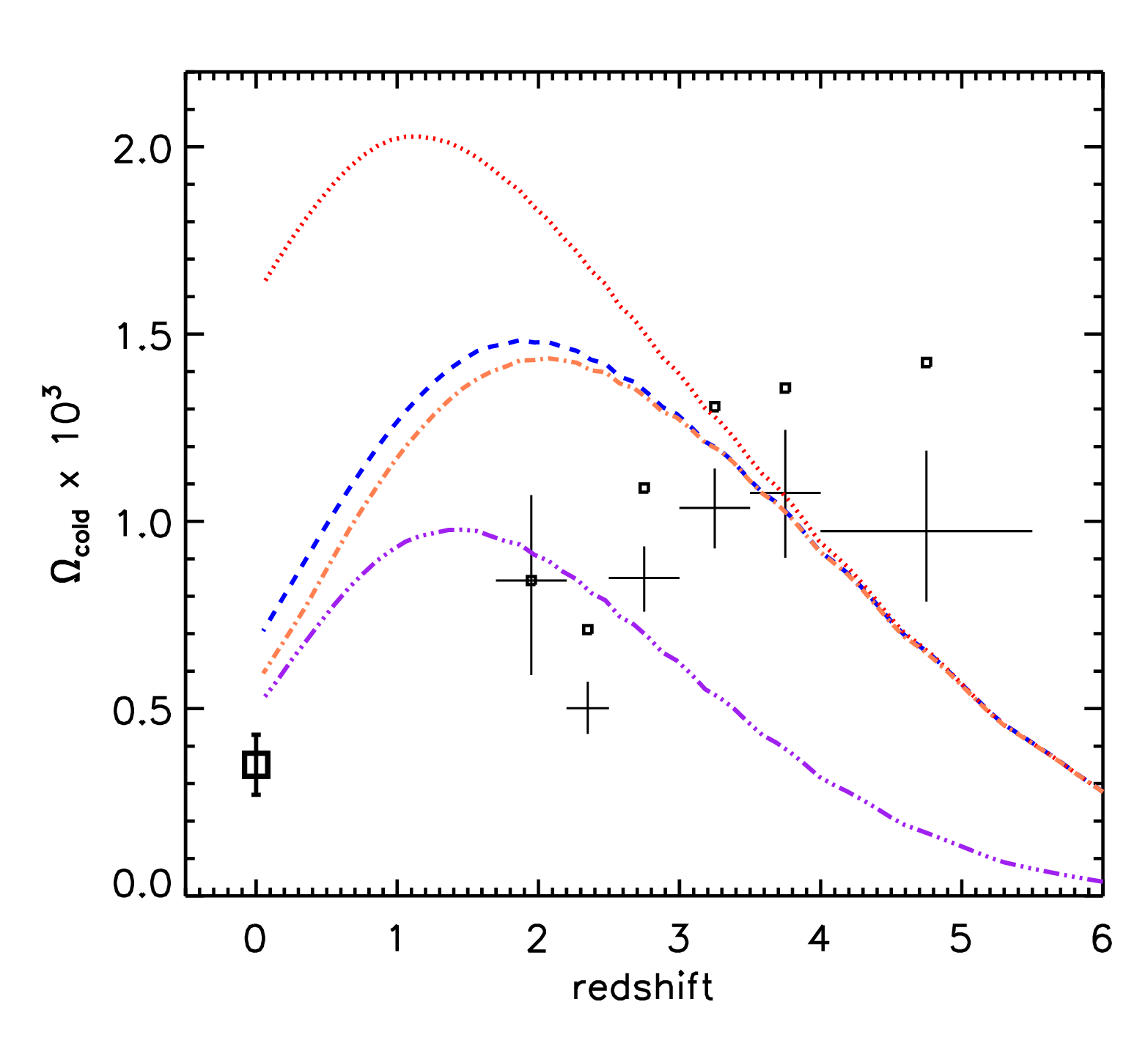}{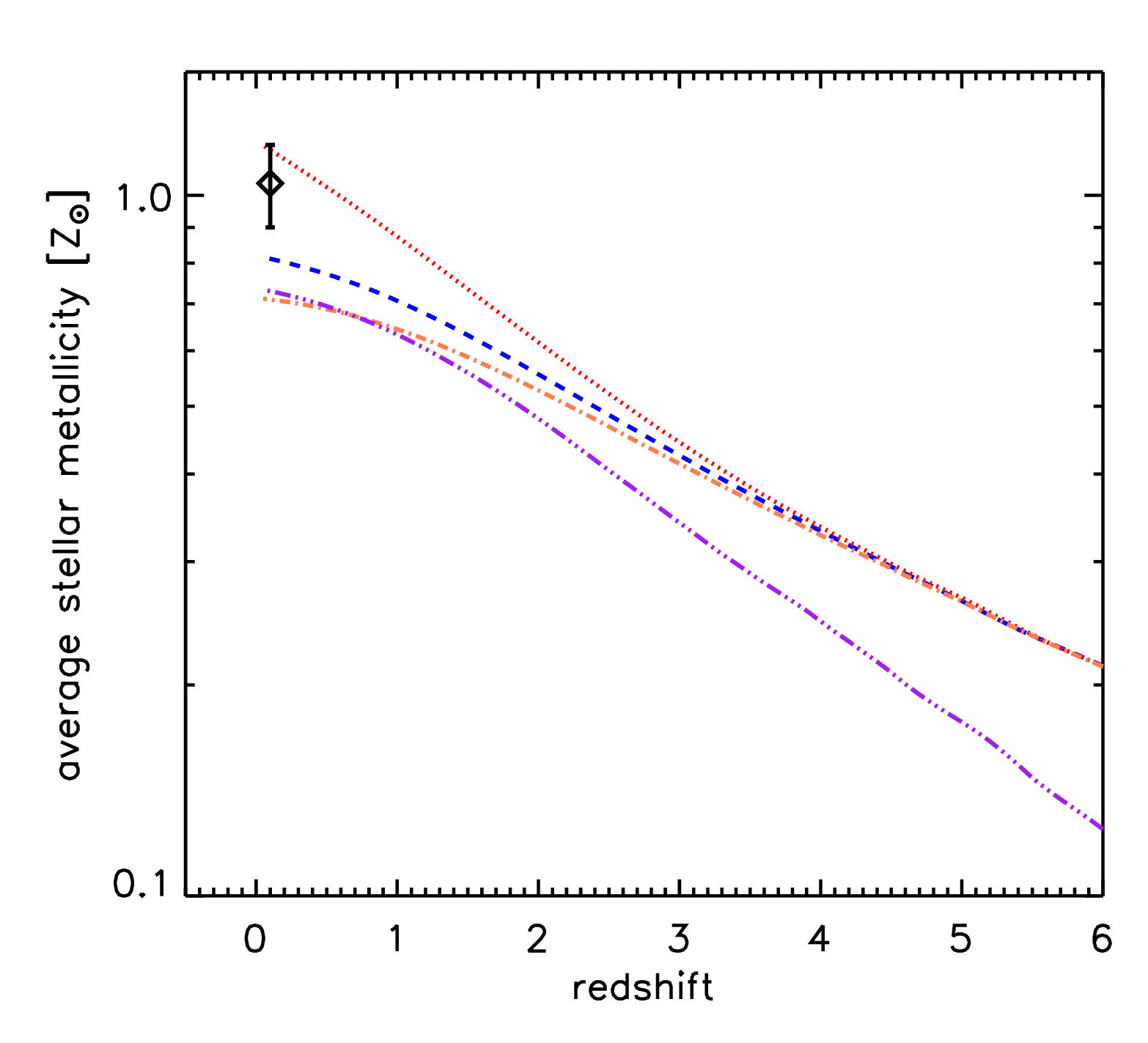}
\end{center}
\caption{\small In all panels, dotted (red), dot-dashed (orange), and
  dashed (blue) lines show predictions from the C-\LCDM\ no AGN FB
  model, the HQ model, and the fiducial (isothermal Bondi) model,
  respectively. The triple-dot dashed (purple) line shows the WMAP3
  model.  Top left: Star formation rate density as a function of
  redshift. The upper set of thicker lines shows the total SFR in the
  models, and the lower set of thin lines shows the SFR due to
  bursts. Symbols show the compilation of observational results of
  \protect\citet{hopkins:04}, converted to a Chabrier IMF. The thick
  solid (gray) curve is the fit to the observational compilation
  presented by \protect\citet{hopkins_beacom:06}. The thin (gray)
  solid broken curve shows observational estimates from GALEX
  \protect\citep{schiminovich:05}.  Top right: The integrated global
  stellar mass density as a function of redshift. Symbols show
  observational estimates from \protect\citet[][$z\sim0$]{bell:03b},
  \citet[][$z\sim0.2$--1; diamonds]{borch:06}, and
  \protect\citet[][$z\sim0.2$--3.5; filled circles]{fontana:06}. The
  thick (gray) curve shows the fit to the observational compilation
  presented in \protect\citet{wilkins:08}.  Bottom left: The mass
  density of cold gas in units of the critical density $\Omega_{\rm
    cold} \equiv \rho_{\rm cold}/\rho_{\rm crit}$. Symbols show the
  observational estimates at $z\sim 0$ from the blind \HI\ survey of
  \protect\citet{zwaan:05}, and from Damped Lyman-$\alpha$ systems
  (crosses) at high redshift \protect\citep{prochaska:05}. Small open
  squares also include the contribution from lower column density
  absorption systems (Lyman-limit systems). Bottom right: The
  mass-weighted average metallicity of stars $\langle Z_{\rm star}
  \rangle \equiv \rho_Z /\rho_{\rm star}$ as a function of redshift.
  The diamond symbol at $z\sim0.1$ shows the observational estimate
  from SDSS galaxies from \protect\citet{gallazzi:07}.
\label{fig:history}}
\end{figure*}

In this paper, we have focussed mainly on predictions of the
present-day ($z\sim 0$) properties of galaxies. We plan to explore the
predictions for the properties of high-redshift galaxies in detail in
future papers. However, in this section we present predictions for the
global histories of several important (and observationally accessible)
components of galaxies: the star formation rate, stars, cold gas,
metals, and black holes.

\subsubsection{Dependence on Cosmology: the WMAP3 model}

In this section we introduce a new model, which has the same recipes
for galaxy and BH formation as our fiducial isothermal Bondi model,
but adopts the cosmological parameters (see Table~\ref{tab:cosmo})
from the 3-year analysis of the WMAP data reported in
\citet{spergel:07}. We will refer to this as the WMAP3
model \footnote{While this paper was in the refeering process, new
  results for the cosmological parameters derived from the five year
  WMAP data, combined with distance estimates from Type Ia SNae and
  Baryon Acoustic Oscillations, were posted on astro-ph
  \citep{komatsu:08}. The new estimates of the parameters most
  relevant to our results, $\sigma_8=0.817 \pm 0.026$ and
  $n_s=0.960^{+0.14}_{-0.013}$, are intermediate between the values
  adopted in our C-\LCDM\ and WMAP3 models, though somewhat closer to
  the WMAP3 values.}.  For our purposes, the most important
differences between the Concordance \LCDM\ (C-\LCDM) model that we
have been using so far and the WMAP3 model is that WMAP3 has a lower
value of $\sigma_8$, the normalization of the primordial power
spectrum, and the primordial power spectrum also has a ``tilt'' ($n_s
= 0.96$) while C-\LCDM\ has a scale-free initial power spectrum
($n_s=1$). This results in less power on small scales, and hence later
structure formation in WMAP3 relative to C-\LCDM. In order to
reproduce the $z=0$ observations as before, we re-tuned the star
formation efficiency, radio mode heating efficiency, and scattering
parameters (we used $A_{\rm Kenn} = 1.67 \times 10^{-4}$, $\kappa_{\rm
  radio}=6.0 \times 10^{-3}$, and $f_{\rm scatter}=0.2$), but left the
other parameters the same. After this re-tuning, the WMAP3 model
produces nearly indistinguishable results from the C-\LCDM\ model for
all of the quantities that we have shown so far \citep[see
  also][]{wang:07}.

\subsubsection{The Global Star Formation and Mass Assembly History}

Fig.~\ref{fig:history} (top left) shows the global star formation rate
density of all galaxies predicted in the three C-\LCDM\ models: the
model with no AGN FB, the Halo Quenching model, and the fiducial
isothermal Bondi model.  A compilation of observational estimates is
also shown. The no AGN FB model overpredicts the amount of star
formation at low redshift, and the SFR does not decline as rapidly as
the data indicate, while both the HQ model and the isothermal Bondi
models produce very good agreement with the observations at $z
\lesssim 2$. We can see that the effect of the AGN heating starts to
become significant only at $z\lesssim4$. At higher redshifts, the AGN
heating has little or no impact on the global SFR, because most star
formation is taking place in relatively small mass halos, which are
not affected by the ``radio mode'' heating, as we have already
discussed. It is also interesting that the simple HQ model produces
such similar results to the fiducial isothermal Bondi model, even over
a large range in redshift. At $z\gtrsim 4$, the C-\LCDM\ models
predict more star formation than the observational compilation of
\citet{hopkins_beacom:06}, but agree with the higher estimates in the
literature \citep[e.g.][]{steidel:99,giavalisco:04}. The WMAP3 model
predicts a much more rapidly declining SFR at $z\ga2$, somewhat lower
than the observations but well within the observational errors.

The lower set of lines shows the star formation contributed by
merger-triggered bursts. The contribution due to bursts is much lower
than in our previous models (SPF01), for several reasons. (1) Our new
treatment of dynamical friction and tidal stripping and destruction
means that many low-mass satellites take longer than a Hubble time to
merge, or are tidally destroyed before they can merge. Therefore the
number of minor mergers is lower. Also, we do not include
satellite-satellite mergers here, which were included in our previous
models. (2) Our newly calibrated burst efficiencies in minor mergers,
which are based on a greatly expanded and improved set of hydrodynamic
simulations, are lower than in our previous models (3) Our Kennicutt
star formation law causes the star formation efficiencies in quiescent
disks to increase with increasing redshift. As we already showed in
SPF01, this leaves less cold gas fuel for bursts, leading to a
decreased contribution to the SFR from bursts.

In the top right panel of Fig.~\ref{fig:history} we show the
complementary quantity $\rho_{\rm star}$, the integrated cosmic
stellar mass density. All of the C-\LCDM\ models predict a
significantly \emph{earlier} assembly of stars in galaxies than
observations of high redshift galaxies indicate. The mass density of
long-lived stars in both our fiducial and the HQ model is a factor of
$\sim 3$ higher at $z\sim2$ and a factor of $\sim 2$ higher at $z \sim
1$ than the observations. However, the WMAP3 model produces excellent
agreement with the stellar mass density as a function of redshift. We
note here that this tension in the model results (i.e. that the
C-\LCDM\ model fits the SFR history data better, while the WMAP3 model
provides a better fit to the stellar mass density) is connected with a
possible inconsistency between the two data sets that has been noted
recently in several papers
\citep[e.g.][]{hopkins_beacom:06,fardal:07,wilkins:08,dave:08}. One
possible resolution of this tension can be obtained if the stellar IMF
has changed with time, and was more top-heavy at high redshift (we
return to this issue in the Discussion).
 
\subsubsection{Evolution of Cold Gas and Metals}

In the bottom left panel of Fig.~\ref{fig:history} we investigate
another complementary quantity, the mass density of gas that has
cooled but not yet formed stars. In our models, all of this cold gas
is assumed to reside in galactic disks.  One can compare the model
predictions with the total mass density of \HI\ gas from blind
\HI\ surveys at $z\sim0$ \citep{zwaan:05}, as well as with estimates
of cold gas at high redshift from Damped Lyman-$\alpha$ systems
\citep[e.g.][and references therein]{prochaska:05}. Note that all the
observational estimates shown here are for atomic gas only, and do not
include the contribution from molecular gas. Therefore, these
observations are lower limits for the model predictions of cold gas.
The C-\LCDM\ models are consistent with the observations to $z\sim 4$,
but are somewhat low at $z\sim5$ (note however that
\citet{prochaska:05} ``caution the reader that the results at $z>4$
should be confirmed by higher resolution observations''). However,
$\Omega_{\rm cold}$ in the WMAP3 model is much lower than the
observations at $z\gtrsim3$, by about a factor of two at $z=3.5$ and
an order of magnitude at $z\sim5$.

In the last (bottom right) panel of Fig.~\ref{fig:history} we show the
stellar-mass weighted, globally averaged metallicity as a function of
redshift predicted by the no AGN FB model, the HQ model, and the
isothermal Bondi models. We can compare this with the results of the
observational analysis of SDSS galaxies by \citet{gallazzi:07}, who
find a mass-weighted average stellar metallicity at $z\sim 0$ of about
solar. The no AGN FB model overshoots this value, while the models
with AGN FB slightly underestimate it. \citet{gallazzi:07} showed that
the Millennium simulations, based on the semi-analytic model of
\citet{croton:06}, give very similar results to our models --- they
underproduce the stellar metallicity at $z=0$ by about 20--30 percent.
The models predict early enrichment, with the mean stellar mass at
$z\sim 6$ about 25\% of solar, and 50\% of solar at
$z\sim2.5$. Because of the difficulty of obtaining an unbiased stellar
mass weighted global mean metallicity at high redshift, we do not
attempt a quantitative comparison with observations, but
qualitatively, these results seem consistent with the relatively high
metallicities detected in high redshift galaxies
\citep[e.g.][]{erb:06}. As expected, the WMAP3 model predicts somewhat
lower metallicity at high redshift.

\begin{figure*} 
\begin{center}
\plottwo{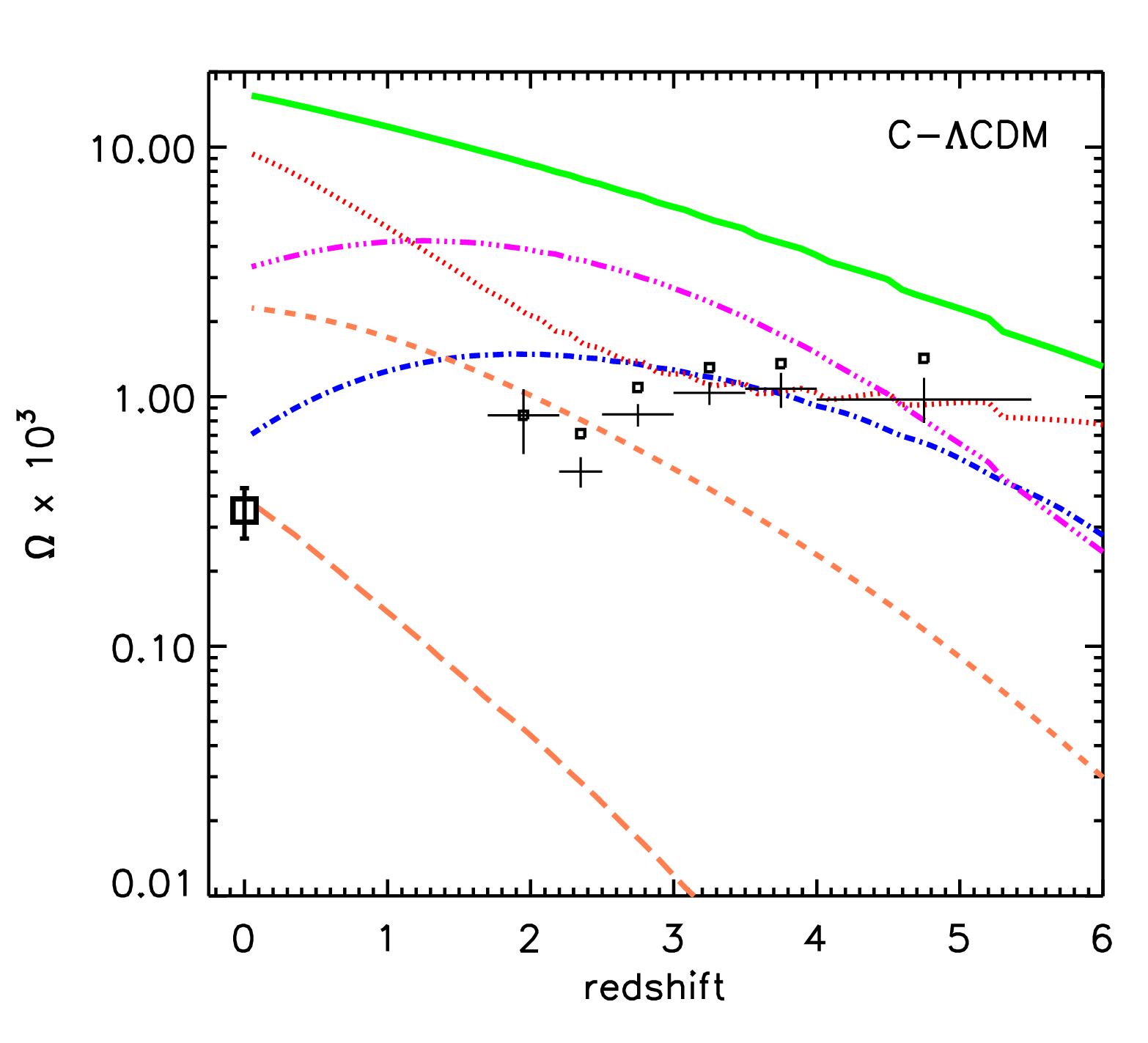}{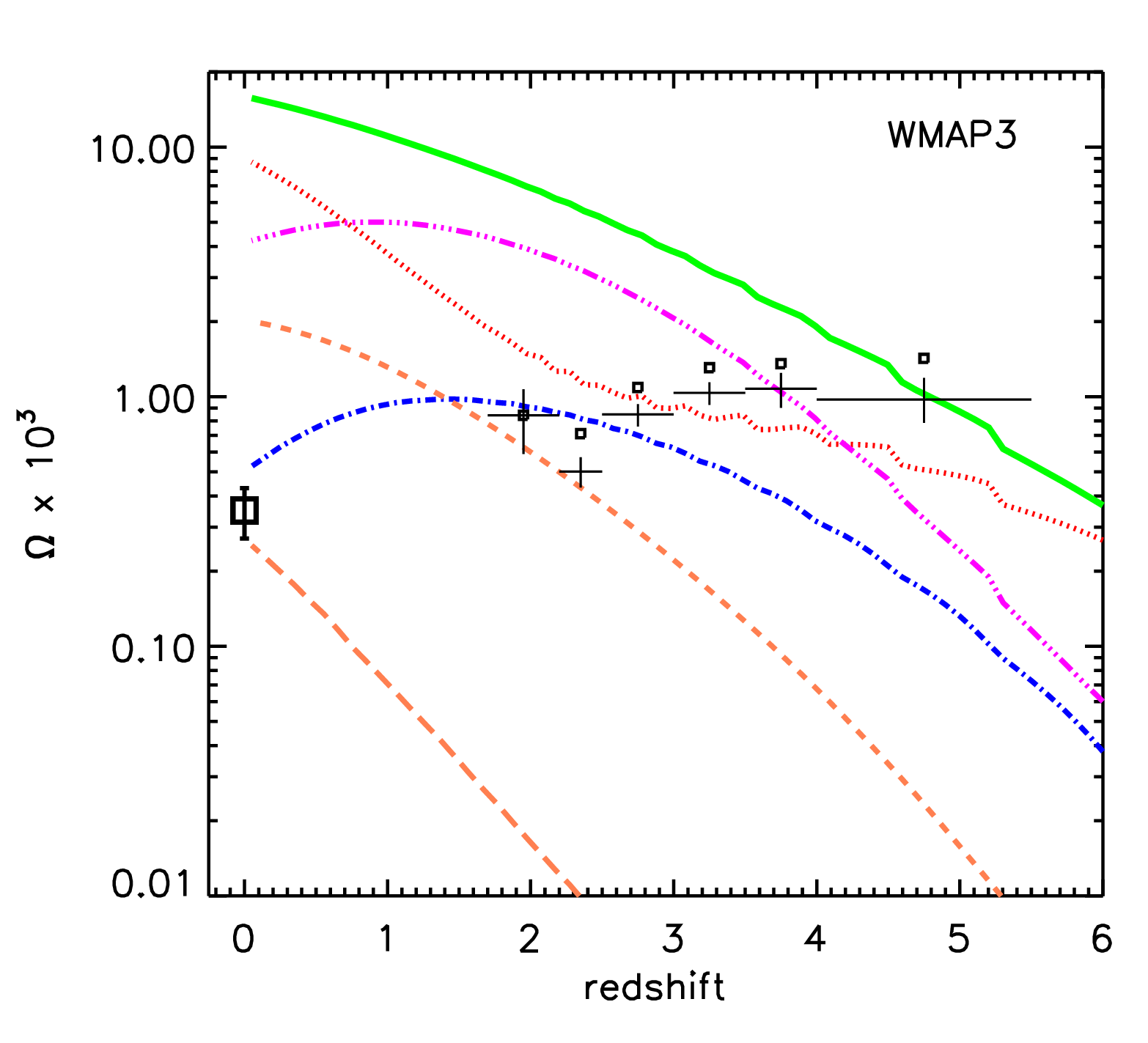}
\end{center}
\caption{\small The global density in units of the critial density of
  various components as a function of redshift predicted by the
  isothermal Bondi model (Left: Concordance \LCDM\ with $f_{\rm
    scatter}=0.4$; Right: WMAP3 with $f_{\rm scatter}=0.2$). From the
  highest to the lowest curve at $z=0$, we show: the universal baryon
  fraction times the mass contained in virialized halos above our mass
  resolution ($10^{10} \msun$; solid green); shock-heated hot gas in
  halos (dotted red); gas in the ``intergalactic medium'' (see text;
  triple-dot-dashed magenta); stars in galaxies (short dashed orange);
  cold gas in galaxies (dot-dashed blue); and stars in Diffuse Stellar
  Halos (DSH) around galaxies (long dashed orange). We repeat the
  observational estimates of $\Omega_{\rm cold}$ from
  Fig.~\protect\ref{fig:history} (squares and crosses).
\label{fig:omega_bar}}
\end{figure*}

\subsubsection{The Baryon Budget and its Evolution}

In Fig.~\ref{fig:omega_bar}, we provide the answer to the question
``where are the baryons'' in our fiducial isothermal Bondi model, in
both the C-\LCDM\ and WMAP3 cosmologies. All quantities in
Fig.~\ref{fig:omega_bar} are shown in units of the critical
density. We show the mass density in collapsed dark matter halos above
our mass resolution of $M_{\rm vir} = 10^{10} \msun$ multiplied by the
universal baryon fraction, which represents all of the baryons that
are available for cooling at a given redshift in our model. We also
show the mass density of hot gas in dark matter halos, and of warm/hot
diffuse gas that has either been prevented from collapsing into halos
by the photoionizing background or has been ejected by supernova
feedback. These two components (warm/hot diffuse gas and hot gas in
halos) dominate the baryon budget at all redshifts, in agreement with
observational constraints at low redshift
\citep[e.g.][]{fukugita_peebles:04} and the predictions of numerical
hydrodynamic simulations \citep[][and references
  therein]{bertone:08}. Finally, we show the baryons in cold gas in
galactic disks, in stars in the disks and bulges of galaxies, and in
``diffuse stellar halos'' (DSH) around galaxies. Cold gas dominates
over stars until $z\sim 2$, where stars begin to dominate.  Our model
predicts that stars in DSH comprise about fifteen percent of the mass
of stars in galaxies at $z\sim0$.

From this plot we can see that, in the WMAP3 model, there is just enough
baryonic material in collapsed halos that \emph{can} cool at $z\gtrsim
3$ to account for the DLAS data. If some of the gas that is ejected
from galaxies in our model instead remains in galactic disks in the
form of cold gas, perhaps this model could be reconciled with the DLAS
data. Note, however, that there is no room for more cold gas in
galaxies at $z=0$, so in order to solve the problem, the gas needs to
be retained at high redshift but ejected or prevented from cooling at
low redshift. One can think of plausible physical reasons that
supernova-driven winds might have more difficulty escaping galaxies at
high redshift, for example, the winds might stall against the
higher-density IGM that is presumably present at these early epochs.

\section{Discussion and Conclusions}
\label{sec:conclude}

We have presented a new semi-analytic model for the self-consistent
evolution of galaxies, black holes and AGN in the framework of the
Cold Dark Matter model of structure formation. Our models are built on
those described in SP99, SPF2001, and subsequent papers, but we
present several important new ingredients here:

\noindent (i) Improved modelling of tidal stripping and destruction of
orbiting sub-halos and of dynamical friction and satellite merger
timescales.

\noindent (ii) Improved modelling of disk sizes, including realistic
DM halo profiles and the effect of ``adiabatic contraction''.

\noindent (iii) A more realistic recipe for star formation in
quiescent disks, based on the empirical Kennicutt Law and including a
surface density threshold for star formation.

\noindent (iv) Updated modelling of the efficiency and timescale of
starbursts, based on an extensive suite of numerical hydrodynamic
simulations of galaxy mergers.

\noindent (v) Tracking of a ``diffuse stellar halo'' component, which is built
up of tidally destroyed satellites and stars scattered in mergers.

\noindent (vi) A self-regulated model for black hole growth and
``bright mode'' accretion ``lightcurves'' based on numerical merger
simulations with AGN feedback.

\noindent (vii) Galaxy-scale AGN-driven winds, based on numerical
merger simulations.

\noindent (viii) Fueling of black holes with hot gas via Bondi accretion,
regulated according to the isothermal cooling flow model of
\citet{nulsen_fabian:00}.

\noindent (ix) Heating by radio jets, calibrated against observations
of X-ray cavities in cooling flow clusters.

We explored the predictions of our new models for a broad range of
physical properties of galaxies, groups, and clusters at $z\sim
0$. One key result is an explanation of the origin of the
characteristic shape of the galaxy stellar mass or luminosity
function. We cast this in terms of the ``star formation efficiency
function'' (i.e., the fraction of baryons in stars as a function of
host halo mass), which can be derived empirically by mapping between
(sub)-halo mass and stellar mass such that the observed stellar mass
function is reproduced \citep{moster:08,wang:06}. We found that, in
our models, the shape of this function arises from the fact that
supernova-driven winds can more efficiently heat and expell gas in
lower mass halos, while radio-mode heating by AGN is more efficient in
higher mass halos, because these halos have larger mass black holes
and these black holes can accrete hot gas more efficiently. The peak
in this function at $M_{\rm vir} \sim 10^{12} \msun$ occurs because
these halos are too massive for supernova-driven winds to escape
easily, and do not have massive enough black holes or a large enough
virial temperature to fuel efficient radio-mode heating.

We showed that our model also reproduces the stellar mass function as
a function of galaxy morphology (at least for galaxies more massive
than $\sim 10^{10} \msun$), the cold gas fractions of disk galaxies as
a function of stellar mass and the cold gas mass function, the stellar
mass-metallicity relation for galaxies, and the BH mass vs. spheroid
mass relation. 

Our model for heating by radio jets is similar in concept, but
different in detailed implementation, to other models that have been
presented in the literature. We tested the basic assumption of our
model for fueling of BH by hot gas, the isothermal cooling flow model
proposed by \citet{nulsen_fabian:00}, using observations of central
temperatures and densities in hot X-ray emitting gas in nine systems
by \citet{allen:06}, and found consistency. We further compared the
jet power required to solve the overcooling/massive galaxy problem in
our models with direct measurements of jet power from the energetics
of bubbles detected in X-ray gas around elliptical galaxies
\citep{allen:06,rafferty:06}. We found that our required jet powers
lie above the observations of \citet{allen:06} but agree with the
higher values obtained by \citet{rafferty:06}. 

We compared the results of our fiducial model with a very simple
implementation of AGN heating, in which we simply switched off cooling
in halos more massive than $\sim 10^{12} \msun$ (the ``Halo
Quenching'' or HQ model). We found that this model produced very
similar results to our fiducial model for the $z=0$ stellar mass
fraction as a function of halo mass ($f_{\rm star}(M_{\rm halo})$),
the stellar mass function, the cold gas fractions and cold gas mass
function, and the global star formation and stellar mass assembly
histories.

We investigated whether the same models which provided a good match to
the stellar mass function also reproduced the distribution of
(specific) star formation rates as a function of stellar mass. We
found that our fiducial model qualitatively reproduces the main
features of the observed distribution of SSFR vs. stellar mass: the
models produce a star forming sequence, which dominates at low stellar
masses ($\lesssim 2-3 \times 10^{10}\msun$) and a quenched population
at high stellar masses. The transition mass is naturally predicted by
the model, and again corresponds to the halo mass scale where AGN
heating becomes effective. This is in marked contrast to models
without AGN feedback, in which all massive galaxies are actively
star-forming.  

However, a more detailed comparison with the observations indicates
that the low-mass star-forming sequence in the models occurs at too
low a specific star formation rate, and does not have the right slope
(our star forming sequence is flat in SSFR vs. stellar mass space,
rather than tilted such that low mass galaxies have high SSFR, as the
observations indicate). This problem is present in all of our models,
and is independent of AGN feedback and robust to the star formation
recipe and the values of our free parameters. It may be a symptom of
the same malady that is responsible for producing low-mass galaxies
with older ages than those estimated from absorption lines in the
spectra of nearby galaxies.

In the Halo Quenching model, the quenching is clearly too sharp a
function of stellar mass. Essentially all galaxies more massive than
$m_{\rm star} \sim 10^{11} \msun$ are completely quenched, while the
observations indicate that massive galaxies have a wide range of
specific star formation rates, from small to moderate amounts of
recent star formation to activity levels that place them on the
star-forming sequence.

We investigated the cosmic histories of the major baryonic components
of the Universe as predicted in our models: star formation and stellar
mass, cold gas, warm/hot gas, and metals. We found that
our fiducial Concordance \LCDM\ model produces very good agreement
with the global star formation rate density at $z\lesssim 2$, but
predicts a higher and flatter SF history at $2 \lesssim z \lesssim 6$
than most observations indicate.  We found that our prediction of the
global SFR density is apparently not affected by AGN feedback at
redshifts above $z\gtrsim 3$, because most star formation is taking
place in small mass halos.

All of the C-\LCDM\ models predict a significantly larger amount of
mass in long-lived stars in galaxies at redshifts $z \gtrsim 0.5$, by
about a factor of two at $z\sim1$ and a factor of three at $z\sim 2$
for the fiducial and HQ models. The fact that we reproduce the stellar
mass density at $z\sim0$ (by construction) then implies that there is
not enough evolution in the galaxy population between $z\sim1$ and
$z\sim0$ in our models with AGN feedback. This problem does not seem
to be specific to our models, but is common to all the recently
published models in which AGN feedback is implemented in a similar way
\citep{cattaneo:06,croton:06,bower:06}. Indeed, one can see from the
predictions of the global SFR history presented in these works
(e.g. Fig.~5 of Croton et al. 2006, Fig.~8 of Bower et al. 2006) that
these models all produce very similar predictions for the global SF
history and for the integrated stellar mass density.

The WMAP3 model, in which structure formation occurs later due to the
reduced power on small scales, has a much lower and more steeply
declining global star formation rate at $z \gtrsim 2$ (about an order
of magnitude lower than C-\LCDM\ at $z=6$). The SFR is lower than the
observational compilation of \citet{hopkins_beacom:06} at these
redshifts by about 0.2--0.3 dex, but still well within the
observational errors. In contrast to the C-\LCDM\ model, the stellar
mass density predicted by the WMAP3 model is in excellent agreement
with observational estimates from $z\sim 4$--0. The delayed SF and
stellar mass assembly history in WMAP3 relative to C-\LCDM\ has also
been illustrated by \citet{wang:07}.

We also compared the predicted global mass density of cold gas in
galactic disks $\Omega_{\rm cold}$ as a function of redshift in our
models with estimates of \HI\ gas mass density at $2 \lesssim z
\lesssim 4.5$ from Quasar Absorption Systems (Damped Lyman-$\alpha$
systems and Lyman limit systems). We found that our C-\LCDM\ models
had no difficulty producing enough cold gas in disks up to $z\sim 4$,
where the observations are the most secure, but the WMAP3 model did
not fare so well. The predicted values of $\Omega_{\rm cold}$ in this
model were too low by a factor of 2--10 at $z \gtrsim 3$.

We analyze the redshift-dependent breakdown of all the baryons in our
fiducial model into each of the various components that we track: hot
gas in halos, warm/hot diffuse gas in the IGM, cold gas in galactic
disks, stars in galaxies, and stars in Diffuse Stellar Halos (DSH). We
find that hot gas in halos and warm/hot gas in the IGM dominate at all
redshifts, in agreement with the predictions of numerical cosmological
simulations \citep[][and references therein]{bertone:08} and with the
observational baryon census at low redshift
\citep{fukugita_peebles:04}. Therefore, in order to produce more cold
gas at high redshift, we either require more efficient cooling of hot
gas or less efficient reheating and ejection of cold gas by supernova
feedback. Since our models are already overproducing stars at high
redshift, the latter is probably a more promising solution. Our models
predict efficient early metal enrichment, with the average stellar
mass-weighted stellar metallicity reaching 25\% of solar at $z\sim6$
and 50\% of solar at $z\sim 2.5$. 

The picture of the cosmic build-up of stars that we see in our models
is interesting in the context of a problem pointed out in several
recent papers
\citep[e.g.][]{hopkins_beacom:06,fardal:07,wilkins:08,dave:08}: when
the best available observational estimate of the dust- and
incompleteness corrected SFR density is integrated, accounting for gas
recycling under the assumption of a universal stellar IMF, the mass of
long-lived stars is overestimated by about a factor of 2--3. These
authors suggest that a possible resolution to this problem is a
non-universal IMF, which was more top-heavy at high redshift. In the
context of the two models we have considered, the C-\LCDM\ models
predict early structure formation, accompanied by a lot of early star
formation. We might be able to reconcile these models with all the
data if in fact the IMF was top-heavy at early times, so that most of
the star formation that we see does not produce long-lived stars. On
the other hand, the WMAP3 model implies later structure formation, and
hence less star formation at high redshift. This provides an alternate
means of reconciling observations of the SF history and stellar mass
assembly history, which requires only that the current observational
estimates of the SFR at $z \gtrsim 2$ are too high by about a factor
of two to three. While the WMAP3 picture seems more attractive in many
respects, it is a concern that the WMAP3 models do not seem to be able
to account for DLAS at $z\gtrsim 3$, again as a consequence of the
reduced small scale power. Better constraints on the amount of cold
gas at high redshift from new facilities such as ALMA could help to
resolve this question.

In agreement with previous work, we have shown that the inclusion of
AGN feedback in semi-analytic models can plausibly solve the
over-cooling problem, the massive galaxy problem, and the star
formation quenching problem in the local Universe --- a huge step
forward. However, we have also shown that several potentially serious
discrepancies still remain, and we have argued that these
discrepancies are not peculiar to our implementation, but are common
to all of the CDM-based semi-analytic models currently on the
market. These discrepancies are connected with low-mass galaxies ($m_*
\lesssim {\rm few} \times 10^{10} \msun$), however, in the picture
that we are developing, small galaxies grow into massive galaxies, and
the growth of galaxies and AGN are intimately interconnected, so these
problems on small-scales may indicate or cause more pervasive
problems. It is likely that these problems are connected to the
modelling of cooling, star formation and/or supernova feedback, or
possibly to the CDM power spectrum on small scales. \emph{Direct} AGN
feedback\footnote{By this we mean feedback by an AGN within the galaxy
itself. Indirect feedback from AGN in external galaxies, e.g. via
pre-heating, may be a promising solution
\citep[e.g.][]{scannapieco:04}} is probably not the solution.

In this paper, we have focussed on predictions of the physical
properties of galaxies at $z\sim0$ and the global evolution of the
major baryonic components of the Universe over time. In a planned
series of papers, we will investigate the predictions of the models we
have presented here for multi-wavelength, observable properties
(e.g. UV through FIR luminosities and colors) of galaxies at low and
high redshift, and examine in more detail the distribution functions
and scaling relations of intrinsic galaxy properties (e.g. stellar
mass, star formation rate, metallicity, etc) at high redshift. In
addition, we will explore the predictions of our models for AGN
properties as a function of redshift and environment, and the
relationship between AGN and their host galaxies. 

\section*{Acknowledgments}
\begin{small}

We would like to thank E. Bell, B. Panter, and D. Schiminovich for
providing us with their data in electronic form. We warmly thank
B. Allgood, E. Bell, J. Bromley, D. Croton, G. de Lucia, A. Dekel,
M. Elvis, S. Faber, A. Fabian, A. Kravtsov, C. Martin, L. Moustakas,
P. Natarajan, H.-W. Rix, S. Trager, R. Wechsler, and A. Walen for
discussions that contributed to this work, and S. Allen, A. Gonzalez,
and S. Zibetti for help interpreting their observational results. We
also thank E. Bell and S. Trager for careful readings of an earlier
draft of the manuscript, and B. Moster for providing us with his
results in advance of publication. RSS thanks the ITC at the CfA for
hospitality. BER gratefully acknowledges support from a Spitzer
Fellowship through a NASA grant administrated by the Spitzer Science
Center. This work was supported in part by a grant from the W.M. Keck
Foundation.

\end{small}

\bibliographystyle{mn} 
\bibliography{mn-jour,agnfb}

\end{document}